\documentclass[oneside,a4paper, reqno,11pt]{amsart}

\usepackage{extsizes}
\usepackage{amsmath, amsfonts, amsthm, amssymb}
\usepackage[english]{babel}

\usepackage[utf8]{inputenc}
\usepackage{lmodern}
\usepackage[x11names]{xcolor} 
\usepackage{graphicx}
\usepackage{float}
\usepackage{braket}
\usepackage[stable, bottom]{footmisc}
\usepackage{wrapfig}
\usepackage{sidecap}
\usepackage{wasysym} 
\usepackage[left=2.75cm,right=2.75cm,bottom=3cm,top=3cm]{geometry}
\usepackage{scalerel}  

\usepackage{hyperref}
\hypersetup{
	colorlinks=true,
	linkcolor=Red4, 
	citecolor=blue
}

\usepackage{mathtools}
\usepackage{subfig}
\usepackage[shortlabels]{enumitem}
\usepackage{caption}
\usepackage{hyperref}
\usepackage{ textcomp }
\usepackage{relsize}

\usepackage{xparse}

\NewDocumentCommand{\overarrow}{O{=} O{\downarrow} m}{%
	\overset{\makebox[0pt]{\begin{tabular}{@{}c@{}}#3\\[0pt]\ensuremath{#2}\end{tabular}}}{#1}
}
\NewDocumentCommand{\underarrow}{O{=} O{\downarrow} m}{%
	\underset{\makebox[0pt]{\begin{tabular}{@{}c@{}}\ensuremath{#2}\\[0pt]#3\end{tabular}}}{#1}
}
\DeclareMathSymbol{\shortminus}{\mathbin}{AMSa}{"39}

\newcommand{\scd}{\mathcal{S}(\C^d)}
\newcommand{\R}{{\mathbb R}}
\newcommand{\C}{{\mathbb C}}
\newcommand{\N}{{\mathbb N}}
\newcommand{\I}{{\mathrm i}}
\newcommand{\aand}{\textrm{\ \  and \  }}
\newcommand{\e}{\operatorname{e}}

\newcommand{\tr}{\operatorname{tr}}

\newcommand{\spann}{\operatorname{span}}
\newcommand{\conv}{\operatorname{conv}}
\newcommand{\diag}{\operatorname{diag}}
\newcommand{\Arg}{\operatorname{Arg}}
\newcommand{\sgn}{\operatorname{sgn}}
\newcommand{\Int}{\operatorname{Int}}

\usepackage{tikz}
\usetikzlibrary{arrows}

\usetikzlibrary{decorations.pathmorphing}
\usetikzlibrary{positioning}
\newcommand{\AxisRotatorj}[1][rotate=0]{
	\tikz [x=0.25cm,y=0.60cm,line width=.2ex,-stealth,#1] \draw (0,0) arc (160:-160:1 and 1);
}
\newcommand{\AxisRotatord}[1][rotate=0]{
	\tikz [x=0.25cm,y=0.60cm,line width=.1ex,-stealth,#1] \draw (0,0) arc (130:-140:1 and 1);
}	
\usepackage{caption}
\usepackage{epstopdf}
\newtheorem{theorem}{Theorem}
\newtheorem*{theorem*}{Theorem}
\newtheorem{corollary}[theorem]{Corollary}
\newtheorem*{corollary*}{Corollary}
\newtheorem{lemma}[theorem]{Lemma}
\newtheorem*{lemma*}{Lemma}

\newtheorem{proposition}[theorem]{Proposition}
\newtheorem{definition}[theorem]{Definition}
\newtheorem*{definition*}{Definition}
\theoremstyle{definition}
\newtheorem{remark}[theorem]{Remark}
\newtheorem*{remark*}{Remark}
\newtheorem{obs}[theorem]{Observation}
\newtheorem*{obs*}{Observation}

\usetikzlibrary{shapes,decorations}
\numberwithin{equation}{section}
\numberwithin{theorem}{section}

\usepackage[backend=bibtex,firstinits=true,citestyle=numeric,bibstyle=numeric,maxbibnames =100,maxcitenames=5,isbn=false]{biblatex}
\renewbibmacro{in:}{}

\makeatletter
\def\l@subsection{\@tocline{2}{0pt}{2.15pc}{5pc}{}}
\makeatother
\addto\captionsenglish{}

\usepackage[foot]{amsaddr}
\makeatletter
\renewcommand{\email}[2][]{%
	\ifx\emails\@empty\relax\else{\g@addto@macro\emails{,\space}}\fi%
	\@ifnotempty{#1}{\g@addto@macro\emails{\textrm{(#1)}\space}}%
	\g@addto@macro\emails{#2}%
}

\begin{document}
	
	\thispagestyle{empty}
	
	\title{   
			Eight types of qutrit dynamics generated by unitary evolution combined with 2\scalebox{0.75}{\raisebox{0.25ex}{+}}1 projective measurement
	}
	
	\author{  
		Anna Szczepanek
	}	
	
	\address{	\scriptsize   Institute of Mathematics, Jagiellonian University, \L ojasiewicza 6, 30-348 Krak\'{o}w, Poland}
	\email{\scriptsize anna.szczepanek@uj.edu.pl}
	\begin{abstract}
		We classify the Markov chains that can be  generated on the 
		set of quantum states by a unitarily evolving $3$-dim quantum system (qutrit) that is repeatedly measured with a projective measurement (PVM) consisting of one rank-$2$ projection and one rank-$1$ projection. 
		The dynamics of such a system can be visualized as taking place on the union of a Bloch ball and a single point, which correspond to the respective projections. 
		The classification is given in terms of the eigenvalues of the $2 \times 2$ matrix that describes the dynamics arising on the Bloch ball, i.e., on the $2$-dim subsystem. We also express this classification as the partition of the numerical range of the unitary operator that governs the evolution of the system. As a result, one can easily determine which of the eight possible chain types can be generated with the help of any given unitary. 
		
		\vspace{1mm}
		
		
		\vspace{1mm}
		

	\end{abstract}
	\maketitle

	\vspace{-2mm}
	
	\section{Introduction}
	\pagestyle{plain}
	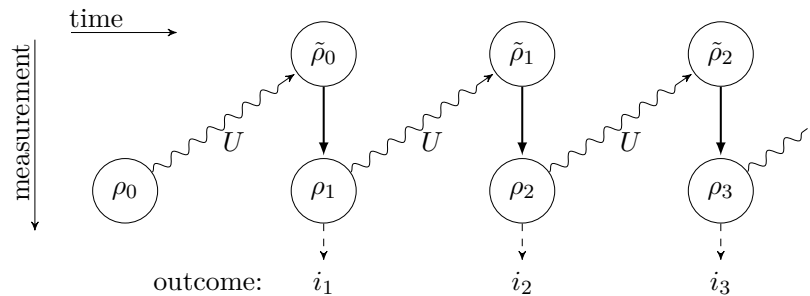
\begin{figure}[b]
		\vspace{-5mm} 
		\captionsetup{width=0.8\linewidth}
		\begin{center}
			\scalebox{0.95}{		\begin{tikzpicture}
					[>=stealth',shorten >=1pt,auto,node distance=0.995cm,
					bubbles/.style={circle,font=\sffamily\normalsize,  minimum size=0.9cm}]
					Bottom
					\node[bubbles] (st33) at (-1.25, 0) { };
					\node[bubbles,draw] (st0) at (1.5, 0) {$\tilde{\rho}_0$};
					\node[bubbles,draw] (st1) at (4.25, 0) {$\tilde{\rho}_1$};
					\node[bubbles,draw] (st2) at (7, 0) {$\tilde{\rho}_2$};
					\node[bubbles] (st3) at (8.6, -0.75) {};
					
					\node[bubbles,draw] (p33) [below=of st33] {${\rho}_0$};
					\node[bubbles,draw] (p0) [below=of st0] {${\rho}_1$};
					\node[bubbles,draw] (p1) [below=of st1] {${\rho}_2$};
					\node[bubbles,draw] (p2) [below=of st2] {${\rho}_3$};
					
					Top
					\node (w0) at (1.5, -3.2)  {$i_1$};
					\node (w1) at (4.25, -3.2) {$i_2$};
					\node (w2) at (7, -3.2) {$i_3$};
					
					\path[->,every node/.style={font=\sffamily\normalsize}]	
					(p33) edge[decorate, decoration={snake, segment length=3mm, amplitude=0.75mm,post length=2mm}] node[below]  {$\ \ U$} (st0)
					(p0) edge[decorate, decoration={snake, segment length=3mm, amplitude=0.75mm,post length=2mm}] node[below]  {$\ \  U$} (st1)
					(p1)  edge[decorate, decoration={snake, segment length=3mm, amplitude=0.75mm,post length=2mm}] node[below]  {$\ \ U$} (st2);
					
					\path[every node/.style={font=\sffamily\normalsize}]	
					(p2)  edge[decorate, decoration={snake, segment length=3mm, amplitude=0.75mm}] node[below]  {} (st3);
					
					\path[->,dashed,every node/.style={font=\sffamily\normalsize}]	
					(p0) edge  (w0)
					(p1) edge  (w1)
					(p2) edge  (w2);
					
					Arrows
					\draw [-latex,thick](st0) -- (p0);
					\draw [-latex,thick](st1) -- (p1);
					\draw [-latex,thick](st2) -- (p2);

					\draw[->] (-2,0.3) -- (-0.5,0.3);
					
					\node (time) at (-1.65, 0.5) {\normalsize{time}};
					\node (res) at (-0.1, -3.175) {\normalsize{outcome:}};
					\node[rotate=90]  (res) at (-2.7, -0.95) {\normalsize{measurement}};
					\draw[->] (-2.5 ,0.2) -- (-2.5,-2.5);
			\end{tikzpicture}}	
		\end{center}
		\vspace{-4mm} 
		\captionsetup{width=0.685\linewidth} 	
		\caption{Repeatedly measured quantum system that between each two consecutive  measurements evolves in accordance to a unitary operator  $U$.  State dynamics  $(\rho_0, \rho_1, \rho_2 \ldots)$ is Markovian.}
		\label{fig1} 
		\vspace{-1mm} 
	\end{figure}

	Consider performing successive (isochronous) measurements  on a $d$-dimensional ($d \geq 2$) quantum mechanical system that  between each two
	consecutive  measurements undergoes deterministic time evolution governed by a unitary operator (Fig. \ref{fig1}). We model the joint evolution of such a system with a \emph{Partial Iterated Function System} (PIFS), which is a notion that slightly (yet significantly) generalizes that of an Iterated Function System (IFS) with place-dependent probabilities. The Markov chain that is generated on the set of quantum states     corresponds to the so-called \emph{discrete quantum trajectories}, see, e.g., \cite{AttPel11,Benetal17,Kum06,Lim10,MaaKum06}. The sequences of measurement outcomes that are emitted by the system need not, however, be Markovian (this was first noted in \cite{BecGra92}), but can be described by a hidden Markov model. The presence of long-term correlations between the outcomes can be interpreted as the system encoding in its current state some information about the outcomes of previous measurements. 
	
	Suppose a unitarily evolving system is repeatedly measured with a rank-1 POVM, i.e., with a measurement that consists of (suitably rescaled) one-dimensional orthogonal projections. Then to each possible outcome there corresponds a single post-measurement state. {(We assume the standard L\"{u}ders instrument is in use.)} 
	It follows that the Markov  chain generated by this system can be easily recovered from the sequences of emitted outcomes. In consequence, symbolic dynamics is Markovian as well, and so long-term correlations cannot  form in the sequences of outcomes.  
	Therefore, for a unitarily evolving and repeatedly measured  quantum system to have potential for information storage it is necessary that the measurement contain operators of ranks higher than one. Then the sequences of outcomes can be non-Markovian  and the system gains the ability to exhibit long-term correlations between outcomes. Crutchfield \& Wiesner dedicated  a series of papers  \cite{CruWie08, Monetal11, Wie10, CruWie08a, CruWie10} to the phenomenon of \emph{intrinsic quantum computation}, i.e., the way in which quantum systems store and manipulate information. Among other things, they investigated a specific example of a  unitarily evolving three-dimensional quantum system (\emph{qutrit})  which is  repeatedly measured with a measurement consisting of two projectors, one of which is of rank two (and so the other necessarily of rank one). 
	
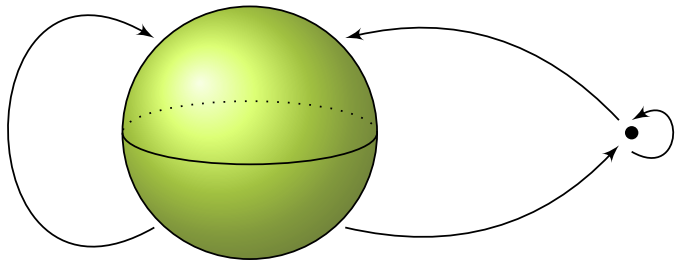
\begin{figure}[b]
		\vspace{-11mm}
		\scalebox{1.675}{
			\begin{tikzpicture}
				\shade[ball color = lime!90, opacity = 0.8] (0,0) circle (1cm);
				\draw (0,0) circle (1cm);
				\draw (-1,0) arc (180:360:1 and 0.25);
				\draw[dotted] (1,0) arc (0:180:1 and 0.25);
				
				\fill[fill=black] (3,0) circle (1.5pt);
				\node (z) at (2.7, -0.1) {};	
				
				\draw[-latex', bend right=30] (2.9,0.1) to node[sloped,bend right=-190,  above, inner sep = 2pt] {} (0.75,0.75);	
				\draw[-latex', bend right=120,  looseness=5] (3,-0.15) to node[bend right=120, inner sep = 2pt,looseness=5] {} (3,0.1);
				\draw[-latex', bend right=30  ] (0.75,-0.75)  to node[sloped,bend right=-190,  above, inner sep = 2pt] {} (2.9,-0.1);	
				\draw[-latex', bend left=120,  looseness=3] (-0.75,-0.75) to node[bend left=120,  inner sep = 2pt,looseness=3] {} (-0.75,0.75);
				
		\end{tikzpicture}		}
		\captionsetup{width=0.7\linewidth}
		\vspace{-10mm}
		\caption{Qutrit measured with a PVM consisting of  one \mbox{rank-$2$} projection and one \mbox{rank-$1$} projection: the ball \&  point system.}\label{oneball}
	\end{figure}

	The class of quantum measurements that consist exclusively of projections (PVMs) is distinguished by the fact that to each measurement outcome there corresponds a distinct subsystem whose 
	dimension is equal to the rank of the projection corresponding to this outcome. In consequence, the dynamics of the system can be thought of as taking place on the union of pairwise disjoint Bloch bodies. 
	In particular, the above-mentioned qutrit system can be visualized as the union of a~Bloch ball (corresponding to the rank-$2$ projection) and a~point (corresponding to the rank-$1$ projection), and so we refer to it as the \emph{ball} \& \emph{point system}, see \mbox{Fig. \ref{oneball}}.

	The aim of this paper is to classify the Markov chains that the ball \& point system can generate on the set of quantum states. We distinguish eight types of chains, and each type is constituted by chains that show qualitatively the same behaviour with respect to the underlying Bloch ball and, in consequence, share the same limiting properties. The resulting classification can serve as a stepping stone to deriving an explicit formula for quantum dynamical entropy of the ball \& point system via Blackwell integral entropy formula  \cite{Black}, see also \cite[Thm. 5.5]{Slo03}.  
	The main step in obtaining this classification is to examine the types of dynamics that arises on the two-dimensional subsystem, i.e., on the Bloch ball, when the deterministic time evolution of the system gets intertwined with the process  of measurement.
	Also, we show how this classification can be transferred onto the numerical  
	range of the unitary operator that governs the time evolution of the  ball \& point  system.  
	As a by-product, it follows that to determine which of the eight possible  types of Markov chains can actually be generated by the ball \& point  systems evolving in accordance to a given unitary, it is sufficient to have a look at the numerical range of this unitary.
	In summary, we obtain the following classifying theorems:
	
	\vspace{-2.0mm}
	
	\begin{enumerate}[leftmargin=*]
		\itemsep=-0.5mm 
		\item[I.] classification of the non-unitary ball dynamics: of its fixed points (Obs.~\ref{fixed}) and generic trajectories (Obs.~\ref{traj});

		\item[II.]  classification of the Markov chains in terms of the eigenvalues of the matrix that describes the ball dynamics (Thm.~\ref{chainsclass});
		
		\item[III.] 
		c lassification of the Markov chains as a partition of the numerical range of the unitary operator that governs the time evolution of the  ball \& point  system (Thm.~\ref{klasyfikacjanatrojkacie}).  
	\end{enumerate}  	
	
	\vspace{-2.0mm}

	This paper is organized as follows. 
	The preliminary Section \ref{secPrel} lays down the basic framework, recalling the mathematical description of quantum theory and providing the definition of Partial Iterated Function Systems.
	In Section \ref{secQPIFS} we define the PIFS that models a unitarily evolving  and  repeatedly measured quantum system. The final  Section \ref{secBall} is dedicated to a thorough study of the ball \& point qutrit system. In Subsection~\ref{secChains} we classify the  types of Markov chains that such a system can generate in terms of the eigenvalues of the matrix describing the dynamics induced by the system on the Bloch ball. 
	In Subsection \ref{secNumrange} we investigate how this classification is reflected in the numerical range of the unitary operator. This numerical range is, generically, a triangle spanned by the eigenvalues of the unitary in question  and its subset corresponding to the \mbox{so-called} elliptic chains  turns out to be contained in a cubic curve. We examine this curve more closely in Subsection~\ref{secEll} and identify it as the Musselman cubic (Remark~\ref{secMuss}). Lastly, in Subsection~\ref{secCW} we discuss the specific unitary operator investigated by Crutchfield \& Wiesner and expand on their results by identifying all types of Markov chains that the ball \& point system governed by this unitary can generate.

	\section{Preliminaries} 
	\label{secPrel} 
	
	\subsection{Quantum states \& measurements}
	Fix $d \geq 2$. The set of $d$-dimensional \emph{quantum states} is defined as
	% \vspace{-0.5mm}
	$
	\mathcal{S}(\C^d) :=  \{\rho \in \mathcal{L}(\C^d)\colon  \rho \geq 0,\  \tr\rho = 1\}$, where $\mathcal{L}(\C^d)$ denotes the space of (bounded) linear maps on $\C^d$. 
	The extreme points of $\mathcal{S}(\C^d)$  form the set $\mathcal{P}(\C^d)$ of \emph{pure states}.  
	It follows that $\mathcal{P}(\C^d)$ is the set of one-dimensional orthogonal projections on $\C^d$ and   \mbox{$\mathcal{S}(\C^d)=\conv\mathcal{P}(\C^d)$}. 
	For ${w} \in \C^{d} \setminus \{0\}$   we let $\rho_w$ denote the pure state corresponding to $w$, i.e., the orthogonal projection on $\spann\{w\}$. Clearly, $\mathcal{P}(\C^d)$ can be put in one-to-one correspondence with rays in $\C^d$, i.e.,  with the complex projective Hilbert space $\mathbb{CP}^{d-1}$.
	
	Let $\mathcal{U}(\C^d)$ denote the set of unitary operators on $\C^d$. Deterministic time evolution of a quantum system  is said to be governed  by $U \in \mathcal{U}(\C^d)$ if it is given by the  \emph{unitary channel}  
	\begin{equation}
		\label{channel}
		\Lambda^U\colon \ 	\mathcal{S}(\mathbb{C}^{d})\ni \rho\longmapsto U\rho\,U^{\ast
		}\in\mathcal{S}(\mathbb{C}^{d}).
	\end{equation}
	Unitary channels are examples of \emph{state  automorphisms}, which, following 
	Kadison's approach \cite{kadison1965transformations},
	we define as affine bijections on $\scd$, see also \cite[Sec. 5.3]{dell2015lectures}.  
	These transformations  represent symmetries in quantum formalism, i.e., describe freedom in choosing a particular mathematical representation of physical objects. 
	
	Another class of maps that give rise to state automorphisms is that of antiunitaries. Recall that $W\colon \C^d \to \C^d$ is called an \emph{antiunitary operator} if   $W(u + \gamma v) = Wu + \overline{\gamma}Wv$ and $\braket{Wu|Wv} =\braket{v|u}$ for all $u, v \in \C^d$, $\gamma \in \C$. We let $\overline{\mathcal{U}}(\C^d)$ stand for the set of antiunitary operators on $\C^d$. For  $W \in \overline{\mathcal{U}}(\C^d)$ we define its adjoint $W^*$ via $\braket{u|W^*v} = \overline{\braket{Wu|v}}$, where $u, v \in \C^d$. One can show that  $W^*\in \overline{\mathcal{U}}(\C^d)$ as well as $WW^{*}=\mathbb{I}_d$.
	It follows that 
	$$\Lambda^W\colon \ \scd \ni \rho \longmapsto W\rho\, W^* \in \scd$$  
	is indeed a~state automorphism.\footnote{In contrast to unitary channels, state automorphisms induced by antiunitary operators are not completely positive. Hence, they 
		describe symmetries that are physically unrealisable, e.g., {time reversal}.} It is a well-known result by Kadison that  there are no other state automorphisms but those induced by  unitary or antiunitary operators, see, e.g.,  \cite[p.~101]{dell2015lectures} or \cite[Thm. 2.63]{HeiZim11}.

	\smallskip
	
	Let $k \in \N$ and put $I_k:=\{1, \ldots, k\}$. 
	The \textsl{measurement} of a $d$-dimensional quantum system with 
	$k$ possible outcomes is given
	by a \textsl{positive operator valued measure} (\textsl{POVM}), i.e., a 
	set  of positive semi-definite (non-zero) Hermitian operators $\Pi_{1}, \ldots, \Pi_k$ 
	on $\mathbb{C}^{d}$ that sum to the identity operator, i.e., \begin{equation}
		\label{sumid}\sum
		\limits_{i =1}^k\Pi_{i}=\mathbb{I}_d.\end{equation}
	We say that $\Pi$ is a \emph{projection valued measure} (\emph{PVM})  or a \emph{\mbox{L\"{u}ders--von Neumann} measurement$\,$} \cite{Luders} if  $\Pi_i$  is a projection for every $i\in I_k$. We then have $k \leq d$  and the projections constituting $\Pi$ are necessarily orthogonal as self-adjoint projections on a Hilbert space. 
	Moreover, they are mutually orthogonal, i.e.,   $\Pi_i\Pi_j=0$ for $i,j \in I_k$,  $i \neq j$ \cite[p. 46]{halmos1957}.

	If the state of the system before the measurement  
	is $\rho\in\mathcal{S}(\mathbb{C}^{d})  $, then the \emph{Born rule} \label{BORNRULE} dictates that the probability
	of obtaining the $i$-th outcome is equal to 
	$\operatorname{tr}( \Pi_{i} \rho) 
	$, where $i\in I_k$ \cite{Born1926}. 
	Generically, the measurement process alters the state
	of the system, but the POVM alone is not sufficient to determine the
	post-measurement  state. This can be done by defining a
	\textsl{measurement instrument} (in the sense of Davies and Lewis
	\cite{DavLew70}) compatible with $\Pi$, see also \cite{compendium}, \cite[Ch. 10]{busch2016}, \cite[Ch. 5]{HeiZim11}. 
	Among infinitely many instruments generating the same
	measurement statistics we only consider here  the so-called \textsl{generalised L\"{u}ders instruments}, disturbing the initial state in the minimal way, where the  state transformation reads   
	%\vspace{1mm}
	\begin{equation} 
		% 	\vspace{1mm}
		\label{instr}\rho \longmapsto \frac{\sqrt{\smash[b]{\Pi_i}}\rho \sqrt{\smash[b]{\Pi_i}}}{\tr(\Pi_i\rho)},\end{equation}
	provided that the measurement has yielded the result $i \in I_k$ 
	\cite[p. 404]{DecGra07}, see also  \cite{Barnum2, Barnum}.

	\subsection{Partial Iterated Function Systems} 
	
	Recall that   $I_k:=\{1, \ldots, k\}$. By  $S_k$ we denote the set of all permutation on $I_k$. Let $X$ be an arbitrary set.
	\begin{definition}\textrm{\rm\cite[p. 59]{Slo03}} 
		The triple $(X, \mathsf{F}_i, \mathsf{p}_i)_{i \in I_k}$ is called  a \emph{Partial Iterated Function System (PIFS)} on $X\!$ if 
		$\,\mathsf{p}_i\colon  X \to [0,1]$,   $\sum_{j \in I_k}\mathsf{p}_j=1$, and  $\mathsf{F}_i\colon  \{x \in X \colon  \mathsf{p}_i(x)> 0\} \to X$, where $i \in I_k$. We call $\,\mathsf{F}_1, \ldots, \mathsf{F}_k$ \emph{evolution maps}.
	\end{definition}

	\noindent
	The action of a PIFS transforms a given initial state $x \in X$ into a new state  $\mathsf{F}_i(x)$ with (place-dependent) probability $\mathsf{p}_i(x)$, and the symbol 
	$i$ 
	corresponding to this evolution is then emitted ($i \in I_k$). 
	Thus, the repeated action of a PIFS generates a Markov chain on~$X$ and 
	yields sequences of symbols from $I_k$, which can be modelled by a hidden Markov chain. The 
	probabilities and evolution maps corresponding  to the strings of symbols are defined inductively in the following natural way. 
	Let $n \in \mathbb{N}$, $\iota := (i_1, \ldots, i_n) \in I_k^{n}$ and $j \in I_k$. For $n=1$ both~$\mathsf{p}_\iota$~and~$\mathsf{F}_\iota$ are given. The probability of the system 
	outputting 
	$\iota j:= (i_1, \ldots, i_n, j) \in I_k^{n+1}$ is defined  as  
	\begin{equation} 
		\label{genprob} \mathsf{p}_{\iota j}(x):=
		\left\lbrace\begin{array}{ll}
			\mathsf{p}_j(\mathsf{F}_{\iota}(x))\mathsf{p}_{\iota}(x) & \textrm{ if }\  \mathsf{p}_{\iota}(x) > 0 \\[0.2em]
			0 & \textrm{ if }\ \mathsf{p}_{\iota}(x) = 0	\end{array}\right.
	\end{equation}
	and the corresponding evolution map is defined as 
	$ \mathsf{F}_{\iota j}(x):=\mathsf{F}_j(\mathsf{F}_{\iota} (x))$ if  $\mathsf{p}_\iota(x) >0$.

	For PIFSs acting on the set of quantum states  we have a natural notion of conjugacy defined with the help of state automorphisms:
	\begin{definition} 
		Let $\mathcal{F}=(\scd, \mathsf{F}_i, \mathsf{p}_i)_{i\in I_k}$ and $\mathcal{\widetilde{F}}=(\scd, \tilde{\mathsf{F}}_i, \tilde{\mathsf{p}}_i)_{i\in I_k}$ be PIFSs. Also, let $V \in \mathcal{U}(\C^d) \cup\,  \overline{\mathcal{U}}(\C^d)$. 
		We say that \emph{$\mathcal{\widetilde{F}}$ is  $V$-conjugate to $\mathcal{F}$} if	there exists $\sigma \in S_k$ such that    
		$$\tilde{\mathsf{F}}_i = \Lambda^{V^*} \circ {\mathsf{F}}_{\sigma(i)}\circ \Lambda^V\  \aand\ \  \tilde{\mathsf{p}}_i = \mathsf{p}_{\sigma(i)}\circ \Lambda^V$$   
		for every $i \in I_k$, where $\Lambda^V$  is the state automorphism induced by $V$, i.e., $\Lambda^V(\rho)= V\rho\, V^*$ for $\rho \in  \scd$.
		When there is no need to specify the conjugating map,  we simply say that 
		\emph{$\mathcal{F}$ and $\mathcal{\widetilde{F}}$ are conjugate} and denote this by $\mathcal{F} \sim \mathcal{\widetilde{F}}$. 
	\end{definition}

	The notion of a PIFS generalizes, slightly but significantly, that of an Iterated Function System (IFS) with place-dependent probabilities (see, e.g., \cite{Barnsley1988, peigne1993iterated}) by allowing each evolution map  to remain undefined on the states that have zero probability of being subject to the action of this map. Such a generalization is necessary in considering quantum measurements because the state transformation associated with a given measurement  outcome 
	cannot be defined on the states with zero probability of producing this outcome, see \eqref{instr}.
	IFSs acting on the set of pure quantum states have been examined in the framework of Event Enhanced Quantum Theory (EEQT) \cite{Jadczyk1, Jadczyk2, Jadczyk3}; in particular, IFSs acting on the Bloch sphere were investigated in \cite{Jadczyk7, Jastrzebski}, see also \cite{Jadczyk4, Jadczyk6, Jadczyk5}. A generalization to systems that act in the space of all quantum states was proposed in \cite{Lozinski}.

	\section{Evolution \& measurement quantum PIFSs} 
	\label{secQPIFS}

	Fix a POVM $\Pi=\{\Pi_1, \ldots, \Pi_k\}$ and $U\! \in \mathcal{U}(\C^d)$. In what follows we define the PIFS corresponding to a quantum system that evolves in accordance to $U$ and is repeatedly measured with  $\Pi$. Let $i \in I_k$.
	Taking into account the Born rule and the unitary evolution prior to the measurement, for $\rho \in \mathcal{S}(\C^d)$  we define  the probability of obtaining the outcome $i$ as 
	\vspace{-1mm}
	\begin{equation*}
		\label{pifsp}
		\mathsf{p}_i(\rho):= \tr(\Pi_iU\rho\, U^*),
	\end{equation*} and the evolution map $\mathsf{F}_i$ is defined as the composition of the unitary channel \eqref{channel} with the state transformation due to $\Pi$ described in \eqref{instr}, i.e.,  
	\vspace{1.5mm}
	\begin{equation*}
		\vspace{0.75mm}	
		\label{pifsF}\mathsf{F}_i(\rho):= \frac{\sqrt{\smash[b]{\Pi_{i}}}\,U\rho\, U^*\sqrt{\smash[b]{\Pi_{i}}}}{\tr(\Pi_iU\rho\, U^*)},
	\end{equation*}
	provided that $\mathsf{p}_i(\rho)>0$.
	Clearly,  
	$(\mathcal{S}(\mathbb{C}^{d}), \mathsf{F}_i, \mathsf{p}_i)_{i \in I_k}$ is a PIFS. We shall denote it by $\mathcal{F}_{U,\,\Pi}$ and refer to it as the \emph{PIFS generated by $U\!$ and $\Pi$}. 
	
	The following properties of $\mathcal{F}_{U,\,\Pi}$ will come in handy later on. Let $i \in I_k$. 
	First, we note that  $\mathcal{F}_{U,\,\Pi}$ preserves~$\mathcal{P}(\C^d)$.
	Indeed, let $w$ be a unit vector in $\C^d$ and consider $\rho_w$. It follows that $\mathsf{p}_i(\rho_w)= ||\sqrt{\smash[b]{\Pi_{i}}}Uw||^2$ and, provided that  ${\sqrt{\smash[b]{\Pi_{i}}}\,Uw } \neq 0$, we obtain   
	\begin{equation}\label{pure} 
		\mathsf{F}_i(\rho_w) = \rho_{{\sqrt{\smash[b]{\Pi_{i}}}\,Uw }}.
	\end{equation} 
	Secondly, for $\rho \in \mathcal{S}(\C^d)$  we put 
	% \vspace{-0.5mm}
	$$
	% 	\vspace{-0.5mm}
	\Lambda^{U,\, \Pi}_i(\rho) := \sqrt{\smash[b]{\Pi_{i}}}U\rho \,  U^*\sqrt{\smash[b]{\Pi_{i}}}
	$$
	and observe that  	 
	$\mathsf{p}_i(\rho)$ and $ \mathsf{F}_i(\rho)$
	can be recovered from the value of $\Lambda^{U,\, \Pi}_i(\rho)$ as, respectively, 
	$\tr(\Lambda^{U,\, \Pi}_{i}(\rho))$ and 
	$\Lambda^{U,\, \Pi}_{i}(\rho)/\tr(\Lambda^{U,\, \Pi}_{i}(\rho))$. Hence, it comes as no surprise that PIFS conjugacy can be expressed as follows: 
	\begin{obs}\label{obsconj} Let $U, \tilde{U} \in \mathcal{U}(\C^d)$ and let  $\Pi, \tilde{\Pi}$ be POVMs. 
		Then 
		$\mathcal{F}_{\tilde{U}, \, \tilde{\Pi}}$ is $V$-conjugate to $\mathcal{F}_{U,\,\Pi}$ if and only if  there exists $\sigma \in S_k$ such that  $$\Lambda^{\tilde{U}, \, \tilde{\Pi}}_i =\Lambda^{V^*} \circ \Lambda^{U,\, \Pi}_{\sigma(i)}  \circ \Lambda^V$$ for every $i \in I_k$, where $V \in \mathcal{U}(\C^d) \cup\, \overline{\mathcal{U}}(\C^d)$.
	\end{obs}

	Since physical systems are invariant under (anti)unitary change of coordinates and under phase transformation, one expects both these operations to lead to PIFSs conjugate to the initial one.
	
	\begin{proposition}\label{exconjpifs} We have 
		%	\vspace{-1mm}
		\begin{enumerate}[\rm (i)]
			\item 
			$\mathcal{{F}}_{\e^{\I\gamma}\!{U},\, \Pi} \sim \mathcal{F}_{U,\, \Pi}\:$  for  $\gamma \in \R$;
			
			\item 	$ \mathcal{{F}}_{VUV^*,\, {V\Pi V^*}}  \sim   \mathcal{F}_{U,\, \Pi}\: $  for $V\! \in \mathcal{U}(\C^d) \cup\, \overline{\mathcal{U}}(\C^d)$.
		\end{enumerate}
	\end{proposition}
	\begin{proof}  
%		\phantom{x}
		\begin{enumerate}[(i)]
			\item 	
			Let $\gamma \in \R$ and note that      ${\Lambda}_j^{\e^{\I\gamma}\!U, \,\Pi} = {\Lambda}_j^{U, \:  \Pi}$ for every $j \in I_k$, which, via Obs. \ref{obsconj}, gives	$\mathcal{{F}}_{\e^{\I\gamma}\!{U},\, {\Pi}} \sim \mathcal{F}_{U,\, \Pi}$, as desired.
			
			\item Let $V\! \in \mathcal{U}(\C^d) \cup \overline{\mathcal{U}}(\C^d)$. Put $\tilde{U}:=VUV^*$ and $\tilde{\Pi}:={V\Pi V^*}$, i.e.,  $\tilde{\Pi}_j={V\Pi_j V^*}$ for $j \in I_k$.
			Since $\sqrt{\smash[b]{V\vphantom{^I}\Pi_j V^*}}  = V\sqrt{\smash[b]{\Pi\vphantom{^I}_j}}V^*$,  for every $j \in I_k$  we obtain 
			%	 		\vspace{-1mm}
			\begin{align*}  
				%		 			\vspace{-1mm}
				{\Lambda}_j^{\tilde{U},\, \tilde{\Pi}}(\rho)  
				&
				= 	\sqrt{\smash[b]{ \tilde{\Pi}_{j}}} \tilde{U} \, \rho\: \tilde{U}^* \sqrt{\smash[b]{\tilde{\Pi}_{j}}}	
				\\[0.25em]
				&
				= (V \sqrt{\smash[b]{\Pi_{j}}}^{\phantom{.}} V^*) ( V U  V^*)  \rho \, (V U^* V^*)  (V \sqrt{\smash[b]{\Pi_{j}}} V^*)
				\\[0.25em]& = 
				V  \sqrt{\smash[b]{\Pi_{j}}} U V^*\rho\, V U^* \sqrt{\smash[b]{\Pi_{j}}}  V^*
				\\[0.25em]& = \Lambda^{V}(\Lambda^{U, \,\Pi}_j(\Lambda^{V^*}(\rho))).
			\end{align*}
			It follows from Obs. \ref{obsconj} that   $\mathcal{{F}}_{\tilde{U},\, \tilde{\Pi}}$ is $V^*$-conjugate to $\mathcal{F}_{U,\, \Pi}$, which concludes the~proof. 
			\qedhere
		\end{enumerate} 
	\end{proof}

	From now on we fix the \emph{maximally mixed state} $\rho_*:=\mathbb{I}_d/d$ as the initial state of our system.  The \emph{Markov chain generated by $\mathcal{F}_{U,\, \Pi}$} has the state space $$\mathsf{S}_*:=\bigcup\limits_{n \in \mathbb{N}}\bigcup\limits_{\iota \in I_k^n} \big\{\mathsf{F}_\iota(\rho_*)\colon \mathsf{p}_\iota(\rho_*)>0\big\},$$ its initial distribution $({\pi_*}(\rho)\colon \rho \in \mathsf{S}_*)$ is given by  
	$$
	{\pi_*}(\rho):= \left\lbrace \begin{array}	{ll}
		\mathsf{p}_i(\rho_*)	  & \textrm{if}\  \rho=\mathsf{F}_i(\rho_*) \textrm{ for some } i\in I_k, \\[0.2em]
		0 	 & \textrm{otherwise,}
	\end{array}\right.$$
	and the {transition matrix $\mathsf{P}_* \in \R^{\mathsf{S}_* \times \mathsf{S}_*}$}  reads  \[\mathsf{P}_*(\rho, \varrho):=\hspace{-1mm}\sum_{\substack{i \in {I}_{k}  \\  \mathsf{F}_{i}(\rho) =\varrho}}\hspace{-1mm}\mathsf{p}_i(\rho)\ \ \ \textrm{for}\ \ \ \rho, \varrho \in \mathsf{S}_*.\]

	In what follows we discuss in more detail the Markov chain generated by $\mathcal{F}_{U,\, \Pi}$ when $\Pi=\{\Pi_1, \ldots, \Pi_k\}$ is a PVM.  
	Clearly, the state space $\mathsf{S}_*$ is contained in the union of images of the evolution maps. We therefore let $j \in I_k$ and discuss the domain and image of $\mathsf{F}_j$. We adopt the following notation. 
	For $\Theta$  being a non-trivial $r$-dimensional subspace of $\C^d$ we put
	%  \vspace{-1mm}
	$\mathcal{S}( \Theta):=\{\rho \in   \mathcal{S}( \C^d)\colon \operatorname{im}\rho \subset \Theta\}
	$ and $\mathcal{P}( \Theta):=\mathcal{S}( \Theta) \cap \mathcal{P}( \C^d)$.
	Note  that   $\mathcal{S}( \Theta)=\conv \mathcal{P}( \Theta)$ and $\mathcal{P}( \Theta) \sim \mathbb{P}\Theta$. 
	Moreover, we can identify $\rho \in S(\Theta)$ with  $\rho|_{\Theta}\colon \Theta \to \Theta$, so   $\mathcal{S}( \Theta)\sim\mathcal{S}( \C^{r})$
	as well as $\mathcal{P}( \Theta) \sim \mathcal{P}( \C^r) \sim \C\mathbb{P}^{r-1}$.

	\begin{proposition}  \label{kerf}
		We have 	$\operatorname{dom}\mathsf{F}_j= \mathcal{S}(\C^d)\setminus \mathcal{S}(U^*\operatorname{ker}\Pi_j)$.
	\end{proposition}
	\begin{proof}
		Let   
		$\rho \in \scd$. 
		By spectral decomposition, there exist  
		$\mu_i \in [0,1]$ and  unit vectors  $w_i \in \C^d$, where $i\in  I_d$, such that    $\sum_{i = 1}^d \mu_i=1$   and  	 $\rho=\sum_{i = 1}^d \mu_i\rho_{w_i}$.
		It follows that 
		\begin{equation}\label{probdom}
			\mathsf{p}_{j}(\rho)=\sum_{i = 1}^d\mu_i\tr(\Pi_jU\rho_{w_i}U^*)
			=\sum_{i = 1}^d\mu_i ||\Pi_jUw_i||^2.
		\end{equation} 	
		In consequence, $\mathsf{p}_{j}(\rho)=0$ iff  for each $i \in I_d$ we have 
		$w_i\in \operatorname{ker}(\Pi_jU)$ or $\mu_i= 0$.
		Since  
		$\operatorname{ker}(\Pi_jU)=	U^*\operatorname{ker}\Pi_j$, we conclude that $\rho \notin \operatorname{dom}\mathsf{F}_j$ iff    $\rho_{w_i} \in  \mathcal{P}(U^*\operatorname{ker}\Pi_j)$ for each $i \in I_d$ such that $\mu_i \neq 0$, which is equivalent to $\rho \in \conv \mathcal{P}(U^*\operatorname{ker}\Pi_j)=\mathcal{S}(U^*\operatorname{ker}\Pi_j)$, as desired.
	\end{proof}

	\begin{proposition} 
		\label{imf} 
		We have 	$\operatorname{im}\mathsf{F}_j = \mathcal{S}(\operatorname{im}\Pi_j)$  
		and $\operatorname{im}(\mathsf{F}_j|_{\mathcal{P}(\C^d)}) = \mathcal{P}(\operatorname{im}\Pi_j)$.
	\end{proposition}
	
	\begin{proof} 
		
		Since 
		$\operatorname{im}(\mathsf{F}_j|_{\mathcal{P}(\C^d)}) \subset \mathcal{P}(\operatorname{im}\Pi_j)$, see \eqref{pure}, it follows easily (via spectral decomposition) that   
		$\operatorname{im}\mathsf{F}_j \subset \mathcal{S}(\operatorname{im}\Pi_j)$.  
		To prove that the converse inclusions also hold, we first show  that $\mathsf{F}_j(\rho) = U \rho\, U^*$ if $\rho \in \mathcal{S}(U^*\operatorname{im}\Pi_j )$, i.e., 
		the action of $\mathsf{F}_j$  on  $\mathcal{S}(U^*\operatorname{im}\Pi_j )$  coincides with that of the unitary channel $\Lambda^U$.  	
		Let $\rho \in \mathcal{S}(U^*\operatorname{im}\Pi_j)$. Clearly, $\rho \in \operatorname{dom}\mathsf{F}_j$.	 Put $r_j:=\operatorname{rank}\Pi_j$. By spectral decomposition,   $\rho=\sum_{i=1}^{r_j}\mu_i\rho_{w_i}$  with unit vectors $w_i \in U^*\operatorname{im}\Pi_j$ and 
		$\mu_i \in [0,1]$ such that  $\sum_{i=1}^{r_j} \mu_i=1$.    	
		For each $i \in I_{r_j}$ we have $Uw_i \in \operatorname{im}\Pi_j$,   so  $\Pi_jUw_i=Uw_i$. Hence, 
		$\mathsf{p}_{j}(\rho) =	\sum_{i=1}^{r_j}\mu_i ||Uw_i||^2=1,
		$ see \eqref{probdom}, and so 
		$\mathsf{F}_j(\rho)=
		\sum_{i=1}^{r_j}\mu_i U\rho_{w_i} U^* 
		=U\rho\, U^*$, as desired.
		It remains to observe that 
		$ \label{bijj}  
		\mathcal{S}(U^* \operatorname{im} \Pi_j) \ni \rho   \mapsto \mathsf{F}_j(\rho)=U\rho\,U^* \in  \mathcal{S}(\operatorname{im} \Pi_j)$
		% \end{equation*} 
		is bijective and both it and its inverse preserve pure states. Thus,  $\mathcal{S}(\operatorname{im}\Pi_j)=\operatorname{im}(\mathsf{F}_j|_{\mathcal{S}(U^* \operatorname{im} \Pi_j )}) \subset \operatorname{im}\mathsf{F}_j$
		and 
		$ \mathcal{P}(\operatorname{im}\Pi_j)=\operatorname{im}(\mathsf{F}_j|_{\mathcal{P}(U^* \operatorname{im} \Pi_j )}) \subset \operatorname{im}(\mathsf{F}_j|_{\mathcal{P}(\C^d)})$, which concludes the proof.
	\end{proof} 
	Hence, as $\operatorname{im}\Pi_j \sim \C^{r_j}$,  we have 
	$\operatorname{im}\mathsf{F}_j \sim  \mathcal{S}(\C^{r_j})$, where $r_j:=\operatorname{rank}\Pi_j$ and $j \in I_k$. 
	Moreover,  $\operatorname{im}\Pi_1, \ldots, \operatorname{im}\Pi_k$ are pairwise orthogonal, so $\operatorname{im}\mathsf{F}_1, \ldots, \operatorname{im}\mathsf{F}_k$ are pairwise disjoint. Therefore,   $\mathsf{S}_*$ 
	is a countable subset of the union of pairwise disjoint Bloch bodies whose {dimensions are determined by the ranks} of the projections constituting $\Pi$.  For instance,   a~three-dimensional quantum system (\emph{qutrit}) measured with a PVM 
	consisting of one \mbox{rank-$2$} projection and one \mbox{rank-$1$} projection   can be visualized as  the union of a~Bloch ball and a~point.

	As for the initial distribution $\pi_*$ of this chain, in the first step of the evolution  we obtain  
	%\vspace{-1mm}
	\begin{equation}\label{firststepppp}
		\mathsf{p}_j(\rho_*)=\operatorname{tr}(\Pi_j\rho_*) = {r_j}/{d}\  \aand\  \mathsf{F}_j(\rho_*)={\Pi_j}/{r_j} \sim {\mathbb{I}_{r_j}}/{r_j},\end{equation} 
	where again  $r_j:=\operatorname{rank}\Pi_j$ for  $j \in I_k$. It follows that 
	${\pi_*}(\rho)= r_i/d$ if $\rho=\Pi_i/r_i$ for some $i \in I_k$, and 
	${\pi_*}(\rho)=0$ otherwise. 
	That is, a projective measurement performed on the maximally mixed state yields the locally maximally mixed state of the subsystem corresponding to  the measurement result with probability proportional to the dimension of this subsystem.

	\section{Ball \& point qutrit system}\label{secBall}

	\subsection{Classification of Markov chains}\label{secChains}
	Fix $U \in \mathcal{U}(\C^3)$ and a PVM $\Pi=\{\Pi_1, \Pi_2\}$ such that $\operatorname{rank}\Pi_1=2$,   $\operatorname{rank}\Pi_2=1$.
	In this section, we classify the Markov chains that $\mathcal{F}_{U,\, \Pi}$ can generate on $\mathcal{S}(\C^3)$.

	Let $\mathsf{z}$ stand for a unit vector from $\C^3$ such that $\operatorname{im}\Pi_2=\operatorname{span}{\{\mathsf{z}\}}$, i.e., $\operatorname{im}\mathsf{F}_2=\{\rho_\mathsf{z}\}$.
	Clearly, we have $\Pi_1=\mathbb{I}_3-\rho_\mathsf{z}$,  so $\Pi=\Pi(\mathsf{z})$ is  fully determined by $[\mathsf{z}]\in \mathbb{CP}^2$.  Putting $\Theta:=\spann\{\mathsf{z}\}^\bot$, we see that  $\Theta  =\operatorname{im}\Pi_1=\operatorname{ker}\Pi_2$ and $\spann\{\mathsf{z}\}=\operatorname{im}\Pi_2=\operatorname{ker}\Pi_1$; hence,  Propositions~\ref{kerf} \& \ref{imf} give  $\operatorname{dom}\mathsf{F}_1=\mathcal{S}(\C^3) \setminus \{\rho_{U^*\mathsf{z}}\}$ and $\operatorname{dom}\mathsf{F}_2 =\mathcal{S}(\C^3) \setminus \mathcal{S}(U^*\Theta)$, as well as  
	$\operatorname{im}\mathsf{F}_1=\mathcal{S}(\Theta) \sim \mathcal{S}(\C^2)$ and
	$\operatorname{im}\mathsf{F}_2=\{\rho_\mathsf{z}\}$.
	Thus, the dynamics of the system  takes place on the union of the Bloch ball and the point   representing $\operatorname{im}\mathsf{F}_1$ and $\operatorname{im}\mathsf{F}_2$,~respectively. 
	
\begin{figure}[h]
		\scalebox{1.00}{		\begin{tikzpicture}
				\shade[ball color = lime!90, opacity = 0.7] (0,0) circle (2cm);
				\draw (0,0) circle (2cm);
				\draw (-2,0) arc (180:360:2 and 0.5);
				\draw[dotted] (2,0) arc (0:180:2 and 0.5);

				\fill[fill=black] (4,0) circle (2.25pt);
				\fill[fill=black] (2,2.2) circle (2.25pt);
				
				\draw [->,blue, dashed,>=stealth'] (1.9,2.09) -- (0.1,0.1) node[pos=0.7, above, black] {$\frac 23$};
				\draw [->,blue, dashed,>=stealth'] (2.06,2.09) -- (3.9,0.1) node[pos=0.7, above, black] {$\frac 13$};
				
				\node (m) at (-0.3, -0.1) {$\rho_{\mathsf{m}}$};
				\node (z) at (3.7, -0.1) {$\rho_{\mathsf{z}}$};
				\node (gw) at (2.3, 2.5) {$\rho_*$};

				\fill[fill=black] (0,0) circle (1.75pt);
				
		\end{tikzpicture}	}
		%  \vspace{-1mm}	
		\caption{First step of the evolution of the ball \& point   system}	\label{firststep}
		%	 \vspace{-1mm}	
	\end{figure}
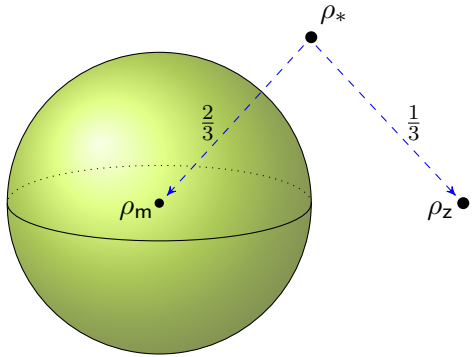

	We put  $\rho_{\mathsf{m}}$   for $\frac 12  \Pi_1 \sim \frac 12 \mathbb{{I}}_2$, i.e., $\rho_{\mathsf{m}}$ is the maximally mixed state of $\mathcal{S}(\Theta)$, and so it occupies the centre of the Bloch ball representing $\mathcal{S}(\Theta)$. 
	In the first step of the evolution of this system we obtain $\mathsf{p}_1(\rho_*)= \frac 23$ and $\mathsf{p}_2(\rho_*)= \frac 13$, while the post-measurement states read 
	$\mathsf{F}_1(\rho_*)=\rho_{\mathsf{m}}$ and $\mathsf{F}_2(\rho_*)=\rho_\mathsf{z}$, see \eqref{firststepppp} and Fig. \ref{firststep}. Consequently,  the initial distribution of the chain generated by $\mathcal{F}_{U,\,\Pi}$ reads  $\pi_* =\frac 23\delta_{\rho_{\mathsf{m}}}+\frac13\delta_{\rho_{\mathsf{z}}}$, where $\delta_{\rho}$ is the Dirac-delta measure   at $\rho \in \mathcal{S}(\C^d)$.

	In what follows, we show that the state space of this chain has a~fairly simple structure; namely, it 
	consists of the states generated by iterating  $\mathsf{F}_1$ on $\rho_{\mathsf{z}}$ and on $\rho_{\mathsf{m}}$.  
	Indeed, we obviously have   $\{\mathsf{F}_1(\rho_*),\mathsf{F}_2(\rho_*)\}=\{\rho_{\mathsf{m}},\rho_{\mathsf{z}}\}$; let us therefore consider $\iota=(i_1, \ldots, i_n)\in  {I}^n_2$ such that $\mathsf{p}_\iota(\rho_*)>0$, where $n \geq 2$. 
	There are three cases to be considered, depending on the presence and position of `2' in $\iota$. Firstly, if $i_{n}=2$, then
	%	\vspace{-1mm} 
	$
	% 	 \vspace{-1mm} 
	\mathsf{F}_\iota(\rho_*)=\mathsf{F}_2\circ \mathsf{F}_{(i_1, \ldots, i_{n-1})}(\rho_*)=\rho_{\mathsf{z}}.$ \pagebreak  
	Secondly, if there exists $j \in I_{n-1}$ such that $i_{j}=2$ and $i_{l}=1$ for every $l \in \{j+1, \ldots, n\}$, i.e., we have
	$\iota=(i_1, \ldots, i_{j-1}, 2, \underbrace{ 1, \ldots, 1}_{n-j})$,
	then  
	% 	\vspace{-1mm} 
	\[
	% 	\vspace{-1mm} 
	\mathsf{F}_\iota(\rho_*)=\mathsf{F}_{{(\underbrace{\scriptstyle 1, \ldots, 1}_{n-j})}}\circ \mathsf{F}_2\circ \mathsf{F}_{(i_1, \ldots, i_{j-1})}(\rho_*)= \mathsf{F}_{\textbf{1}^{n-j}}(\rho_{\mathsf{z}}),\] 
	where $\textbf{1}^r$ denotes the string consisting of $r$ consecutive 1's, i.e., 
	$\textbf{1}^r:=({1}, \ldots, 1) \in {I}^r_2$, $r \in \N$.  
	The last case is that of  $\iota=\textbf{1}^n$, which gives  $\mathsf{F}_{\iota}(\rho_*)= \mathsf{F}_{\textbf{1}^{n-1}}(\rho_{\mathsf{m}})$.
	It follows that  
	% 	 	\vspace{-1mm}
	$$
	%  		\vspace{-1mm}
	\mathsf{S}_*= \{\rho_{\mathsf{z}},\rho_{\mathsf{{m}}}\}\cup  \{\mathsf{F}_{\textbf{1}^r}(\rho_{\mathsf{z}}) \colon 
	\mathsf{p}_{\textbf{1}^r}(\rho_{\mathsf{z}}) >0,\: r \in \N\} \cup  \{\mathsf{F}_{\textbf{1}^r}(\rho_{\mathsf{m}}) \colon \mathsf{p}_{\textbf{1}^r}(\rho_{\mathsf{m}}) >0,\: r \in \N\}.$$
	{Let $\mathfrak{n}(\rho)$   denote the maximum number of iterations of  $\mathsf{F}_1$  that can be applied to $\rho \in \mathcal{S}(\C^d)$. If there exists $s\in \mathbb{N}$ such that 
		$\mathsf{p}_{\textbf{1}^s}(\rho) = 0$, then   $\mathsf{p}_{\textbf{1}^q}(\rho) = 0$ for every $q>s$. In this case we set   
		$\mathfrak{n}(\rho):=
		\max\{r \in \N \colon \mathsf{p}_{\textbf{1}^r}(\rho) > 0\}
		$; otherwise, we set  $\mathfrak{n}(\rho):=\infty$.} 
	We obtain 
	% 	 	\vspace{-4mm}
	\begin{equation}\label{statespace}
		%  	\vspace{-1mm} 
		\mathsf{S}_{*}= \left\lbrace\mathsf{F}^n_{1}(\rho_{\mathsf{z}})\colon n=0, \ldots, \mathfrak{n}(\rho_{\mathsf{z}})\right\rbrace \cup  \left\lbrace\mathsf{F}^n_{1}(\rho_{\mathsf{m}})\colon n=0, \ldots, \mathfrak{n}(\rho_{\mathsf{m}})\right\rbrace.
	\end{equation} 
	
	%\end{document}
	
	To further investigate $\mathsf{S}_*$, it is convenient to consider   
	separately the case of $\mathfrak{n}(\rho_{\mathsf{z}})=0$,  i.e.,  when $\mathsf{p}_1(\rho_{\mathsf{z}})=0$, and so the Bloch ball cannot be accessed from $\rho_{{\mathsf{z}}}$.  
	Note that if the system occupies $\rho_{\mathsf{z}}$, then it goes to the ball with probability $\mathsf{p}_1(\rho_{\mathsf{z}})=||\Pi_1U\mathsf{z}||^2$ or  stays in $\rho_{\mathsf{z}}$ with complementary probability $\mathsf{p}_2(\rho_{\mathsf{z}})=||\Pi_2U\mathsf{z}||^2=|\braket{\mathsf{z}|U\mathsf{z}}|^2$. We put $\omega:=\braket{\mathsf{z}|U\mathsf{z}}$. By $\sigma(A)$ we denote the set of eigenvalues of a linear map $A$.
	
	\begin{proposition}\label{unitarycond} 		The following conditions are equivalent: 	
		\vspace{-1.75mm}
		\begin{enumerate}[\rm (i), leftmargin=20mm]
			\itemsep-0.33mm
			\item  $\mathfrak{n}(\rho_{\mathsf{z}}) =0$;
			\item $|\omega|=1$;
			\item $\mathsf{z}$ is an eigenvector of $U$;
			\item $\Pi_1U\big|_\Theta$ is unitary;
			\item $\omega \in \sigma(U)$.			
		\end{enumerate}
	\end{proposition}
	
	\begin{proof} 
		
		\noindent
		\begin{enumerate}[leftmargin=33mm]	
			\itemsep-0.125mm
			
			\item	[(i) $\Leftrightarrow$ (ii)]
			$\mathfrak{n}(\rho_{\mathsf{z}}) =0$ iff $\mathsf{p}_1(\rho_{\mathsf{z}})=0$ iff  $\mathsf{p}_2(\rho_{\mathsf{z}})=|\omega|^2=1$.
			
			\item [(i) $\Leftrightarrow$ (iii)]	
			$\mathfrak{n}(\rho_{\mathsf{z}})=0$ iff $\mathsf{p}_1(\rho_{\mathsf{z}})=0$ iff  $\rho_{\mathsf{z}} \notin \operatorname{dom}\mathsf{F}_1=\mathcal{S}(\C^3) \setminus \{\rho_{U^*\mathsf{z}}\}$ iff  $\rho_{\mathsf{z}}=\rho_{U^*\mathsf{z}}$ iff $U^*\mathsf{z} \in \spann\{\mathsf{z}\}$ iff $\mathsf{z}$ is an eigenvector of $U$.
			
			\item [(iii) $\Leftrightarrow$ (iv)]  
			Since $U$ has an orthonormal eigenbasis and $\Theta=\operatorname{im}\Pi_1$ is the orthogonal complement of $\spann\{\mathsf{z}\}$, it follows that		
			$\mathsf{z}$ is an eigenvector of $U$ iff $\Theta$~is spanned by two eigenvectors of $U$ iff $\Theta$ is $U$-invariant, which is in turn equivalent to $\Pi_1U|_\Theta=U|_\Theta$.  This 
			%			last condition  
			implies the unitarity of  $\Pi_1U|_\Theta$. We now  show that the converse implication also holds. 
			Assume that  $\Pi_1U|_\Theta$ is unitary and let $w  \in \Theta$. It follows that  $||\Pi_1Uw||=||w||=||Uw||$, thus also   $||\Pi_2Uw||=0$. Hence,  $Uw \in \operatorname{ker}\Pi_2=\Theta$. Since $\Pi_1$ acts on $\Theta$ as the identity, we have $\Pi_1Uw=Uw$. Therefore,  $\Pi_1U|_\Theta=U|_\Theta$, as desired.
			
			\item	[(iii) $\Rightarrow$ (v)] 
			Let $\mu \in \C$ be an eigenvalue of $U$ associated with $\mathsf{z}$.
			Clearly,   $U\mathsf{z}=\mu \mathsf{z}$ implies that $\omega=\braket{\mathsf{z}|U\mathsf{z}}= \mu \in \sigma(U)$, as desired.

			\item	[(v) $\Rightarrow$ (ii)] Every eigenvalue of $U$ is of unit modulus. \qedhere
		\end{enumerate}
	\end{proof}

	\medskip
	
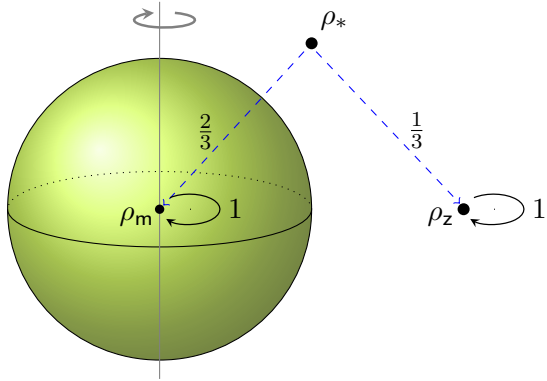
\begin{figure}[h]
		\captionsetup{width=0.75 \linewidth}
		\vspace{-4mm}
		\begin{center}	
			\scalebox{1}{		\begin{tikzpicture}
					\shade[ball color = lime!80, opacity = 0.8] (0,0) circle (2cm);
					\draw (0,0) circle (2cm);
					\draw (-2,0) arc (180:360:2 and 0.5);
					\draw[dotted] (2,0) arc (0:180:2 and 0.5);

					\fill[fill=black] (4,0) circle (2.25pt);
					\fill[fill=black] (2,2.2) circle (2.25pt);
					
					\draw [->,blue, dashed] (1.9,2.09) -- (0.05,0.05) node[pos=0.7, above, black] {$\frac 23$};
					\draw [->,blue, dashed] (2.06,2.09) -- (3.95,0.075) node[pos=0.7, above, black] {$\frac 13$};
					
					\node (m) at (-0.3, -0.1) {$\rho_{\mathsf{m}}$};
					\node (z) at (3.7, -0.1) {$\rho_{\mathsf{z}}$};
					\node (gw) at (2.3, 2.5) {$\rho_*$};
					
					\draw[gray] (0,-2.25)  -- (0,2.75)  node [pos=0.95] {\AxisRotatorj[rotate=-90,x=0.1cm,y=0.4cm,gray]};
					
					\draw[black] (4.4,0.0)  -- (4.4,0)  node [pos=0.6] {\AxisRotatord[rotate=0,x=0.4cm,y=0.2cm]};	
					\draw[black] (0.4,0)  -- (0.4,0.0)  node [pos=0.6] {\AxisRotatord[rotate=0,x=0.4cm,y=0.2cm]};	
					
					\fill[fill=black] (0,0) circle (1.75pt);
					
					\node (11) at (5,0) {$1$};
					\node (12) at (1, 0) {$1$};
					
			\end{tikzpicture}}
		\end{center}
		\vspace{-6mm}	
		\caption{Dynamics of the ball \& point system in the case of  $\mathfrak{n}(\rho_{\mathsf{z}})=0$
		}\label{unitdynpic} 
		%\vspace{-3mm}
	\end{figure}

	\noindent
	$\bullet$ \label{unitarycase}\textbf{Case of $\mathfrak{n}(\rho_{\mathsf{z}})=0$}.
	It  follows from Prop. \ref{unitarycond} that both $\spann\{{\mathsf{z}}\}$ and its orthogonal complement $\Theta$ are  invariant subspaces of $U$.
	Thus, intuitively,  there is no interaction between these two parts of the system as they  are subject to separate unitary dynamics (one of which  is  trivial) induced by the suitable restrictions of $U$. In consequence, the first measurement causes the system to get trapped into one of these subsystems. Furthermore, unitary dynamics on $\mathcal{S}(\Theta)$ corresponds to a rotation of the Bloch ball,  see \cite[Ch. 3, Sec. 5]{cornwell1984group},  \cite[Example~2.51]{HeiZim11}. Obviously, every rotation fixes the centre of the ball. Hence, if the first measurement sends the system to the ball, it arrives at a fixed point 
	$\rho_{\mathsf{m}}$  
	and loops there infinitely, see Fig.~\ref{unitdynpic}.
	Formally, in addition to $\mathsf{p}_2(\rho_{\mathsf{z}})=1$ and $\mathfrak{n}(\rho_{\mathsf{z}})=0$, we have  $\mathsf{p}_1(\rho_{\mathsf{m}})=1$ along with $\mathsf{F}_{1}(\rho_{\mathsf{m}})=\rho_{\mathsf{m}}$ and $\mathfrak{n}(\rho_{\mathsf{m}})=\infty$, because  $\Pi_1U|_\Theta=U|_\Theta$ gives  $\Lambda^{U,\, \Pi}_1(\rho_{\mathsf{m}}) = \Pi_1U\rho_{\mathsf{m}}\,U^*\Pi_1=U\rho_{\mathsf{m}}\,U^*=\rho_{\mathsf{m}}$.
	We conclude that $\mathsf{S}_{*}=\{\rho_{\mathsf{z}}, \rho_{\mathsf{m}}\}$ and $\mathsf{P}_{*}=\mathbb{I}_{\mathsf{S}_{*}}$.
	
	\noindent
	
	\smallskip
	
\begin{figure}[t]
		\captionsetup{width=0.65\linewidth}
		\scalebox{1}{ \hspace{-10mm}	\begin{tikzpicture}	  
				\shade[ball color = lime!90, opacity = 0.8] (0,0) circle (2cm);
				\draw (0,0) circle (2cm);
				\draw (-2,0) arc (180:360:2 and 0.5);
				\draw[dotted] (2,0) arc (0:180:2 and 0.5);
				\fill[fill=black!70] (0,0) circle (1.75pt);
				
				\fill[fill=black] (5,0) circle (2.25pt);
				
				\fill[fill=black] (1.94,0.45) circle (1.75pt);
				\fill[fill=black] (1.8,0.9) circle (1.75pt);
				\fill[fill=black] (1.6,1.2) circle (1.75pt);
				\fill[fill=black] (1.3,1.5) circle (1.75pt);
				
				\fill[fill=black!70] (0.2,0.3) circle (1.75pt);
				\fill[fill=black!70] (0.6,0.8) circle (1.75pt);
				\fill[fill=black!70] (0.8,1.2) circle (1.75pt);
				
				\node[font=\small] (m) at (-0.3, -0.1) {$\rho_{\mathsf{m}}$};
				\node[font=\small] (z) at (5.4, 0) {$\rho_{\mathsf{z}}$};
				\node[font=\footnotesize] (v) at (2.985,0.48) {$\rho_{\mathsf{v}}=\mathsf{F}_1(\rho_{\mathsf{z}})$};
				\node[font=\footnotesize] (m2) at (0.8,0.2) {$\mathsf{F}_1(\rho_{\mathsf{m}})$};

				\draw [->,>=stealth] (4.89, -0.04)  to [out=200,in=-60] (1.99,0.4) ;
				\draw [->,>=stealth] (1.94,0.45)  to [out=0,in=-14] (1.875,0.87) ;
				\draw [->,>=stealth] (1.8,0.9)  to [out=30,in=-20] (1.675,1.19) ;
				\draw [->,>=stealth]  (1.6,1.2)  to [out=60,in=-5] (1.375,1.5) ;
				\draw [->,>=stealth]  (1.3,1.5)   to [out=80,in=-20] (1.12,1.74) ;
				
				\draw [->,>=stealth,color=black!70] (0,0)  to [out=120,in=210] (.135,0.3);
				\draw [->,>=stealth,color=black!70] (0.2,0.3)  to [out=120,in=190] (0.535,0.8) ;
				\draw [->,>=stealth,color=black!70] (0.6,0.8)  to [out=120,in=200] (0.735,1.2) ;
				\draw [->,>=stealth,color=black!70] (0.8,1.2)  to [out=120,in=210] (0.9,1.4) ;
		\end{tikzpicture}}
		\vspace{-1mm}
		\caption{Potential state space $\mathsf{S}_*$ in the case of  $\mathfrak{n}(\rho_{\mathsf{z}})\geq 1$    (the arrows represent the action of $\mathsf{F}_1$)  
		}\label{pictraj}
		\vspace{-2mm}
	\end{figure}
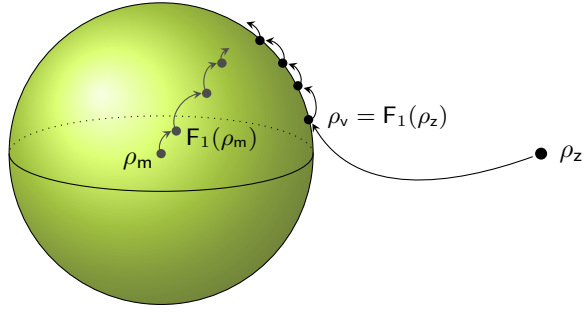

	\noindent
	$\bullet$ \textbf{Case of $\mathfrak{n}(\rho_{\mathsf{z}})\geq 1$}. From Prop. \ref{unitarycond} we have $U\mathsf{z} \notin \spann\{\mathsf{z}\}$, and so 
	$\Pi_1U\mathsf{z} \neq 0$. Putting ${\mathsf{v}}:=\Pi_1U\mathsf{z}$, we see that  $\rho_{\mathsf{v}}=\mathsf{F}_1(\rho_{\mathsf{z}})$ is the pure state in $\mathcal{S}(\Theta)$ at which the system arrives from $\rho_{\mathsf{z}}$ with probability $\mathsf{p}_1(\rho_{\mathsf{z}})=||\mathsf{v}||^2 >0$.
	Thus, \eqref{statespace} takes the form
	% \vspace{-1mm}
	\begin{align*} 
		% \vspace{-1mm}
		\mathsf{S}_{*}&=  \{\rho_{\mathsf{z}}\} \cup \left\lbrace\mathsf{F}^n_{1}(\rho_{\mathsf{v}})\colon n=0, \ldots, \mathfrak{n}(\rho_{\mathsf{v}})\right\rbrace  \cup \left\lbrace\mathsf{F}^n_{1}(\rho_{\mathsf{m}})\colon n=0, \ldots, \mathfrak{n}(\rho_{\mathsf{m}})\right\rbrace,
	\end{align*} 
	see also Fig. \ref{pictraj}. Clearly, if  the result of the first measurement is `1', then the system embarks on the trajectory of $\rho_{\mathsf{m}}$ under $\mathsf{F}_{1}$; otherwise, it goes to $\rho_{\mathsf{z}}$, from which the trajectory of $\rho_{\mathsf{v}}$ can be accessed. 
	Generically, in each step the system can either follow the trajectory in the Bloch ball or go to $\rho_{\mathsf{z}}$. (In Prop. \ref{rhou} we show that there exists exactly one state in  $\mathcal{S}(\Theta)$  from which the system cannot go to $\rho_{\mathsf{z}}$ and this state is pure.) 
	Note that after the first visit in $\rho_{\mathsf{z}}$  the system can explore the ball only along the trajectory of $\rho_{\mathsf{v}}$; in particular, the trajectory of $\rho_{\mathsf{m}}$, once left, cannot be re-entered.

	In order to establish $\mathsf{S}_*$, we need to determine  the trajectories of $\rho_{\mathsf{m}}$ and $\rho_{\mathsf{v}}$ under 
	$\mathsf{F}_1$. 
	To this end, we need  more  insight into how  $\mathsf{F}_1$ acts on $\mathcal{S}(\Theta)$. Recall that 
	\begin{equation}
		\label{dynA}\mathsf{F}_1\,\big|_{\mathcal{S}(\Theta)}\colon\ \ \ \mathcal{S}(\Theta)\setminus\{\rho_{{U^*\mathsf{z}}}\} \ni
		\rho \longmapsto 
		\frac{A {\rho}  A^{*}}{\tr(A	{\rho}  A^{*})} \in \mathcal{S}(\Theta),
	\end{equation} 	
	where $A:=\Pi_1U|_\Theta$. Note that 
	$\mathsf{F}_1$ is defined on the whole of $\mathcal{S}(\Theta)$ iff   $\rho_{U^*\mathsf{z}} \notin \mathcal{S}(\Theta)$ iff $U^*\mathsf{z} \notin \Theta$ iff $\omega \neq 0$. Moreover, 
	$\mathsf{F}_1|_{\mathcal{S}(\Theta)}$ is invertible  iff $A$ is invertible, and its inverse then  reads $\rho \mapsto 
	\frac{A^{-1}{\rho} \, (A^{-1})^*}{\tr(A^{-1}	{\rho} \, (A^{-1})^*)}$. We have already pointed out that $\mathsf{F}_1$  preserves pure states, see \eqref{pure}, and now we can easily see that so does its inverse. 
 	
	\begin{proposition} \label{Ainvertibleomegazero}
		$A$ is   invertible iff  $\omega \neq 0$
	\end{proposition}
	
	\begin{proof}
		We show that $0 \in \sigma(A)$ iff    $\omega = 0$.
		Assume $0 \in \sigma(A)$ and let $v \in \Theta$ be an eigenvector of $A$ associated with null eigenvalue. It follows that $Uv \in \operatorname{ker}\Pi_1=\spann\{\mathsf{z}\}$, and so  $U^*\mathsf{z}   \in \Theta$. We   get 
		$\omega=\braket{U^*\mathsf{z}|\mathsf{z}}=0$, as desired.
		Conversely, $\omega=0$ gives $U^*\mathsf{z} \in \Theta$; hence,  $\Pi_1U(U^*\mathsf{z})=0$, and so  $A$ has a~null eigenvalue, which concludes the proof.  
	\end{proof}
	
	\begin{corollary}\label{f1bijective}
		$\mathsf{F}_1|_{\mathcal{S}(\Theta)}$ and $\mathsf{F}_1|_{\mathcal{P}(\Theta)}$ are bijections iff $\omega \neq 0$.
	\end{corollary}

	\begin{corollary}\label{ninfinity}
		If $\omega \neq 0$, then 
		$\mathfrak{n}(\rho_{\mathsf{v}})=\mathfrak{n}(\rho_{\mathsf{m}})=\infty$.
	\end{corollary}

	The trajectories of $\rho_{\mathsf{m}}$ and $\rho_{\mathsf{v}}$ under   $\mathsf{F}_1$ may degenerate since either of these  states may coincide with a fixed point of $\mathsf{F}_1$, where the system would loop regardless of the overall dynamics of the ball, or (in the case of $\omega = 0$) with the sole point where $\mathsf{F}_1$ is not defined, which would cause the system to get projected to $\rho_{\mathsf{z}}$ with unit probability. 
	We say that $\rho \in \mathcal{S}(\Theta)$ is a  \emph{{singular point}} of $\mathsf{F}_1$ if  $\mathsf{F}_1(\rho)=\rho$ or $\rho \notin \operatorname{dom} \mathsf{F}_1$; otherwise, we  call   $\rho$  a~\emph{generic point} of $\mathsf{F}_1$. Singular points can be easily characterised in the case of pure states: 
	\begin{obs}\label{fixedobs} 
		Let $w \in \C^d\setminus \{0\}$. Then 
		$\rho_w\in \mathcal{P}(\Theta)$  is a singular point of $\mathsf{F}_1$  iff $w$ is an eigenvector of $A$. 
		In that case, denoting the eigenvalue of $A$ associated with $w$ by $\lambda$, we obtain  
		$\mathsf{p}_1(\rho_w)=|\lambda|^2$,  so 
		$\mathsf{F}_1(\rho_w)=\rho_w$ if $\lambda \neq 0$, and $\rho_w \notin \operatorname{dom} \mathsf{F}_1$ if $\lambda = 0$.
	\end{obs}

	The following three steps lead to the characterization of the trajectories of $\rho_{\mathsf{v}}$ and $\rho_{\mathsf{m}}$ under $\mathsf{F}_1$, and then to the classification of the Markov chains that can be generated by $\mathcal{F}_{U,\,\Pi}$ in the case of $A$ being non-unitary. First, we establish some further properties of~$A$. Next, we discuss the types of dynamics that $\mathsf{F}_1$ can induce on the Bloch ball, and for each of these types we describe its generic trajectories and  singular points. Finally, for each type we decide whether $\rho_{\mathsf{v}}$ or $\rho_{\mathsf{m}}$ is a generic or singular point of $\mathsf{F}_1$.

	\medskip

	\noindent
	{$\RHD$ \bf{Step I. Properties of $\boldsymbol{A}$.}} 
	{The following considerations include the case of $A$ being unitary.} 
	Recall that  $\omega:=\braket{\mathsf{z}|U\mathsf{z}}$ and $\mathsf{v}:=\Pi_1U\mathsf{z}$. 
	Let  ${u_1}, {u_2}$ be any two vectors that span~$\Theta$. Fixing 
	$\{{u_1}, {u_2}, \mathsf{z}\}$ as the basis of $\C^3$, we obtain
	the matrix representation
	%\vspace{-1mm}
	\begin{equation}
		\label{trunc}
		U \sim  \begin{bmatrix}
			a_{11}  & a_{12} & v_1  \\ 
			a_{21} & a_{22} & v_2  \\
			*  &  *  & \omega \end{bmatrix},
		%$}
	\end{equation}
	where  $a_{ij},v_j \in \C$ with $i,j \in \{1, 2\}$. Observe that $\begin{bsmallmatrix}
		a_{11}  & a_{12}   \\
		a_{21} & a_{22}   \end{bsmallmatrix} 
	$ and $[v_1,v_2]^T$ represent $A$ and $\mathsf{v}$, respectively, in the basis $\{{u_1}, {u_2}\}$ of $\Theta$.
	
	Obviously,  the  eigenvalues of $A$ depend only on $\tr A$ and $\det A$. Moreover, we have 
	% \vspace{-1mm}
	\begin{equation}\label{trdet}
		\tr A= \tr U-\omega \ \aand \ \det A=\overline{\omega}\det U.\end{equation} 
	Indeed, the trace formula is obvious by \eqref{trunc}, while that for the determinant can be deduced from the following
	\begin{theorem*}{\rm \cite[Thm. 2.3]{Zhang}}
		Let 	$V=
		\begin{bsmallmatrix}
			V_1   & V_2  \\ 
			V_3 & V_4
		\end{bsmallmatrix}$ 
		be a complex block matrix, where $V_1$, $V_4$ are square matrices. If $V$ is invertible and 
		$\,V^{-1}=
		\begin{bsmallmatrix}
			Y_1   & Y_2  \\ 
			Y_3 & Y_4
		\end{bsmallmatrix}
		$ is partitioned conformally to $V$,  
		then  $\det V_1 = \det  Y_4 \det V$.
	\end{theorem*}

	\noindent
	If $V$ in the above theorem is unitary, then $V^{-1}=V^*$,  so 
	$\det V_1=\overline{\det V_4}\det V$. This particular result is due to Muir \cite[p. 339]{Muir26}, see also \cite{Salaff}. In our situation the submatrix $V_4$ is of size $1 \times 1$, in which case the proof simplifies further, see \cite[p. 172]{Zhang}.

	The next theorem allows us to determine the set ${\mathsf{s}}(A)$ of  singular values of $A$.

	\begin{theorem*}
		{\rm\cite[Thm. 6.3]{Zhang}}
		Let 
		$V=
		\begin{bsmallmatrix}
			V_1   & V_2  \\ 
			V_3 & V_4
		\end{bsmallmatrix},
		$
		where $V_1\in \C^{n \times n}$ and $V_4\in \C^{m \times m}$, be a unitary block matrix. If $m = n$, then $V_1$ and $V_4$ have the
		same singular values. If $n > m$ and the singular values of $V_4$ read
		$s_1, \ldots , s_m$, then the singular values of $V_1$ read $s_1, \ldots , s_m,
		\underbrace{1, \ldots , 1}_\text{n-m}$.
	\end{theorem*} 		
	
	\vspace{-1mm}
	
	\noindent 
	Applying the above theorem to $U$ with $n=2$ and $m=1$, we obtain $\mathsf{s}(A)=\{|\omega|,1\}$. 
	
	\begin{remark}\label{remarknumrange}
		Note that $\sigma(A)$ and $\mathsf{s}(A)$ depend on $\mathsf{z} \in \C^3$ only via \mbox{$\omega=\braket{\mathsf{z}|U\mathsf{z}}\in \C$}. 	We further investigate this fact in Subsection \ref{secNumrange}.
	\end{remark}
	
	Clearly,   {$\max\{|\lambda_1|, |\lambda_2|\} \leq  ||A|| = \max \mathsf{s}(A) = 1$}. 
	Further, from the determinant formula in \eqref{trdet} we obtain $|\lambda_1\lambda_2|=|\omega|$, and so 
	$|\omega|=1$ iff $|\lambda_1|=|\lambda_2|=1$. Thus, via Prop. \ref{unitarycond} we get 
	\begin{proposition}\label{unit11}  $A$ is unitary iff $|\lambda_1|=|\lambda_2|=1.$
	\end{proposition}
	
	%		\vspace{-1mm}
	
	In what follows we consider a  Schur form of $A$. (Recall that a \emph{Schur form} of $V\in \mathcal{L}(\C^d)$ is an upper triangular matrix which represents $V$ in an~orthonormal basis comprising an eigenvector of $V$ and which has the eigenvalues of $V$ as diagonal entries.) Let us put $e_1$ for a normalised eigenvector of $A$ associated with $\lambda_1$ and let $w \in \Theta$ be a unit vector orthogonal to $e_1$.
	With respect to the orthonormal basis $\{e_1, w\}$ of $\Theta$ we have 
	%		\vspace{-1mm}
	\begin{equation}	
		\label{AAdagg}
		A \sim \begin{bsmallmatrix}
			\lambda_1\vphantom{^T} & x \\ 0 & \lambda_2\vphantom{^T}
		\end{bsmallmatrix} \aand A^* \sim \begin{bsmallmatrix}
			\overline{\lambda_1}& 0\\
			\overline{x} & \overline{\lambda_2}
		\end{bsmallmatrix},\end{equation} where $x:=\braket{e_1|Aw}$. Obviously, $w$ is an eigenvector of $A$ iff $x=0$. 
	From \eqref{AAdagg} we obtain  
	\begin{equation}\label{tr1}	
		\tr (A^*A)=
		|\lambda_1|^2+|\lambda_2|^2+|x|^2.
	\end{equation}
	On the other hand, the singular values of $A$ are the square roots of the  eigenvalues of $A^*A$ and 
	$\mathsf{s}(A)=\{|\lambda_1\lambda_2|, 1\}$. Thus,
	%  \vspace{-1mm}
	\begin{equation}\label{tr2}		
		% \vspace{-1mm}
		\tr (A^*A)=|\lambda_1\lambda_2|^2 + 1.
	\end{equation}
	Combining \eqref{tr1} \& \eqref{tr2}, we get 
	$|\lambda_1|^2+|\lambda_2|^2+|x|^2=1+|\lambda_1\lambda_2|^2$. It follows that   
	\begin{equation}	
		\label{xschur}
		|x|^2=(1-|\lambda_1|^2)(1-|\lambda_2|^2),\end{equation} 
	and so $x=0$ iff $\max\{|\lambda_1|, |\lambda_2|\}=1$. As a result, we obtain

	\begin{proposition}\label{orthbasis}
		$A$ is normal  (i.e., $A$ has an orthonormal eigenbasis) iff \mbox{$\max\{|\lambda_1|, |\lambda_2|\}=1$}. 
	\end{proposition}

	Next, we briefly consider the case of $A$ having a double eigenvalue. Namely, if $\lambda_1 = \lambda_2$ and $A$ is not defective (i.e., it has two linearly independent eigenvectors), then $A=\lambda_1 \mathbb{I}_\Theta$ and every orthonormal basis of $\Theta$ constitutes an~orthonormal eigenbasis of $A$.
	With the help of Prop.~\ref{unit11} \& Prop.~\ref{orthbasis} we obtain   
	\begin{proposition} \label{Adef} If $\sigma(A) = \{\lambda_1\}$, then  
		$A$ is defective iff $|\lambda_1|<1$. Otherwise, $A$ is unitary.
	\end{proposition}

	\medskip

	\noindent
	$\RHD$ {\bf{Step II. Ball dynamics.}} \label{ballsect} 
	Let $[v]$ stand for the equivalence class of $v \in \C^d\setminus  \{0\}$ in $\mathbb{CP}^{d-1}$. Clearly, spectral decomposition theorem and the linearity of $	\Lambda^{U,\, \Pi}_1$ allow us to deduce the dynamics generated by $\mathsf{F}_1$ on $\mathcal{S}(\Theta)$, see \eqref{dynA}, from the action of $\mathsf{F}_1$ on $\mathcal{P}(\Theta)$.
	Using \eqref{pure}, we can translate   $\mathsf{F}_1|_{\mathcal{P}(\Theta)}$  into the following  (partial) map
	% \vspace{-1mm} 
	\[\mathcal{A} \colon \ 
	%\vspace{-1mm}
	\mathbb{P}\Theta \ni [w] \longmapsto [Aw] \in \mathbb{P}\Theta \ \ \textrm{ if }\ \ A{w}\neq 0.\]
	In what follows we first briefly discuss the dynamics induced by $\mathcal{A}$ on $\mathbb{P}\Theta \sim \C\mathbb{P}^1$, and then analyse how it extends to $\mathcal{S}(\Theta)$.
	%\smallskip
	
\begin{figure}[b]\captionsetup[subfigure]{labelformat=empty}	 
		\vspace{-5mm}
		\tikzset{every picture/.style={line width=0.75pt}}   
		\begin{center}{\centering
				\subfloat[{(a) loxodromic } ]
				{ \hspace{2mm}	\scalebox{0.33}{	
						\begin{tikzpicture}[x=0.75pt,y=0.75pt,yscale=-1,xscale=1]
							\shade[ball color = lime!80, opacity = 0.85] (351.25,252.36) circle (6.3cm);
							\draw  [line width=0.75]  (112.01,252.36) .. controls (112.01,120.63) and (219.12,13.84) .. (351.25,13.84) .. controls (483.38,13.84) and (590.49,120.63) .. (590.49,252.36) .. controls (590.49,384.1) and (483.38,490.89) .. (351.25,490.89) .. controls (219.12,490.89) and (112.01,384.1) .. (112.01,252.36) -- cycle ;
							\draw  [draw opacity=0][fill={rgb, 255:red, 3; green, 86; blue, 251 }  ,fill opacity=1 ] (363,57.5) .. controls (363,54.12) and (365.75,51.38) .. (369.14,51.38) .. controls (372.53,51.38) and (375.28,54.12) .. (375.28,57.5) .. controls (375.28,60.88) and (372.53,63.62) .. (369.14,63.62) .. controls (365.75,63.62) and (363,60.88) .. (363,57.5) -- cycle ;
							\draw  [draw opacity=0][fill={rgb, 255:red, 3; green, 86; blue, 251 }  ,fill opacity=1 ] (388.5,487.42) .. controls (388.5,484.04) and (391.25,481.29) .. (394.64,481.29) .. controls (398.03,481.29) and (400.78,484.04) .. (400.78,487.42) .. controls (400.78,490.8) and (398.03,493.54) .. (394.64,493.54) .. controls (391.25,493.54) and (388.5,490.8) .. (388.5,487.42) -- cycle ;
							\draw [color={rgb, 255:red, 0; green, 0; blue, 0 }  ,draw opacity=1 ][line width=1.5]    (112.67,247.5) .. controls (263.15,324.06) and (508.53,274.27) .. (576.5,181.09) ;
							\draw [shift={(578.5,178.25)}, rotate = 484.38] [fill={rgb, 255:red, 0; green, 0; blue, 0 }  ,fill opacity=1 ][line width=0.08]  [draw opacity=0] (11.61,-5.58) -- (0,0) -- (11.61,5.58) -- cycle    ;
							\draw [shift={(356.77,280.68)}, rotate = 533] [fill={rgb, 255:red, 0; green, 0; blue, 0 }  ,fill opacity=1 ][line width=0.08]  [draw opacity=0] (11.61,-5.58) -- (0,0) -- (11.61,5.58) -- cycle    ;
							\draw [color={rgb, 255:red, 0; green, 0; blue, 0 }  ,draw opacity=1 ][line width=1.5]    (146,131) .. controls (248.96,238.91) and (510.7,198.82) .. (513.01,83.52) ;
							\draw [shift={(513,80)}, rotate = 448.54] [fill={rgb, 255:red, 0; green, 0; blue, 0 }  ,fill opacity=1 ][line width=0.08]  [draw opacity=0] (11.61,-5.58) -- (0,0) -- (11.61,5.58) -- cycle    ;
							\draw [shift={(348.17,193.05)}, rotate = 536.05] [fill={rgb, 255:red, 0; green, 0; blue, 0 }  ,fill opacity=1 ][line width=0.08]  [draw opacity=0] (11.61,-5.58) -- (0,0) -- (11.61,5.58) -- cycle    ;
							\draw [color={rgb, 255:red, 0; green, 0; blue, 0 }  ,draw opacity=1 ][line width=1.5]    (236.5,42.75) .. controls (252.26,114.66) and (359.71,135.38) .. (414.53,93.94) ;
							\draw [shift={(417,92)}, rotate = 500.83] [fill={rgb, 255:red, 0; green, 0; blue, 0 }  ,fill opacity=1 ][line width=0.08]  [draw opacity=0] (11.61,-5.58) -- (0,0) -- (11.61,5.58) -- cycle    ;
							\draw [color={rgb, 255:red, 0; green, 0; blue, 0 }  ,draw opacity=1 ][line width=1.5]    (396.2,481.7) .. controls (396.39,472.82) and (423.37,465.09) .. (445.91,469) ;
							\draw [shift={(449.8,469.8)}, rotate = 193.57] [fill={rgb, 255:red, 0; green, 0; blue, 0 }  ,fill opacity=1 ][line width=0.08]  [draw opacity=0] (11.61,-5.58) -- (0,0) -- (11.61,5.58) -- cycle    ;
							\draw [color={rgb, 255:red, 0; green, 0; blue, 0 }  ,draw opacity=1 ][line width=1.5]    (267.5,475.25) .. controls (309.08,438.62) and (456.02,405.67) .. (548.22,383.66) ;
							\draw [shift={(551,383)}, rotate = 526.55] [fill={rgb, 255:red, 0; green, 0; blue, 0 }  ,fill opacity=1 ][line width=0.08]  [draw opacity=0] (11.61,-5.58) -- (0,0) -- (11.61,5.58) -- cycle    ;
							\draw [shift={(405.09,419.36)}, rotate = 524.3] [fill={rgb, 255:red, 0; green, 0; blue, 0 }  ,fill opacity=1 ][line width=0.08]  [draw opacity=0] (11.61,-5.58) -- (0,0) -- (11.61,5.58) -- cycle    ;
							\draw [color={rgb, 255:red, 0; green, 0; blue, 0 }  ,draw opacity=1 ][line width=1.5]    (143.33,370.08) .. controls (373.01,373.05) and (463.19,342.29) .. (583.35,295.1) ;
							\draw [shift={(587,293.67)}, rotate = 518.52] [fill={rgb, 255:red, 0; green, 0; blue, 0 }  ,fill opacity=1 ][line width=0.08]  [draw opacity=0] (11.61,-5.58) -- (0,0) -- (11.61,5.58) -- cycle    ;
							\draw [shift={(369.32,357.59)}, rotate = 531.46] [fill={rgb, 255:red, 0; green, 0; blue, 0 }  ,fill opacity=1 ][line width=0.08]  [draw opacity=0] (11.61,-5.58) -- (0,0) -- (11.61,5.58) -- cycle    ;
						
							\draw [color={rgb, 255:red, 0; green, 0; blue, 0 }  ,draw opacity=1 ][line width=1.5]    (412.5,97.25) .. controls (444.33,58.33) and (403.33,19.08) .. (366,26.25) ;
					
							\draw [color={rgb, 255:red, 0; green, 0; blue, 0 }  ,draw opacity=1 ][line width=1.5]    (360.02,28.42) .. controls (341.22,36.82) and (347.76,53.57) .. (363,57.5) ;
							\draw [shift={(358.2,29.3)}, rotate = 337.58000000000004] [fill={rgb, 255:red, 0; green, 0; blue, 0 }  ,fill opacity=1 ][line width=0.08]  [draw opacity=0] (11.61,-5.58) -- (0,0) -- (11.61,5.58) -- cycle    ;
							
						\end{tikzpicture} 	
					}
				}
				\hspace{3mm}
				\subfloat[{(b) elliptic} ]
				{	\scalebox{0.33}{	
						\begin{tikzpicture}[x=0.75pt,y=0.75pt,yscale=-1,xscale=1]
							\shade[ball color = lime!80, opacity = 0.85] (351.25,252.36) circle (6.3cm);

							\draw  [line width=0.75]  (112.01,252.36) .. controls (112.01,120.63) and (219.12,13.84) .. (351.25,13.84) .. controls (483.38,13.84) and (590.49,120.63) .. (590.49,252.36) .. controls (590.49,384.1) and (483.38,490.89) .. (351.25,490.89) .. controls (219.12,490.89) and (112.01,384.1) .. (112.01,252.36) -- cycle ;
							\draw  [draw opacity=0][fill={rgb, 255:red, 3; green, 86; blue, 251 }  ,fill opacity=1 ] (363,57.5) .. controls (363,54.12) and (365.75,51.38) .. (369.14,51.38) .. controls (372.53,51.38) and (375.28,54.12) .. (375.28,57.5) .. controls (375.28,60.88) and (372.53,63.62) .. (369.14,63.62) .. controls (365.75,63.62) and (363,60.88) .. (363,57.5) -- cycle ;
							\draw  [draw opacity=0][fill={rgb, 255:red, 3; green, 86; blue, 251 }  ,fill opacity=1 ] (388.5,487.42) .. controls (388.5,484.04) and (391.25,481.29) .. (394.64,481.29) .. controls (398.03,481.29) and (400.78,484.04) .. (400.78,487.42) .. controls (400.78,490.8) and (398.03,493.54) .. (394.64,493.54) .. controls (391.25,493.54) and (388.5,490.8) .. (388.5,487.42) -- cycle ;
							\draw [color={rgb, 255:red, 0; green, 0; blue, 0 }  ,draw opacity=1 ][line width=1.5]    (112,236) .. controls (262.48,312.56) and (494.31,291.44) .. (584.32,229.88) ;
							\draw [shift={(587,228)}, rotate = 504.4] [fill={rgb, 255:red, 0; green, 0; blue, 0 }  ,fill opacity=1 ][line width=0.08]  [draw opacity=0] (11.61,-5.58) -- (0,0) -- (11.61,5.58) -- cycle    ;
							\draw [shift={(349.58,285.22)}, rotate = 539.85] [fill={rgb, 255:red, 0; green, 0; blue, 0 }  ,fill opacity=1 ][line width=0.08]  [draw opacity=0] (11.61,-5.58) -- (0,0) -- (11.61,5.58) -- cycle    ;
							\draw [color={rgb, 255:red, 0; green, 0; blue, 0 }  ,draw opacity=1 ][line width=1.5]    (157,121) .. controls (238.18,237.82) and (483.04,227.23) .. (549.05,127.06) ;
							\draw [shift={(551,124)}, rotate = 481.45] [fill={rgb, 255:red, 0; green, 0; blue, 0 }  ,fill opacity=1 ][line width=0.08]  [draw opacity=0] (11.61,-5.58) -- (0,0) -- (11.61,5.58) -- cycle    ;
							\draw [shift={(353.93,205.4)}, rotate = 180.96] [fill={rgb, 255:red, 0; green, 0; blue, 0 }  ,fill opacity=1 ][line width=0.08]  [draw opacity=0] (11.61,-5.58) -- (0,0) -- (11.61,5.58) -- cycle    ;
							\draw [color={rgb, 255:red, 0; green, 0; blue, 0 }  ,draw opacity=1 ][line width=1.5]    (225.5,50.75) .. controls (241.34,123.02) and (444.87,151.43) .. (487.75,60.78) ;
							\draw [shift={(489,58)}, rotate = 473.05] [fill={rgb, 255:red, 0; green, 0; blue, 0 }  ,fill opacity=1 ][line width=0.08]  [draw opacity=0] (11.61,-5.58) -- (0,0) -- (11.61,5.58) -- cycle    ;
							\draw [color={rgb, 255:red, 0; green, 0; blue, 0 }  ,draw opacity=1 ][line width=1.5]    (365.2,487.7) .. controls (372.65,467.93) and (395.78,465.22) .. (420.5,475.47) ;
							\draw [shift={(424,477)}, rotate = 204.78] [fill={rgb, 255:red, 0; green, 0; blue, 0 }  ,fill opacity=1 ][line width=0.08]  [draw opacity=0] (11.61,-5.58) -- (0,0) -- (11.61,5.58) -- cycle    ;
							\draw [color={rgb, 255:red, 0; green, 0; blue, 0 }  ,draw opacity=1 ][line width=1.5]    (267.5,475.25) .. controls (308.87,438.8) and (420.58,405.03) .. (496.55,437.47) ;
							\draw [shift={(500,439)}, rotate = 204.73] [fill={rgb, 255:red, 0; green, 0; blue, 0 }  ,fill opacity=1 ][line width=0.08]  [draw opacity=0] (11.61,-5.58) -- (0,0) -- (11.61,5.58) -- cycle    ;
							\draw [shift={(378.06,428.78)}, rotate = 529.38] [fill={rgb, 255:red, 0; green, 0; blue, 0 }  ,fill opacity=1 ][line width=0.08]  [draw opacity=0] (11.61,-5.58) -- (0,0) -- (11.61,5.58) -- cycle    ;
							\draw [color={rgb, 255:red, 0; green, 0; blue, 0 }  ,draw opacity=1 ][line width=1.5]    (144.33,375.08) .. controls (307.35,357.18) and (446.2,352.1) .. (566.36,347.15) ;
							\draw [shift={(570,347)}, rotate = 537.63] [fill={rgb, 255:red, 0; green, 0; blue, 0 }  ,fill opacity=1 ][line width=0.08]  [draw opacity=0] (11.61,-5.58) -- (0,0) -- (11.61,5.58) -- cycle    ;
							\draw [shift={(356.62,357.15)}, rotate = 536.44] [fill={rgb, 255:red, 0; green, 0; blue, 0 }  ,fill opacity=1 ][line width=0.08]  [draw opacity=0] (11.61,-5.58) -- (0,0) -- (11.61,5.58) -- cycle    ;
							\draw [color={rgb, 255:red, 0; green, 0; blue, 0 }  ,draw opacity=1 ][line width=1.5]    (390.12,73.69) .. controls (461.4,53.71) and (403.69,26.05) .. (343,36.25) ;
							\draw [shift={(392.2,73.09)}, rotate = 524.6800000000001] [fill={rgb, 255:red, 0; green, 0; blue, 0 }  ,fill opacity=1 ][line width=0.08]  [draw opacity=0] (11.61,-5.58) -- (0,0) -- (11.61,5.58) -- cycle    ;
							\draw [color={rgb, 255:red, 0; green, 0; blue, 0 }  ,draw opacity=1 ][line width=1.5]    (333.01,40.43) .. controls (313.99,49.34) and (314.27,80.88) .. (389,74) ;
							\draw [shift={(331.15,41.41)}, rotate = 337.58000000000004] [fill={rgb, 255:red, 0; green, 0; blue, 0 }  ,fill opacity=1 ][line width=0.08]  [draw opacity=0] (11.61,-5.58) -- (0,0) -- (11.61,5.58) -- cycle    ;
						\end{tikzpicture}	
					}
				}
				\hspace{2mm}
				\subfloat[{(c) parabolic} ]
				{	\scalebox{0.33}{

						\begin{tikzpicture}[x=0.75pt,y=0.75pt,yscale=-1,xscale=1]
							\shade[ball color = lime!80, opacity = 0.85] (351.25,252.36) circle (6.3cm);

							\draw  [line width=0.75]  (112.01,252.36) .. controls (112.01,120.63) and (219.12,13.84) .. (351.25,13.84) .. controls (483.38,13.84) and (590.49,120.63) .. (590.49,252.36) .. controls (590.49,384.1) and (483.38,490.89) .. (351.25,490.89) .. controls (219.12,490.89) and (112.01,384.1) .. (112.01,252.36) -- cycle ;
							\draw [color={rgb, 255:red, 0; green, 0; blue, 0 }  ,draw opacity=1 ][line width=1.5]    (369.14,57.5) .. controls (380.33,0.83) and (289.33,19.5) .. (369.14,57.5) ;
							\draw [shift={(341.16,22.66)}, rotate = 350.36] [fill={rgb, 255:red, 0; green, 0; blue, 0 }  ,fill opacity=1 ][line width=0.08]  [draw opacity=0] (11.61,-5.58) -- (0,0) -- (11.61,5.58) -- cycle    ;
							\draw [color={rgb, 255:red, 0; green, 0; blue, 0 }  ,draw opacity=1 ][line width=1.5]    (369.14,57.5) .. controls (89.82,219.37) and (202.68,453.27) .. (318.49,488.98) ;
							\draw [shift={(322,490)}, rotate = 195.3] [fill={rgb, 255:red, 0; green, 0; blue, 0 }  ,fill opacity=1 ][line width=0.08]  [draw opacity=0] (11.61,-5.58) -- (0,0) -- (11.61,5.58) -- cycle    ;
							\draw [shift={(197.73,266.37)}, rotate = 280.81] [fill={rgb, 255:red, 0; green, 0; blue, 0 }  ,fill opacity=1 ][line width=0.08]  [draw opacity=0] (11.61,-5.58) -- (0,0) -- (11.61,5.58) -- cycle    ;
							\draw [color={rgb, 255:red, 0; green, 0; blue, 0 }  ,draw opacity=1 ][line width=1.5]    (369.14,57.5) .. controls (409,43) and (460.8,48.4) .. (484,54) ;
							\draw [shift={(426.51,48.22)}, rotate = 358.54] [fill={rgb, 255:red, 0; green, 0; blue, 0 }  ,fill opacity=1 ][line width=0.08]  [draw opacity=0] (11.61,-5.58) -- (0,0) -- (11.61,5.58) -- cycle    ;
							\draw [color={rgb, 255:red, 0; green, 0; blue, 0 }  ,draw opacity=1 ][line width=1.5]    (209.64,61.66) .. controls (268.94,69.94) and (288.08,68.71) .. (369.14,57.5) ;
							\draw [shift={(205,61)}, rotate = 8.13] [fill={rgb, 255:red, 0; green, 0; blue, 0 }  ,fill opacity=1 ][line width=0.08]  [draw opacity=0] (11.61,-5.58) -- (0,0) -- (11.61,5.58) -- cycle    ;
							\draw [color={rgb, 255:red, 0; green, 0; blue, 0 }  ,draw opacity=1 ][line width=1.5]    (369.14,57.5) .. controls (257,318) and (604,232) .. (375.28,57.5) ;
							\draw [shift={(415.93,220.63)}, rotate = 529.15] [fill={rgb, 255:red, 0; green, 0; blue, 0 }  ,fill opacity=1 ][line width=0.08]  [draw opacity=0] (11.61,-5.58) -- (0,0) -- (11.61,5.58) -- cycle    ;
							\draw [color={rgb, 255:red, 0; green, 0; blue, 0 }  ,draw opacity=1 ][line width=1.5]    (369.14,57.5) .. controls (352,174) and (455,127) .. (375.28,57.5) ;
							\draw [shift={(394.43,127.72)}, rotate = 535.06] [fill={rgb, 255:red, 0; green, 0; blue, 0 }  ,fill opacity=1 ][line width=0.08]  [draw opacity=0] (11.61,-5.58) -- (0,0) -- (11.61,5.58) -- cycle    ;
							\draw [color={rgb, 255:red, 0; green, 0; blue, 0 }  ,draw opacity=1 ][line width=1.5]    (369.14,57.5) .. controls (394,45) and (407,28) .. (407,21) ;
							\draw [shift={(391.81,42.51)}, rotate = 320.81] [fill={rgb, 255:red, 0; green, 0; blue, 0 }  ,fill opacity=1 ][line width=0.08]  [draw opacity=0] (11.61,-5.58) -- (0,0) -- (11.61,5.58) -- cycle    ;
							\draw [color={rgb, 255:red, 0; green, 0; blue, 0 }  ,draw opacity=1 ][line width=1.5]    (369.14,57.5) .. controls (494,85) and (560,147) .. (579,183) ;
							\draw [shift={(486.63,100.48)}, rotate = 28.35] [fill={rgb, 255:red, 0; green, 0; blue, 0 }  ,fill opacity=1 ][line width=0.08]  [draw opacity=0] (11.61,-5.58) -- (0,0) -- (11.61,5.58) -- cycle    ;
							\draw [color={rgb, 255:red, 0; green, 0; blue, 0 }  ,draw opacity=1 ][line width=1.5]    (369.14,57.5) .. controls (159.28,112.87) and (141.64,196.57) .. (113.73,249.18) ;
							\draw [shift={(112.01,252.36)}, rotate = 298.97] [fill={rgb, 255:red, 0; green, 0; blue, 0 }  ,fill opacity=1 ][line width=0.08]  [draw opacity=0] (11.61,-5.58) -- (0,0) -- (11.61,5.58) -- cycle    ;
							\draw [shift={(215.34,120.45)}, rotate = 327.40999999999997] [fill={rgb, 255:red, 0; green, 0; blue, 0 }  ,fill opacity=1 ][line width=0.08]  [draw opacity=0] (11.61,-5.58) -- (0,0) -- (11.61,5.58) -- cycle    ;
							\draw [color={rgb, 255:red, 0; green, 0; blue, 0 }  ,draw opacity=1 ][line width=1.5]    (395.5,371.25) .. controls (509,372.75) and (640,224) .. (375.28,57.5) ;
							\draw [shift={(526.08,222.18)}, rotate = 436.73] [fill={rgb, 255:red, 0; green, 0; blue, 0 }  ,fill opacity=1 ][line width=0.08]  [draw opacity=0] (11.61,-5.58) -- (0,0) -- (11.61,5.58) -- cycle    ;
							\draw [color={rgb, 255:red, 0; green, 0; blue, 0 }  ,draw opacity=1 ][line width=1.5]    (369.14,57.5) .. controls (308,136) and (199.5,378.85) .. (395.5,371.25) ;
							\draw [shift={(286.55,242.48)}, rotate = 280.06] [fill={rgb, 255:red, 0; green, 0; blue, 0 }  ,fill opacity=1 ][line width=0.08]  [draw opacity=0] (11.61,-5.58) -- (0,0) -- (11.61,5.58) -- cycle    ;
							\draw  [draw opacity=0][fill={rgb, 255:red, 3; green, 86; blue, 251 }  ,fill opacity=1 ] (363,57.5) .. controls (363,54.12) and (365.75,51.38) .. (369.14,51.38) .. controls (372.53,51.38) and (375.28,54.12) .. (375.28,57.5) .. controls (375.28,60.88) and (372.53,63.62) .. (369.14,63.62) .. controls (365.75,63.62) and (363,60.88) .. (363,57.5) -- cycle ;

						\end{tikzpicture}
					}
				}
				\hspace{-2mm}
			}
		\end{center}	
		\captionsetup{width=0.7\linewidth}	
		\vspace{-4mm}	
		\caption{  M\"obius maps on the Riemann sphere
		}	\label{figmob}
	\end{figure}
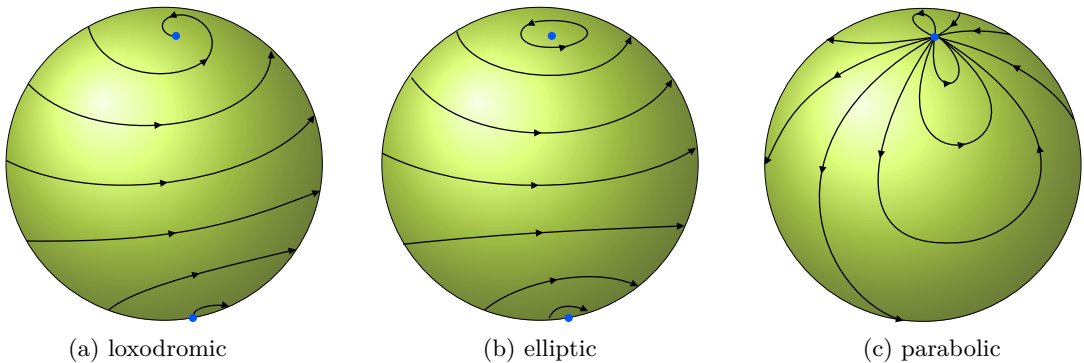

	Recall that the  bijection  $\mathbb{CP}^{d-1}\ni [z] \mapsto [Vz]\in \mathbb{CP}^{d-1}$  induced  by a non-singular map $V\in \mathcal{L}(\C^d)$ is called a \emph{homography}. Hence, if  $0 \notin \sigma(A)$, then  $\mathcal{A}$ is a homography on  \mbox{$\mathbb{P}\Theta \sim \mathbb{CP}^1$}. 
	Homographies on $\mathbb{CP}^1$
	are isomorphic to the group of \emph{M\"obius maps}, which are orientation preserving conformal automorphisms of the Riemann sphere: the homography $[z]\mapsto[Wz]$ on $\mathbb{CP}^1$ induced by  a non-singular matrix $W=\begin{bsmallmatrix}
		a&b\\c&d
	\end{bsmallmatrix}$ corresponds to the  M\"obius {transformation}  $\hat{\C} \ni z \mapsto \frac{az + b}{cz + d} \in \hat{\C},$  see \cite{anderson2005hyperbolic, barnsley2014, Beardon1983, Jadczyk6,  needham1998visual}. 
	Based on the number and type of fixed points,  non-identity   M\"obius maps are classified into three types (see also Fig. \ref{figmob}): 
	\vspace{-1.0mm}
	\begin{itemize}[leftmargin=9.0mm]
		\itemsep=-0.5mm
		\item 
		\emph{loxodromic}: two fixed points, one of which is attractive and the other repulsive;
		\item  \emph{elliptic}: two fixed points and both are neutral
		(i.e., neither attractive nor repulsive);	
		\item \emph{parabolic}: one fixed point, which is attractive (and unstable). 
	\end{itemize}
	
	Now we examine how  the dynamics induced by $\mathcal{A}$ on $\C\mathbb{P}^1$ depends on the eigenvalues of~$A$. Let us adopt the following notation. If $A$ is not defective, 
	then we let  ${{e}_j}$ stand for a normalised eigenvector of $A$ associated with $\lambda_j$ ($j = 1, 2$). 
	Here and henceforth, with no loss of generality, we assume that $|\lambda_2| \leq |\lambda_1|$  and put  $\psi:=\Arg\frac{\lambda_2}{\lambda_1}$, provided that  $0 \notin \sigma(A)$.  
	Clearly, the eigenbasis $\{e_1, e_2\}$ of $A$ is unique (up to  phases) and with respect to it we have $A\sim \diag(\lambda_{1},\lambda_{2})$. 
	If $A$ is defective,    then ${{e}_1}$ will stand for a~normalised eigenvector of $A$ and ${{g}_1}$ 
	for a unit vector orthogonal to $e_1$. It follows that $g_1$ is a~generalised eigenvector  of $A$ since $\operatorname{ker}(A - \lambda_1 \mathbb{I}_2)^2=\C^2$.
	The orthonormal generalised eigenbasis $\{e_1, g_1\}$ of $A$ is unique (up to  phases) and with respect to it we have  $A \sim \begin{bsmallmatrix}\lambda_{1} & x \\ 0 & \lambda_{1}
	\end{bsmallmatrix}$, where $x=\braket{e_1|Ag_1}$ satisfies $|x|=1-|\lambda_1|^2 \in (0,1]$, see   \eqref{xschur} and Prop. \ref{Adef}.

	Note that the fixed points of $\mathcal{A}$ coincide with the eigenspaces of $A$ that are associated with non-zero eigenvalues, and $\mathcal{A}$ is undefined on the eigenspace of $A$ associated with null eigenvalue, provided that $0 \in \sigma(A)$.
	We exclude the  case of  $A$ generating trivial  (identity) dynamics, which  happens when   $A\sim \lambda_1 \mathbb{I}_{\Theta}$ with $\lambda_1 \neq 0$. This is equivalent to $\sigma(A)=\{\lambda_1\}$ and $|\lambda_1|=1$, see  Prop. \ref{Adef}, and $A$ is then necessarily unitary.
	There are three cases to be considered: 
	\begin{enumerate}[leftmargin=11.5mm]\itemsep=-0.5mm
		\item[({G})]   the generic case of  $A$ being non-trivial, non-singular, and non-defective, i.e., $ 0 \notin\sigma(A)$ and $\lambda_1 \neq \lambda_2$;
		\item[({D})] the case of $A$ being defective   and non-singular, i.e.,   $\sigma(A)=\{\lambda_1\}$ and   $0<|\lambda_1|<1$;
		\item[({S})] the case of $A$ being singular, i.e., $ 0 \in\sigma(A)$.
	\end{enumerate}
	
	First, we examine Cases ({G}) and ({D}), while Case ({S}) will be investigated separately later on (see p. \pageref{pageS}).
	Let $n, m \in \N$, $m \neq n$.

	\begin{enumerate}[leftmargin=7.5mm]
		\item[({G})] 
		$\mathcal{A}$ has two fixed points   \mbox{$[{{e}_1}]$, $[{{e}_2}]$}.  
		For ${w}=c_1{{e}_1}+c_2{{e}_2}$, where $c_1, c_2 \in \C\setminus\{0\}$, we have 
		\begin{equation*}
			{A}^{n}{w}=c_1\lambda_1^n{{e}_1}+c_2\lambda_2^n{{e}_2},
		\end{equation*}
		which gives 
		\begin{equation*}
			\mathcal{A}^{n}{[w]}=c_1{[{e}_1]}+c_2 \big|\tfrac{\lambda_2}{\lambda_1}\big|^n  \exp({\I\psi n})[{e}_2],
		\end{equation*}
		so we obtain the following subcases:
		\begin{enumerate}[leftmargin=9.5mm]\itemsep=1mm
			\item[({G1})]
			if $|\lambda_1|>|\lambda_2|$, then 
			$\mathcal{A}^{m}{[w]}\neq\mathcal{A}^{n}{[w]}$   and 	$\mathcal{A}^n{[w]} \overset{n \to \infty}{\longrightarrow}  {[ {e}_1]}$;
			thus, $[{{e}_1}]$ is attractive. The dynamics generated by  $\mathcal{A}$ is {loxodromic}; 
			\item[({G2})]
			if $|\lambda_1|=|\lambda_2|$, then 
			both fixed points $[{{e}_1}]$ and $[{{e}_2}]$ of $\mathcal{A}$ are neutral and the dynamics generated by  $\mathcal{A}$ is  elliptic. Every non-fixed point is 
			almost periodic\footnote{Recall that in the case of compact metric spaces, a point is \emph{almost periodic} iff it is \emph{Birkhoff-recurrent} iff the~closure of its trajectory is a minimal set of the dynamical system under consideration,  
			see \cite[p. 38]{ellis2014automorphisms}. This result goes back to Birkhoff \cite[pp. 198-199]{birkhoff1927dynamical}, see also  \mbox{\cite[pp. 71, 129]{de1993elements}} or \cite[pp. 169-170]{vries2014topological}, as well as  \cite[p.~30]{gottschalk1955topological} for historical background.
			}
			and it is periodic  iff  $\psi$ is commensurable with~$\pi$. Hence, we  distinguish two further subcases:
			\begin{enumerate}[leftmargin=12mm]\itemsep=2mm
				\item[({G2i)}]	 if  $\psi\notin \mathbb{Q}\pi$, then  $\mathcal{A}^{m}{[w]}\neq\mathcal{A}^{n}{[w]}$  and $\{\mathcal{A}^n{[w]}\colon n \in \mathbb{N}\}$ is a~dense subset of $\{c_1{[ {e}_1]}+c_2\e^{\I\gamma}{[ {e}_2]} \colon  \gamma\in \R\}$;

				\item[({G2ii})] if  $\psi\in \mathbb{Q}\pi\setminus\{0\}$, then $\mathcal{A}^{m}{[w]}=\mathcal{A}^{n}{[w]}$ iff $m=n\hspace{-2mm}\mod \kappa$, where
				$\kappa$ stands for the smallest natural number $k \geq 2$ such that $\exp(\psi\I)$ is a $k$-th root of unity. (That is, $\exp(\psi\I)$ is a primitive $\kappa$-th root of unity.) In the particular case of $\kappa=2$, i.e., $\psi=\pi$, the dynamics generated by  $\mathcal{A}$ is called \emph{circular}. 
			\end{enumerate}
		\end{enumerate}
		
		\item[({D})]  
		$[{{e}_1}]$ is the sole fixed point of $\mathcal{A}$ and the dynamics generated by  $\mathcal{A}$ is {parabolic}.
		Since    ${A}^n{{g}_1}=\lambda_1^{n}{{g}_1}+n	{x}\lambda_1^{n-1}{{e}_1}$, 
		we obtain 
		\vspace{-1mm}
		$$	
		\vspace{-1mm}
		A^nw=(c_1\lambda_1^n+{c}_2n{x}\lambda_1^{n-1}){ {e}_1} +{c}_2\lambda_1^n{ {g}_1}$$ for ${w}=c_1{{e}_1}+{c}_2{{g}_1}$, where $c_1, c_2 \in \C$ and $c_2 \neq 0$.
		%	\end{equation}
		It follows that  
		\begin{equation*}
			\label{evoldef}\mathcal{A}^n{[w]}= \big(\tfrac{c_1}{n}+\tfrac{{x}{c}_2}{\lambda_1} \big){[{e}_1]}+\tfrac{{c}_2}{n}\,{[{g}_1]}\, ,
		\end{equation*}
		so 
		$\mathcal{A}^{m}{[w]}\neq\mathcal{A}^{n}{[w]}$  and $\mathcal{A}^n{[w]} \overset{n \to \infty}{\longrightarrow}  {[ {e}_1]}$.
	\end{enumerate}

	Let us analyse how the action of $\mathcal{A}$ extends  to the whole of the Bloch ball. 
	As one can expect, we shall see that 
	upon extending loxodromic and parabolic M\"obius maps we do not obtain any fixed points in the interior of the ball and the trajectories of all non-fixed points converge to the sole attractive fixed point on the Bloch sphere. An elliptic M\"obius map extends to a dynamical system that has two neutral fixed points on the sphere, and the convex hull of these two states constitutes the set of fixed points of the system, all of which are neutral, while  non-fixed points are almost periodic.

	Let $\rho \in \mathcal{S}(\Theta)\setminus \mathcal{P}(\Theta)$. It is easy to show that there exists a unit vector $\hat{w}\in \Theta$ such that $\rho=\alpha \rho_{e_1}+ (1-\alpha)\rho_{\hat{w}}$ with some $\alpha \in (0,1)$. Let $n, m \in \N$ be such that $m \neq n$. It follows that 
	% 	\vspace{-1mm}
	\begin{align*}
		\tr (A^n \rho A^{* n}) 
		%  \\ & 
		=\alpha |\lambda_1|^{2n} +(1-\alpha) ||A^n \hat{w}  ||^{2n}
		>0;
	\end{align*}
	hence,  putting  $r_n:={||A^n \hat{w}||^{2n}}/{ |\lambda_1|^{2n}}$, we obtain 
	$$\mathsf{F}_1^n(\rho )=
	\frac{ A^n \rho A^{* n} }{ \tr (A^n \rho A^{* n}) }=
	\frac{\alpha}{\alpha+ (1-\alpha)r_n} \rho_{e_1} +  \frac{
		(1-\alpha )r_n }{\alpha+ (1-\alpha)r_n} \rho_{A^n \hat{w}}.$$
	\begin{enumerate}[leftmargin=7.5mm]\itemsep=1mm 
		\item[({G})]	  In the eigenbasis of $A$ we have  $\hat{w}=c_1e_1+c_2e_2$ with some $c_1, c_2 \in \C$, $c_2 \neq 0$, and so  $$||{A}^{n}{\hat{w}}||^{2}=|c_1|^2|\lambda_1|^{2n}+|c_2|^2|\lambda_2|^{2n}+2|\lambda_1 \lambda_2|^{n} \Re ( \overline{c_1}c_2\exp(\I \psi n) \braket{e_1 | e_2}).$$ 
		Observe   that $\rho_{\hat{w}}$ is a fixed point of $\mathsf{F}_1$ iff  $c_1 = 0$, in which case we have  $\rho_{\hat{w}} = \rho_{e_2}$.
		
		\vspace{-1mm}
		
		\begin{enumerate}[leftmargin=10mm]
			\itemsep=1mm 
			\item[({G1})]	  If $|\lambda_1|>|\lambda_2|$, then
			$\mathsf{F}_1^n(\rho) \overset{n \to \infty}{\longrightarrow}  {\rho}_{e_1}$ and $\mathsf{F}_1^{m}(\rho)\neq \mathsf{F}_1^{n}(\rho)$, because
			\begin{itemize}[leftmargin=4mm]\itemsep=1mm
				
				\item if $c_1 \neq 0$,  then  
				$\rho_{A^n \hat{w}} \overset{n \to \infty}{\longrightarrow} \rho_{{e_1}}$ and 
				$\rho_{A^m \hat{w}} \neq \rho_{A^n \hat{w}}$;
				
				\item 	 if $c_1 = 0$, 
				then $r_n = |c_2|^2|{\lambda_2}/{\lambda_1}|^{2n}$, so  $r_n   \overset{n \to \infty}{\longrightarrow} 0$ and  $r_m \neq r_n$.
				
			\end{itemize}

			\item[({G2})]	 If $|\lambda_1|=|\lambda_2|$, then	$r_n = |c_1|^2+|c_2|^2+2\Re(\overline{c_1}c_2 \exp(\I \psi n )\braket{e_1|e_2} )  $. It follows that

			\begin{itemize}[leftmargin=4mm]\itemsep=1mm
				\item  if $c_1 = 0$,   
				then $\rho$ is a fixed point of $\mathsf{F}_1$, because  $\rho_{A^n \hat{w}}=  \rho_{e_2}$ and $r_n=|c_2|^2=1$,  thus also  $\mathsf{F}_1^{n}(\rho)=\rho$;
				\item 	 if $c_1 \neq 0$, 
				then $\rho$ is almost periodic (because so is $\rho_{\hat{w}}$) and its periodicity depends on the commensurability of $\psi$  with $\pi$:
			\end{itemize}  
			
			\begin{enumerate}[leftmargin=16mm] \itemsep=1mm
				\item[({G2i)}]  if   $\psi\notin \mathbb{Q}\pi$, then 
				$\rho$ is almost periodic but not periodic. 	
				In particular, 
				$\mathsf{F}_1^{m}(\rho)\neq \mathsf{F}_1^{n}(\rho)$ and $\left(\mathsf{F}_1^n(\rho)\right)_{n \in \N}$ is non-convergent; 
				
				\item[({G2ii})]   if $\psi\in \mathbb{Q}\pi\setminus\{0\}$ and  $\psi$ is a primitive $\kappa$-th root of unity, then    
				$\rho$ is   periodic of period~$\kappa$.  
				\vspace{-1mm} 
			\end{enumerate}    
		\end{enumerate}
		
		\item[({D})]  
		We have   
		$\mathsf{F}_1^{m}(\rho)\neq \mathsf{F}_1^{n}(\rho)$ and $\mathsf{F}_1^n(\rho_m) \overset{n \to \infty}{\longrightarrow}  
		\rho_{e_1}$, because $\rho_{A^m \hat{w}}\neq \rho_{A^n \hat{w}} $ and  $\rho_{A^n \hat{w}} \overset{n \to \infty}{\longrightarrow}  \rho_{{e_1}}$.
	\end{enumerate}

	\begin{obs} \label{fixedobsss}
		Let $\rho \in \mathcal{S}(\Theta) \setminus \mathcal{P}(\Theta)$. In Cases ({G}) and ({D}), $\rho$  is a fixed point of $\mathsf{F}_1$   iff  $|\lambda_1|=|\lambda_2|$  and $\rho = \alpha {\rho}_{e_1}+(1-\alpha){\rho}_{e_2}$ with some  $\alpha \in (0,1)$, i.e., $\rho \in \conv\{ {\rho}_{e_1}, {\rho}_{e_2}\}$. The matrix representation of $\rho$ with respect to the eigenbasis $\{e_1, e_2\}$ of~$A$ reads 
		%  	\vspace{-1mm}
		\begin{equation*}
			% 	 	  	\vspace{1mm}
			\resizebox{4.5cm}{!}{$
				\begin{bmatrix}
					\alpha  &   \alpha \braket{e_1|e_2}
					\\
					(1-\alpha)\braket{e_2|e_1} & 1-\alpha 
				\end{bmatrix}.	$}
		\end{equation*}	 
	\end{obs}

	\begin{obs}\label{gd1}
		Generic trajectories
		in Cases ({G1}) and ({D}) are qualitatively the same, i.e., infinite and convergent to the sole attractive point $\rho_{e_1}$ of the system, even though the overall dynamics on the ball is different.
	\end{obs}

	It remains to investigate the case of $A$ being singular, i.e., Case (S). \label{pageS} Recall that $\mathcal{A}$ is then  undefined at exactly one point in $\mathbb{P}\Theta$ which {can be identified with} the eigenspace of $A$ associated with null  eigenvalue. There are two subcases, corresponding to the multiplicity of zero as the eigenvalue of $A$:
	%\newpage
	\vspace{-1mm}
	\begin{enumerate}[leftmargin=8.75mm] 
		\itemsep-0.25mm
		\item[\rm({S1})] if $\lambda_1\neq\lambda_2=0$, i.e., $A \sim \diag(\lambda_1, 0)$, then for every $w={c_1e_1+c_2e_2}$ ($c_1, c_2 \in \C$) we obtain ${A}w=c_1e_1$. Consequently, $\mathcal{A}{[e_1]}=[{e}_1]$,  $\mathcal{A}$ is undefined at $[{e}_2]$, and  $\mathcal{A}{[w]}=[{e}_1]$ for every $[w]\notin \{[e_1], [e_2]\}$;
		\item[\rm({S2})]  if $\lambda_1=\lambda_2=0$, i.e., $A \sim \begin{bsmallmatrix}
			0 & 1 \\ 0 & 0
		\end{bsmallmatrix}$, then ${A}{ {e}_1}=0$ and ${A}{ {g_1}}={ {e_1}}$, so for every $w={c_1e_1+c_2g_1}$  ($c_1, c_2 \in \C$)  we obtain ${A}w=c_2e_1$. It follows that  $\mathcal{A}$ has no fixed points, $\mathcal{A}$ is undefined at ${[e_1]}$, and  if $[w]\neq [e_1]$, then $\mathcal{A}{[w]}=[{e}_1]$.			
	\end{enumerate}

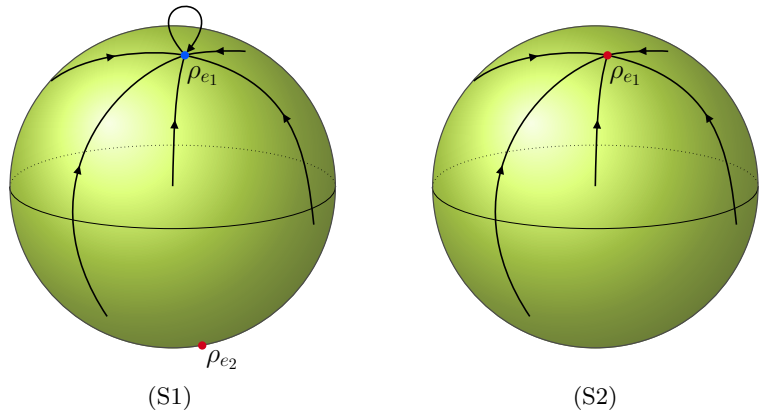
\begin{figure}[h]\captionsetup[subfigure]{labelformat=empty}	 
		\vspace{-8mm}
		\subfloat[{(S1)} ]
		{	
			\scalebox{0.6}{

				\begin{tikzpicture}[x=0.75pt,y=0.75pt,yscale=-1,xscale=1]
					\shade[ball color = lime!80, opacity = 0.85] (246.4,164.7) circle (3.57cm);
					
					\draw[dotted] (111,164.7) arc (180:360:135.5 and 35);
					\draw (381.7,164.7) arc (0:180:135.5 and 35);

					\draw  [color={rgb, 255:red, 74; green, 74; blue, 74 }  ,draw opacity=1 ][line width=0.75]  (111.01,164.65) .. controls (111.01,90.11) and (171.61,29.69) .. (246.38,29.69) .. controls (321.14,29.69) and (381.74,90.11) .. (381.74,164.65) .. controls (381.74,239.19) and (321.14,299.61) .. (246.38,299.61) .. controls (171.61,299.61) and (111.01,239.19) .. (111.01,164.65) -- cycle ;
					
					\draw  [draw opacity=0][fill={rgb, 255:red, 208; green, 2; blue, 27 }  ,fill opacity=1 ] (267.45,297.65) .. controls (267.45,295.74) and (269.01,294.18) .. (270.93,294.18) .. controls (272.85,294.18) and (274.4,295.74) .. (274.4,297.65) .. controls (274.4,299.56) and (272.85,301.11) .. (270.93,301.11) .. controls (269.01,301.11) and (267.45,299.56) .. (267.45,297.65) -- cycle ;
				
					\draw [color={rgb, 255:red, 0; green, 0; blue, 0 }  ,draw opacity=1 ][line width=1.0]    (307,51.05) .. controls (295.5,50.05) and (267.5,51.05) .. (256.5,54.39) ;
					\draw [shift={(281.56,51.09)}, rotate = 357.45] [fill={rgb, 255:red,0 ; green, 0; blue, 0 }  ,fill opacity=1 ][line width=0.08]  [draw opacity=0] (7.29,-3.46) -- (0,0) -- (7.29,3.46) -- cycle    ;
				
					\draw [color={rgb, 255:red, 0; green, 0; blue, 0 }  ,draw opacity=1 ][line width=1.0]    (145.22,75.9) .. controls (173.11,53.41) and (228.13,48.83) .. (256.5,54.39) ;
					\draw [shift={(198.81,54.93)}, rotate = 529.64] [fill={rgb, 255:red, 0; green, 0; blue, 0 }  ,fill opacity=1 ][line width=0.08]  [draw opacity=0] (7.29,-3.46) -- (0,0) -- (7.29,3.46) -- cycle    ;
				
					\draw [color={rgb, 255:red, 0; green, 0; blue, 0 }  ,draw opacity=1 ][line width=1.0]    (246.34,164.1) .. controls (247.18,122.01) and (247.93,80.8) .. (256.5,54.39) ;
					\draw [shift={(248.18,108.5)}, rotate = 453.29] [fill={rgb, 255:red, 0; green, 0; blue, 0 }  ,fill opacity=1 ][line width=0.08]  [draw opacity=0] (7.29,-3.46) -- (0,0) -- (7.29,3.46) -- cycle    ;
				
					\draw [color={rgb, 255:red, 0; green, 0; blue, 0 }  ,draw opacity=1 ][line width=1.0]    (260.84,48.49) .. controls (297.81,0.34) and (212.83,1.84) .. (254.91,52.22) ;
					\draw [shift={(257.75,52.3)}, rotate = 310.56] [fill={rgb, 255:red, 0; green, 0; blue, 0 }  ,fill opacity=1 ][line width=0.08]  [draw opacity=0] (7.29,-3.46) -- (0,0) -- (7.29,3.46) -- cycle    ;
				
					\draw [color={rgb, 255:red, 0; green, 0; blue, 0 }  ,draw opacity=1 ][line width=1.0]    (191.8,273.27) .. controls (123.44,170.73) and (190.6,75.68) .. (256.5,54.39) ;
					\draw [shift={(168.99,146.67)}, rotate = 465.54] [fill={rgb, 255:red, 0; green, 0; blue, 0 }  ,fill opacity=1 ][line width=0.08]  [draw opacity=0] (7.29,-3.46) -- (0,0) -- (7.29,3.46) -- cycle    ;
				 
					\draw [color={rgb, 255:red, 0; green, 0; blue, 0 }  ,draw opacity=1 ][line width=1.0]    (363.76,196.34) .. controls (357.46,137.88) and (349.36,74.01) .. (256.5,54.39) ;
					\draw [shift={(337.93,103.23)}, rotate = 417.41999999999996] [fill={rgb, 255:red, 0; green, 0; blue, 0 }  ,fill opacity=1 ][line width=0.08]  [draw opacity=0] (7.29,-3.46) -- (0,0) -- (7.29,3.46) -- cycle    ;
				
					\draw  [draw opacity=0][fill={rgb, 255:red, 3; green, 86; blue, 251 }  ,fill opacity=1 ] (253.02,54.39) .. controls (253.02,52.48) and (254.58,50.93) .. (256.5,50.93) .. controls (258.42,50.93) and (259.97,52.48) .. (259.97,54.39) .. controls (259.97,56.31) and (258.42,57.86) .. (256.5,57.86) .. controls (254.58,57.86) and (253.02,56.31) .. (253.02,54.39) -- cycle ;

					\draw (272.93,300) node [anchor=north west][inner sep=0.75pt]  [font=\LARGE]  {$\rho _{e_{2}}$};
				
					\draw (256.94,62) node [anchor=north west][inner sep=0.75pt]  [font=\LARGE]  {$\rho _{e_{1}}$};

			\end{tikzpicture}}
		}
		\hspace{8mm}		
		\subfloat[{(S2)}]
		{
			
			\scalebox{0.6}{
				\begin{tikzpicture}[x=0.75pt,y=0.75pt,yscale=-1,xscale=1]
					
					\shade[ball color = lime!80, opacity = 0.85] (246.4,164.7) circle (3.57cm);
					
					\draw[dotted] (111,164.7) arc (180:360:135.5 and 35);
					\draw (381.7,164.7) arc (0:180:135.5 and 35);	
					
						\draw  [color={rgb, 255:red, 74; green, 74; blue, 74 }  ,draw opacity=1 ][line width=0.75]  (111.01,164.65) .. controls (111.01,90.11) and (171.61,29.69) .. (246.38,29.69) .. controls (321.14,29.69) and (381.74,90.11) .. (381.74,164.65) .. controls (381.74,239.19) and (321.14,299.61) .. (246.38,299.61) .. controls (171.61,299.61) and (111.01,239.19) .. (111.01,164.65) -- cycle ;
			
					\draw [color={rgb, 255:red, 0; green, 0; blue, 0 }  ,draw opacity=1 ][line width=1.0]    (307,51.05) .. controls (295.5,50.05) and (267.5,51.05) .. (256.5,54.39) ;
					\draw [shift={(281.56,51.09)}, rotate = 357.45] [fill={rgb, 255:red, 0; green, 0; blue, 0 }  ,fill opacity=1 ][line width=0.08]  [draw opacity=0] (7.29,-3.46) -- (0,0) -- (7.29,3.46) -- cycle    ;
			
					\draw [color={rgb, 255:red, 0; green, 0; blue, 0 }  ,draw opacity=1 ][line width=1.0]    (145.22,75.9) .. controls (173.11,53.41) and (228.13,48.83) .. (256.5,54.39) ;
					\draw [shift={(198.81,54.93)}, rotate = 529.64] [fill={rgb, 255:red, 0; green, 0; blue, 0 }  ,fill opacity=1 ][line width=0.08]  [draw opacity=0] (7.29,-3.46) -- (0,0) -- (7.29,3.46) -- cycle    ;
				
					\draw [color={rgb, 255:red, 0; green, 0; blue, 0 }  ,draw opacity=1 ][line width=1.0]    (246.34,164.1) .. controls (247.18,122.01) and (247.93,80.8) .. (256.5,54.39) ;
					\draw [shift={(248.18,108.5)}, rotate = 453.29] [fill={rgb, 255:red, 0; green, 0; blue, 0 }  ,fill opacity=1 ][line width=0.08]  [draw opacity=0] (7.29,-3.46) -- (0,0) -- (7.29,3.46) -- cycle    ;
				
					\draw [color={rgb, 255:red, 0; green, 0; blue, 0 }  ,draw opacity=1 ][line width=1.0]    (191.8,273.27) .. controls (123.44,170.73) and (190.6,75.68) .. (256.5,54.39) ;
					\draw [shift={(168.99,146.67)}, rotate = 465.54] [fill={rgb, 255:red, 0; green, 0; blue, 0 }  ,fill opacity=1 ][line width=0.08]  [draw opacity=0] (7.29,-3.46) -- (0,0) -- (7.29,3.46) -- cycle    ;
				 
					\draw [color={rgb, 255:red, 0; green, 0; blue, 0 }  ,draw opacity=1 ][line width=1.0]    (363.76,196.34) .. controls (357.46,137.88) and (349.36,74.01) .. (256.5,54.39) ;
					\draw [shift={(337.93,103.23)}, rotate = 417.41999999999996] [fill={rgb, 255:red, 0; green, 0; blue, 0 }  ,fill opacity=1 ][line width=0.08]  [draw opacity=0] (7.29,-3.46) -- (0,0) -- (7.29,3.46) -- cycle    ;
				
					\draw  [draw opacity=0][fill={rgb, 255:red, 208; green, 2; blue, 27 }  ,fill opacity=1 ] (253.02,54.39) .. controls (253.02,52.48) and (254.58,50.93) .. (256.5,50.93) .. controls (258.42,50.93) and (259.97,52.48) .. (259.97,54.39) .. controls (259.97,56.31) and (258.42,57.86) .. (256.5,57.86) .. controls (254.58,57.86) and (253.02,56.31) .. (253.02,54.39) -- cycle ;

					\draw (256.94,62) node [anchor=north west][inner sep=0.75pt]  [font=\LARGE]  {$\rho _{e_{1}}$};
					\draw (272.93,300) node [anchor=north west][inner sep=0.75pt]  [font=\LARGE]  {$\phantom{\rho _{e_{2}}}$};	
					
				\end{tikzpicture}
			}
		}

		\captionsetup{width=0.8\linewidth}
		\caption{
			The action of $\mathsf{F}_1$	on $\mathcal{S}(\Theta)$	in the case of $A$ being singular \\
			%		The red dot indicates the sole state at which $\mathsf{F}_1$ is not defined. \\
			{(S1)} $\mathcal{S}(\Theta)\setminus\{\rho_{e_2}\}$ gets collapsed to
			$\rho_{e_1}$, which is a fixed point of $\mathsf{F}_1$; \\ {(S2)} $\mathcal{S}(\Theta)\setminus\{\rho_{e_1}\}$ gets collapsed to $\rho_{e_1}$, at which $\mathsf{F}_1$ is not defined. 
		}\label{nullballs}
	\end{figure}

	\noindent
	Extending the action of $\mathcal{A}$ to $\mathcal{S}(\Theta)$, we easily obtain (see also Fig. \ref{nullballs}):
	\vspace{-1mm}
	\begin{enumerate}[leftmargin=8.75mm] 	
		\itemsep-0.25mm
		\item[\rm({S1})]	if  $\lambda_1\neq \lambda_2=0$, then ${ \operatorname{dom} \mathsf{F}_1|_{\mathcal{S}(\Theta)}=\mathcal{S}(\Theta)\setminus\{\rho_{e_2}\}}$. 
		For $\rho\neq \rho_{e_2}$ we have 
		$\mathsf{F}_1(\rho)
		=\rho_{e_1}$. 
		\item[\rm({S2})]  if $\lambda_1=\lambda_2=0$, then ${\operatorname{dom} \mathsf{F}_1|_{\mathcal{S}(\Theta)}=\mathcal{S}(\Theta)\setminus\{\rho_{e_1}\}}$.
		For $\rho \neq \rho_{e_1}$ we have $\mathsf{F}_1(\rho) = \rho_{e_1}$; in particular, $\mathsf{F}_1^{n}$ is undefined on $\rho$ for $n\geq 2$.
	\end{enumerate}

	\pagebreak

	For the sake of convenience, we now reassemble the results that concern the fixed points  and generic trajectories of $\mathsf{F}_1|_{\mathcal{S}(\Theta)}$ in the case of $A$ being  non-unitary.

	\begin{obs} \label{fixed} Fixed points of {$\mathsf{F}_1|_{\mathcal{S}(\Theta)}$}  if $A$ is not unitary.
		\vspace{-1mm}	
		\begin{enumerate}[leftmargin=15mm]
			\itemsep-0.25mm
			\item[{ ({G1})}] If $0<|\lambda_2|<|\lambda_1| \leq 1$, then $\operatorname{fix}\mathsf{F}_1|_{\mathcal{S}(\Theta)}=\{ {\rho}_{e_1}, {\rho}_{e_2}\}$.

			\item[{({G2})}] If $0< |\lambda_2|=|\lambda_1|< 1 $ with $\psi \neq 0$, then $\operatorname{fix}\mathsf{F}_1|_{\mathcal{S}(\Theta)}=\conv\{ {\rho}_{e_1}, {\rho}_{e_2}\}$.
			
			\item [{({D})}] If $\lambda_1=\lambda_2$ with $0<|\lambda_1|<1$, then   $\operatorname{fix}\mathsf{F}_1|_{\mathcal{S}(\Theta)}=\{ {\rho}_{e_1}\}$.

			\item[{ ({S1})}] If $\lambda_2=0$ and  $0 < |\lambda_1|\leq 1$, then  $\operatorname{fix}\mathsf{F}_1|_{\mathcal{S}(\Theta)}=\{ {\rho}_{e_1}\}$.
			
			\item[{ ({S2})}] If $\lambda_1=\lambda_2=0$,  then 	 $\operatorname{fix}\mathsf{F}_1|_{\mathcal{S}(\Theta)}=\varnothing$.
		\end{enumerate}   
	\end{obs}

	\begin{obs} \label{traj} Generic trajectories under {$\mathsf{F}_1|_{\mathcal{S}(\Theta)}$} if $A$ is not unitary.
		\\	Let $\rho \in  \operatorname{dom} \mathsf{F}_1|_{\mathcal{S}(\Theta)} \setminus   \operatorname{fix}\mathsf{F}_1|_{\mathcal{S}(\Theta)}$. 
		\vspace{-1mm}
		\begin{enumerate}[leftmargin=15mm]
			\itemsep-0.5mm
		
			\item[({G1})] If $0<|\lambda_2|<|\lambda_1| \leq 1$, then $\rho$ has an infinite trajectory convergent to $\rho_{e_1}$.

			\item[({G2i})] If $0< |\lambda_2|=|\lambda_1|<  1$ and  $\psi\notin \pi\mathbb{Q}$, then the trajectory of $\rho$ is  almost periodic but not periodic, so infinite and non-convergent. 
			
			\item[({G2ii})]   If $0 < |\lambda_2|=|\lambda_1|< 1$ and   $\psi\in \pi\mathbb{Q}\setminus \{0\}$ is a primitive $\kappa$-th root of unity, then $\rho$ has a periodic trajectory with period $\kappa$. 
			
			\item[({D})] If $\lambda_1=\lambda_2$ with $0<|\lambda_1|<1$, then $\rho$ has an infinite trajectory convergent to $\rho_{e_1}$ (cf. Obs. \ref{gd1}).

			\item[({S1})] If $\lambda_2=0$ and  $0 < |\lambda_1|\leq 1$, then $\mathsf{F}_1(\rho)=\rho_{e_1}\in \operatorname{fix}\mathsf{F}_1|_{\mathcal{S}(\Theta)}$.
			
			\item[({S2})]   If $\lambda_1=\lambda_2=0$, then $\mathsf{F}_1(\rho)=\rho_{e_1}\notin \operatorname{dom} \mathsf{F}_1|_{\mathcal{S}(\Theta)}$.
		\end{enumerate}  
	\end{obs}

	\noindent
	$\RHD$ {\bf{Step III. {Singularity.}}}
	The last step in determining the trajectories of $\rho_{\mathsf{v}}$ and $\rho_{\mathsf{m}}$ under $\mathsf{F}_1$ is to decide for each of these states whether it is a generic or singular point of ${\mathsf{F}_1}$, i.e., whether or not it belongs to $\operatorname{dom} \mathsf{F}_1|_{\mathcal{S}(\Theta)}\setminus \operatorname{fix}\mathsf{F}_1|_{\mathcal{S}(\Theta)}$.
	Throughout this step we assume that  $A$ is non-unitary.
	\begin{proposition}\label{genericrm}
		If $A$ is not unitary, then $\rho_\mathsf{m}$ is a generic point of ${\mathsf{F}_1}$.
	\end{proposition}
	\begin{proof}
		First, recall that there is at most one state in $\mathcal{S}(\Theta)$ at which  ${\mathsf{F}_1}$ is undefined  and this state is pure, see \eqref{dynA}, so  $\rho_{\mathsf{m}} \in \operatorname{dom}\mathsf{F}_1|_{\mathcal{S}(\Theta)}$
		
		Next, by contradiction, suppose that $\rho_{\mathsf{m}} \in \operatorname{fix}\mathsf{F}_1|_{\mathcal{S}(\Theta)}$. 
		Recall that, in the case of $A$ being non-unitary, a state from 
		$\mathcal{S}(\Theta) \setminus \mathcal{P}(\Theta)$
		can be a fixed point of  ${\mathsf{F}_1}$ only if $0<|\lambda_1|=|\lambda_2|< 1$ and $\psi \neq 0$, see  Obs.~\ref{fixed}. In addition,     $\rho_{\mathsf{m}} \sim 	\scaleto{\tfrac{1}{2}}{11pt} \begin{bsmallmatrix}
			1  & 0   \\
			0 & 1   \end{bsmallmatrix}$  with respect to every basis of $\Theta$, so if $\rho_{\mathsf{m}} \in \operatorname{fix}\mathsf{F}_1|_{\mathcal{S}(\Theta)}$, then Obs.~\ref{fixedobsss} implies that $\braket{e_1|e_2}=0$. Hence,  $A$ has an orthogonal eigenbasis, thus also $|\lambda_1|=1$, see Prop.~\ref{orthbasis}. In consequence,   $|\lambda_1|=|\lambda_2|=1$, which is a contradiction.  Therefore, $\rho_{\mathsf{m}} \notin \operatorname{fix}\mathsf{F}_1|_{\mathcal{S}(\Theta)}$, as claimed.
	\end{proof}

	Next, recall   from Obs. \ref{fixedobs}  that  $\rho_{\mathsf{v}}$ is a singular point of ${\mathsf{F}_1}$ iff $\mathsf{v}=\Pi_1U\mathsf{z}$ is an eigenvector of $A:=\Pi_1U|_\Theta$. 
	In the next theorem we express this condition in terms of the eigenvalues of $A$. (Recall that they are assumed to be  ordered as $|\lambda_2| \leq |\lambda_1|$.) The following lemma is \cite[Lemma 1]{projection} adapted to the current context. 
	For completeness, we include this lemma along with its proof.  
	
	\pagebreak
	
	\begin{lemma}\label{lemmal1} 
		If $|\lambda_1|=1$, then ${e_1}$ is an eigenvector of $U\!$ associated with $\lambda_1$. 
	\end{lemma}
	\begin{proof} 
		Clearly, we have   $1=||U{e_1}|| ^2 = ||\Pi_1U{e_1}||^2+||\Pi_2U{e_1}||^2=|\lambda_1|^2+||\Pi_2U{e_1}||^2$. Hence, if $|\lambda_1|=1$, then  $\Pi_2U{e_1}=0$. Thus, $U{{e_1}} \in \Theta$ and so $U{{e_1}}=\Pi_1U{{e_1}}= \lambda_1{{e_1}}$, as desired. 
	\end{proof}

	\begin{proposition}\label{eigen}
		If  $A$ is not unitary, then $\mathsf{v}$ is an eigenvector of $A$  iff $|\lambda_1|=1$. \mbox{In that  case} the eigenvalue corresponding to $\mathsf{v}$ is equal to $\lambda_2$. 
	\end{proposition}
	
	\begin{proof}We consider two cases, corresponding to whether or not $A$ is defective.
		\begin{itemize} \itemsep=-1mm
			\item	Assume that $A$ is not defective. 
			\begin{itemize}		
				\item[($\Leftarrow$)] 
				Assume that $|\lambda_1|=1$. It follows from Lemma \ref{lemmal1} that ${e_1}$ is an eigenvector of $U$ associated with $\lambda_1$. Let  ${w} \in \Theta$ complete $\{{e_1},{\mathsf{z} }\}$ to  an orthonormal basis of $\C^3$. 
				We deduce from Prop.~\ref{orthbasis} that ${w}$ is an eigenvector of $A$ associated with $\lambda_2$. 
				Clearly,   $\spann\{{w},{\mathsf{z} }\}$ is invariant under $U$,  so, in particular,  $U\mathsf{z} \in \spann\{{w},{\mathsf{z} }\}$. It follows that 
				$\Pi_1U{\mathsf{z} } \in \Pi_1(\spann\{{w},\mathsf{z}\})= \spann\{{w}\}$. 
				Moreover,  	
				$\Pi_1U{\mathsf{z} }\neq 0$ since $||\Pi_1U{\mathsf{z} }||^2= 1-|\omega|^2$ and $|\omega|\neq 1$, because $A$ is not unitary, see Prop. \ref{unitarycond}. 
				Hence, $\mathsf{v} = \Pi_1U{\mathsf{z} }$ is an eigenvector of~$A$ and it is associated with $\lambda_2$, as required.

				\item[($\Rightarrow$)] Assume that $\mathsf{v}$ is an eigenvector of $A$ associated with $\lambda_a$, $a \in \{1,2\}$, i.e.,  we have 	$\Pi_1 U \mathsf{z} =q e_a$ with some $q \in \C \setminus\{0\}$.	Thus, 
				$
				U\mathsf{z}=q e_a+\omega\mathsf{z}.$
				Moreover, since $\Pi_1 U e_a=\lambda_a e_a$,   we have 
				$
				Ue_a=\lambda_a e_a+r\mathsf{z}$ 
				with some   $r \in \C$. Therefore,  $\spann\{{e_a},{\mathsf{z} }\}$ is $U$-invariant. 
				Let  ${w} \in \Theta$ be 
				such that it completes  $\{{e_a}, \mathsf{z}\}$ to an orthonormal basis of $\C^3$. 
				It follows that 
				$\spann\{w\} \subset \Theta$ is   $U$-invariant, and so   
				$w$ is an eigenvector of both  $U$ and $A$ associated with $\lambda_b$, where $b \in \{1,2\}$, $b \neq a$. Hence, $|\lambda_b|=1$ as an eigenvalue of $U$. Since $A$ is not unitary, we have $|\lambda_2|<1$. We conclude that  $b=1$ and $a=2$, i.e.,  $|\lambda_1|=1$ and the eigenvalue of $A$ associated with $\mathsf{v}$ is   equal to $\lambda_2$, as claimed.
				
			\end{itemize}

			\item It remains to show that if $A$ is defective, then $\mathsf{v}$ is not an eigenvector of $A$, cf. Prop.~\ref{Adef}.  
			By contradiction, suppose that $\mathsf{v}$	is an eigenvector of  $A$, i.e.,   
			$\Pi_1 U \mathsf{z}=q{e_1}$ with some 
			$q\in \C\setminus\{0\}$.
			We deduce, arguing as above, that  $\spann\{{e_1},\mathsf{z}\}$ is $U$-invariant, and so the ray in $\Theta$ that is orthogonal to $e_1$ is an eigenspace of $A$.  This contradicts the assumption of $A$ being defective and concludes the proof. 
		\end{itemize}
	\end{proof}
	
	\begin{corollary}
		From Thm. \ref{eigen} it follows that $\rho_{\mathsf{v}}$ is a singular point of ${\mathsf{F}_1}$ iff $|\lambda_1|=1$ and in that case  we have $\rho_{\mathsf{v}}=\rho_{e_2}$,  thus also  $\mathsf{p}_1(\rho_{\mathsf{v}})=|\lambda_2|^2$. Hence,
		%		\vspace{-0.5mm}
		\begin{itemize}\label{corfixedv}
			\item  		$\rho_{\mathsf{v}}\in \operatorname{fix}\mathsf{F}_1|_{\mathcal{S}(\Theta)}$  iff $|\lambda_1|=1$ and $\lambda_2\neq 0$;
			
			\item $\rho_{\mathsf{v}} \notin \operatorname{dom} \mathsf{F}_1|_{\mathcal{S}(\Theta)}$ iff   $|\lambda_1|=1$ and $\lambda_2=0$.
		\end{itemize}
	\end{corollary}
	
	\vspace{1mm}

	\begin{remark}
		Using Obs. \ref{traj}, Prop. \ref{genericrm}, and  Cor. \ref{corfixedv}, we can deduce	$\mathfrak{n}(\rho_{\mathsf{z}})$ and 	$\mathfrak{n}(\rho_{\mathsf{m}})$ in the missing case of $0 \in \sigma(A)$, cf.  Cor. \ref{ninfinity}. Namely, $\mathfrak{n}(\rho_{\mathsf{z}})=\mathfrak{n}(\rho_{\mathsf{v}})+1$ and 
		%		\vspace{-0.5mm}
		\begin{itemize}
			\item  if $\lambda_1 \neq \lambda_2=0$  then  $\mathfrak{n}(\rho_{\mathsf{m}})=\infty$ and (a)  $\mathfrak{n}(\rho_{\mathsf{v}})=\infty$ if $|\lambda_1|<1$;
			\\[0.25em]
			{\phantom{.}} \hspace{57.0mm}	(b)   	$\mathfrak{n}(\rho_{\mathsf{v}})=0$ if $|\lambda_1|=1$;
			\item 	if $\lambda_1 =\lambda_2=0$ then $\mathfrak{n}(\rho_{\mathsf{m}})=1$ and $\mathfrak{n}(\rho_{\mathsf{v}})=1$. 
		\end{itemize}
	\end{remark}

	\pagebreak 
	
	\noindent
	$\RHD$ {\bf{Classification.}}  
	In the following proposition we show that every pair of numbers  from the unit disc in $\C$ can constitute the spectrum of $A$.
	\begin{proposition}\label{alltypes}
		Let $\mu_1, \mu_{2}$ belong to the unit disc in $\C$. Then there exists a~unitary matrix $V \in \C^{3\times 3}$ such that $\mu_1$, $\mu_{2}$ are the eigenvalues of the $2 \times 2$ leading principal submatrix of $V$.
	\end{proposition}
	\begin{proof} Let $\mu_1, \mu_{2} \in \C$ be such that $0 \leq  |\mu_1|,  |\mu_{2}| \leq 1$. 
		Put
		\begin{equation*}
			\widetilde{V}:= \begin{bmatrix}
				|\mu_1|  & \sqrt{(1-|\mu_1|^2)(1-|\mu_2|^2)} &  -|\mu_2|\sqrt{1-|\mu_1|^2} 
				\\[0.25em] 
				0 & |\mu_2|  {\: \e^{\I\Arg(\mu_2)-\I\Arg(\mu_1)}} &    \sqrt{1-|\mu_2|^2}  {\: \e^{\I\Arg(\mu_2)-\I\Arg(\mu_1)}}  \\[0.25em]
				\,\sqrt{1-|\mu_1|^2} &   -|\mu_1|\sqrt{1-|\mu_2|^2} & |\mu_1\mu_2| 
			\end{bmatrix},
		\end{equation*}
		where we adopt  the convention that $\Arg 0=0$. 
		By direct calculation we verify that $\widetilde{V}$ is  unitary. Thus, $V:=\e^{\I\Arg(\mu_1)}\widetilde{V}$ is also unitary. The $2 \times 2$ leading principal submatrix of $V$ reads
		$
		\begin{bsmallmatrix}
			\mu_1  & * \\ 
			0 & \mu_2  \\
		\end{bsmallmatrix},
		$
		so $\mu_1$ and $\mu_2$ are its eigenvalues, as desired.
	\end{proof}
	
	Compiling all of the above results, we can finally classify the Markov chains that can be generated by $\mathcal{F}_{U,\,\Pi}$.
	We know from Prop.~\ref{genericrm} that the trajectory of $\rho_{\mathsf{m}}$ is   given by Obs.~\ref{traj}, provided that $A$ is non-unitary. The trajectory of $\rho_{\mathsf{v}}$ depends on whether $|\lambda_1|$ is of unit length, see Cor. \ref{corfixedv}, which splits Cases {({G1})} \& {({S1})}  into further subcases:
	
	\vspace{-2mm}
	
	\begin{enumerate}[(a)]
		\itemsep=-0.5mm
		\item  if $|\lambda_1|<1$, then $\rho_{\mathsf{v}}$ is generic and its trajectory is   given by Obs. \ref{traj},
		
		\item if $|\lambda_1|=1$, then $\rho_{\mathsf{v}}$ is singular and its trajectory   can be deduced from  Cor. \ref{corfixedv}.
	\end{enumerate}	 
	
	In what follows, the states from $\mathcal{S}(\Theta) \setminus \mathcal{P}(\Theta)$ are referred to as \emph{mixed} states. Recall that if $A$ is non-singular, then $\mathsf{F}_1$ is a bijection on $\mathcal{S}(\Theta)$ and on $\mathcal{P}(\Theta)$, see Prop.~\ref{Ainvertibleomegazero} \&  Cor.~\ref{f1bijective}.  Thus, the trajectory of a pure (resp. mixed) state under $\mathsf{F}_1$  consists entirely of pure (resp. mixed) states.
	
	\begin{theorem} \label{chainsclass} Classification of the types of chains that can be generated by $\mathcal{F}_{U,\,\Pi}$.
		
		\vspace{0.75mm}
		
		\noindent
		In brackets we indicate the case in Obs. \ref{traj} from which a given type originates, along with the subcase  \emph{{(a)}} or \emph{{(b)}}  corresponding to whether $|\lambda_1|<1$ or $|\lambda_1|=1$, respectively.
		See   Fig.~\ref{diag2} for the  transition diagrams.

		\begin{enumerate}[leftmargin=34mm]
			\itemsep-0.25mm
			\item[\sf \underline{generic} {\rm[{G1a+D}]}:] If  $0<|\lambda_2|<|\lambda_1|< 1$ or $\lambda_1=\lambda_2$ with $0<|\lambda_1|< 1$, then $\rho_{\mathsf{v}}$ has an~infinite trajectory over pure states and $\rho_{\mathsf{m}}$ has an infinite trajectory over mixed states; both these trajectories converge~to~$\rho_{e_1}$.
			
			\item[\sf \underline{taupek} {\rm[{G1b}]}:] If $0<|\lambda_2|<|\lambda_1|= 1$, then
			$\rho_{\mathsf{v}}=\rho_{e_2}$ and the system loops there with probability $|\lambda_2|^2$, while $\rho_{\mathsf{m}}$ has an infinite trajectory convergent over mixed states to $\rho_{e_1}$.

			\item[\sf \underline{$\infty$-elliptic} {\rm[{G2i}]}:]  If    $0<|\lambda_2|=|\lambda_1|< 1$ and $\psi\notin \pi\mathbb{Q}$, then $\rho_{\mathsf{v}}$ has an infinite trajectory over pure states and $\rho_{\mathsf{m}}$ has an infinite trajectory over mixed states; both these trajectories are almost periodic.

			\item[ \sf\underline{finite-elliptic} {\rm[{G2ii}]}:] If $0<|\lambda_2|=|\lambda_1|< 1$ and  $\psi\in \pi\mathbb{Q}\setminus\{0\}$, then $\rho_{\mathsf{v}}$ has a periodic trajectory over pure states and $\rho_{\mathsf{m}}$ has a periodic trajectory over mixed states. Their  periods are equal to $\kappa$ such that $\psi$ is a primitive $\kappa$-th root of unity. If  $\kappa=2$, then this chain is called \underline{\sf circular}.

			\item[\sf \underline{null\vphantom{y}} {\rm[{S}]}:]
			If $\lambda_2=0$,
			then the system cannot loop over $\rho_{{\mathsf{z}}}$ since the probability of it going from $\rho_{{\mathsf{z}}}$ to the ball is equal to
			$\mathsf{p}_1(\rho_\mathsf{z})=1-|\omega|^2=1$, see Prop.~\ref{Ainvertibleomegazero}.
			We distinguish three subtypes of null chains:
			\vspace{-0.5mm}
			\begin{enumerate}[\rm (i), leftmargin=25mm]
				\itemsep=0.75mm
				\item[\sf \underline{generic-null} {\rm[{S1a}]}:]  if $0<|\lambda_1|< 1$, then  ${\mathsf{F}_1}(\rho_{\mathsf{m}})={\mathsf{F}_1}(\rho_{\mathsf{v}})=\rho_{e_1}$, and the system loops over  $\rho_{e_1}$ with probability $|\lambda_1|^2$; 
				\item[\sf \underline{taupek-null} {\rm[{S1b}]}:] if $|\lambda_1|=1$, then ${\mathsf{F}_1}(\rho_{\mathsf{m}})=\rho_{e_1}$, and the system loops over  $\rho_{e_1}$ with unit probability, while 
				$\rho_{\mathsf{v}}=\rho_{{e_2}} \notin \operatorname{dom} \mathsf{F}_1|_{\mathcal{S}(\Theta)}$, so the system returns from $\rho_{\mathsf{v}}$ to $\rho_{\mathsf{z}}$ with unit probability;
				\item[\sf \underline{double-\vphantom{y}null} {\rm[{S2}]}:] 
				if $\lambda_1 = 0$, then ${\mathsf{F}_1}(\rho_{\mathsf{m}})={\mathsf{F}_1}(\rho_{\mathsf{v}})=\rho_{e_1}\notin \operatorname{dom} \mathsf{F}_1|_{\mathcal{S}(\Theta)}$, and   the system  goes from $\rho_{{e_1}}$ to $\rho_{\mathsf{z}}$ with unit probability.
			\end{enumerate} 
			\vspace{-1mm}
			\item[\sf \underline{unitary}:] If $|\lambda_2|=|\lambda_1|=1$, then  $\rho_{\mathsf{z}}$ and $\rho_{\mathsf{m}}$ \mbox{have trivial trajectories, see p. \pageref{unitarycase}}.
		\end{enumerate}  
	\end{theorem} 	
	
	Clearly, Prop.~\ref{alltypes} assures that 
	all types of chains listed in Theorem \ref{chainsclass} are realisable. Note that  generic and $\infty$-elliptic chains have isomorphic transition diagrams, even though the trajectories of  $\rho_{{\mathsf{v}}}$ and $\rho_{{\mathsf{m}}}$ have different limiting properties.
	
	{In the case of null chains, direct calculation gives	$\mathsf{p}_2(\rho_\mathsf{v}) = |\lambda_{1}|^2$. Thus, in the transition diagram of the double-null chain  the arrow from $\rho_{\mathsf{v}}$ to $\rho_{\mathsf{z}}$ is not present.   Likewise, in the transition diagrams of generic  and  both kinds of elliptic chains it may happen that 
		one arrow going to $\rho_{\mathsf{z}}$ from some state on the trajectory of $\rho_{\mathsf{v}}$ is actually non-existent, i.e., the corresponding probability is zero: in the following Prop.~\ref{rhou} we show that if $A$ is not unitary, then there exists exactly one state in $\mathcal{S}(\Theta)$ which under the action of $\mathcal{F}_{U,\,\Pi}$ remains with unit probability   in $\mathcal{S}(\Theta)$ and this state is pure.
		Clearly, the potential presence of such a~state in the trajectory of $\rho_{\mathsf{v}}$ does not affect any limiting properties of the chain that are essential in calculating quantum dynamical entropy of $\mathcal{F}_{U,\,\Pi}$ via the Blackwell integral formula. It would, however, allow \mbox{long-term} correlations to appear in the sequence of measurement  outcomes, which could be interpreted as the system exhibiting information storage.}

	\begin{proposition}If $A$ is not unitary, then there exists exactly one state \mbox{$\rho \in \mathcal{S}(\Theta)$} such that $\mathsf{p}_1(\rho)=1$. Moreover, $\rho$
		is pure. \label{rhou}\end{proposition}
	\begin{proof}
		Since $\Theta$ is a two-dimensional subspace of $\C^3$, we have $\dim (\Theta \cap U^*\Theta) \in \{1,2\}$. Clearly,  $\dim (\Theta \cap U^*\Theta) =2$ iff  $\Theta$ is $U$-invariant, which in turn is equivalent to  $A$ being unitary, a contradiction. Therefore, we have  $\dim (\Theta \cap U^*\Theta) = 1$, i.e., there exists exactly one ray in $\Theta$ with the property that its image under $U$ is also contained in $\Theta$. Putting  $u$ for a unit vector which spans this ray, we get $Uu=Au$, thus also  $||Au||=||Uu||=1$.
		Hence, $\rho_u \in \mathcal{P}(\Theta)$ satisfies $\mathsf{p}_1{(\rho_u)}=||Au||^2=1$. 
		
		Suppose now that there exists  $\rho \in \mathcal{S}(\Theta) \setminus \mathcal{P}(\Theta)$ such that $\mathsf{p}_1(\rho)=1$. By spectral decomposition we have  $\rho=\gamma\rho_{a}+(1-\gamma)\rho_{b}$, where   $\gamma \in (0,1)$ and  $\rho_{a}, \rho_{b}\in \mathcal{P}(\Theta)$ are mutually orthogonal. It follows easily that  $1=\mathsf{p}_1(\rho)=\gamma\mathsf{p}_1(\rho_a)+(1-\gamma)\mathsf{p}_1(\rho_b)$ holds only if  $\mathsf{p}_1(\rho_a)=\mathsf{p}_1(\rho_b)=1$, which implies that $\rho_a=\rho_b=\rho_u$. This contradicts the mutual orthogonality of $\rho_a$ and $\rho_b$, concluding the proof.
	\end{proof}

	\newpage

	\begin{figure}[H]
		\vspace{-2mm}
		\scalebox{0.8}{	\begin{tikzpicture}[->,>=stealth',shorten >=1pt,auto,node distance=1.75cm,
				main node/.style={rectangle,draw,font=\sffamily\small,  minimum size=0.75cm,rounded corners}]
				
				\node[ circle] (3){};
				
				\node[circle, draw] (z)[above right of=3] {$\rho_{\mathsf{z}}$} ;	
				\node[ main node] (2z)[above right of=z] {$\rho_{\mathsf{v}}$};
				\node[ main node] (3z)[right of=2z]   {} ;
				\node (4z)[right of=3z] {$\ldots$};
				\node[ main node] (5z)[right of=4z]  {} ;
				\node (6z)[right of=5z] {$\ldots$};
				
				\node[ main node,rounded corners] (i)[below right of=z] {$\rho_{\mathsf{m}}$} ; 
				\node[ main node,rounded corners] (2i)[right of=i] {} ;
				\node (4i)[right of=2i] {$\ldots$};
				\node[ main node,rounded corners] (5i)[right of=4i]  {} ;		
				\node (6i)[right of=5i] {$\ldots$};	
				
				\node (8i)[above right of=6i] {};	
				\node[main node, draw, color=blue]  (7i)[  right of=6z] {$\rho_{e_1}$};

				\path[every node/.style={font=\sffamily\normalsize}]
				(z) edge node  {} (2z)
				(2z) edge node {}  (3z)
				(3z) edge node  {}  (4z)
				(4z) edge node   {}  (5z)
				(5z) edge node   {}  (6z)
				
				(i) edge node[below]  {}  (2i)
				(2i) edge node[below]  {} (4i)
				(4i) edge node[below, pos=0.35]   {}  (5i)
				(5i) edge node[below]   {}  (6i)

				(z) edge [loop above]  node {}(z)
				(2z) edge [bend left=20] node {} (z)
				(3z) edge [bend left=8] node {} (z)
				(5z) edge [bend left=5] node {} (z)
				(i) edge [bend right=18]node  {}(z)
				(2i) edge[bend right=6] node {} (z)
				(5i) edge [bend right=4]node {} (z)  
				
				(6i) edge[color=blue, dashed,bend right=50,pos=0.5] node  {$n \to \infty$}(7i)
				
				(6z) edge[color=blue, dashed,pos=0.75] node  {}(7i)	
				(7i) edge[loop right, color=blue] node {$|\lambda_1|^2$} (7i)	
				;
		\end{tikzpicture}}  
		\vspace{-2mm}
		\caption*{ \footnotesize   generic chain}

		\vspace{6mm}
		
		\scalebox{0.8}{	\begin{tikzpicture}[->,>=stealth',shorten >=1pt,auto,node distance=1.75cm,
				main node/.style={rectangle,rounded corners,draw,font=\sffamily\small,  minimum size=0.75cm}]

				\node[circle, draw] (z)[above right of=3] {$\rho_{\mathsf{z}}$} ;	
				\node[ main node] (2z)[above right of=z] {$\rho_{e_2}$};
				\node (3z)[right of=2z] {} ;	
				\node (4z)[right of=3z] {};
				\node (5z)[right of=4z] {};
				
				\node[ main node,rounded corners] (i)[below right of=z] {$\rho_{\mathsf{m}}$} ; 
				\node[ main node,rounded corners] (2i)[right of=i] {} ;
				\node (4i)[right of=2i] {$\ldots$};
				\node[ main node,rounded corners] (5i)[right of=4i]  {} ;	
				
				\node (8i)[above right of=5i] {};	
				\node (6i)[right  of=8i] {};
				\node[main node, draw, color=blue]  (7i)[  right of=5z] {$\rho_{e_1}$};
				\node (9i)[below of=7i] {$\ldots$};
				
				\path[every node/.style={font=\sffamily\normalsize}]
				
				(z) edge node  {} (2z)
				(2z) edge[loop right] node {$|\lambda_2|^2$}  (2z)
				(2z) edge [bend left=20] node {} (z)
				
				(i) edge node[below]  {}  (2i)
				(2i) edge node[below]  {} (4i)
				(4i) edge node[below, pos=0.35]   {}  (5i)

				(z) edge [loop above]  node {}(z)
				(i) edge [bend right=16]node  {}(z)
				(2i) edge[bend right=6] node {} (z)
				(5i) edge [bend right=8]node {} (z)  
				
				(5i) edge[bend right=20,pos=0.75] node  {} (9i)	
				(9i)edge[color=blue, dashed,bend right=25] node  {$n \to \infty$} (7i)
				(7i) edge[loop right, color=blue] node {} (7i)
				;
				
		\end{tikzpicture}}
		\vspace{-2mm}
		\caption*{\footnotesize taupek chain}
		
		\vspace{7mm}
		
		\scalebox{0.8}{	\begin{tikzpicture}[->,>=stealth',shorten >=1pt,auto,node distance=1.75cm,
				main node/.style={rectangle,draw,font=\sffamily\small,  minimum size=0.75cm,rounded corners}]
				
				\node[ circle] (3){};
				
				\node[circle, draw] (z)[above right of=3] {$\rho_{\mathsf{z}}$} ;	
				\node[ main node] (2z)[above right of=z] {$\rho_{\mathsf{v}}$};
				\node[ main node] (3z)[right of=2z]   {} ;
				\node (4z)[right of=3z] {$\ldots$};
				\node[ main node] (5z)[right of=4z]  {} ;
				\node (6z)[right of=5z] {$\ldots$};
				
				\node[ main node,rounded corners] (i)[below right of=z] {$\rho_{\mathsf{m}}$} ; 
				\node[ main node,rounded corners] (2i)[right of=i] {} ;
				\node (4i)[right of=2i] {$\ldots$};
				\node[ main node,rounded corners] (5i)[right of=4i]  {} ;		
				\node (6i)[right of=5i] {$\ldots$};		
				\node  (9)[ below right of=6z] {};
				
				\path[every node/.style={font=\sffamily\normalsize}]
				(z) edge node  {} (2z)
				(2z) edge node {}  (3z)
				(3z) edge node  {}  (4z)
				(4z) edge node   {}  (5z)
				(5z) edge node   {}  (6z)
				
				(i) edge node[below]  {}  (2i)
				(2i) edge node[below]  {} (4i)
				(4i) edge node[below, pos=0.35]   {}  (5i)
				(5i) edge node[below]   {}  (6i)

				(z) edge [loop above]  node {}(z)
				(2z) edge [bend left=20] node {} (z)
				(3z) edge [bend left=8] node {} (z)
				(5z) edge [bend left=5] node {} (z)
				(i) edge [bend right=18]node  {}(z)
				(2i) edge[bend right=6] node {} (z)
				(5i) edge [bend right=4]node {} (z)  
				(9) edge[loop right, color=white] node {$|\lambda_1|^2$} (9)
				;
		\end{tikzpicture}}  
		\vspace{-2mm}
		\caption*{\footnotesize $\infty$-elliptic chain}
		
		\vspace{7mm}
		
		\scalebox{0.8}{	\begin{tikzpicture}[->,>=stealth',shorten >=1pt,auto,node distance=1.75cm,
				main node/.style={rectangle,rounded corners,draw,font=\sffamily\small,  minimum size=0.75cm}]

				\node[circle, draw] (z)[above right of=3] {$\rho_{\mathsf{z}}$} ;	
				\node[ main node] (2z)[above right of=z] {$\rho_{\mathsf{v}}$};
				\node[ main node] (3z)[right of=2z]   {} ;
				\node (4z)[right of=3z] {$\ldots$};
				\node[ main node] (5z)[right of=4z]  {} ;
				
				\node[ main node, rounded corners] (i)[below right of=z] {$\rho_{\mathsf{m}}$} ; 
				\node[ main node,rounded corners] (2i)[right of=i] {} ;
				\node (4i)[right of=2i] {$\ldots$};
				\node[ main node,rounded corners] (5i)[right of=4i]  {} ;		
				\node  (8)[ right  of=5i] {};	
				\node  (9)[ below right of=5i] {};
				
				\path[every node/.style={font=\sffamily\normalsize}]
				(z) edge node  {} (2z)
				(2z) edge node {}  (3z)
				(3z) edge node  {}  (4z)
				(4z) edge node   {}  (5z)
				(5z) edge[bend right=22] node   {}  (2z)
				
				(i) edge node[below]  {}  (2i)
				(2i) edge node[below]  {} (4i)
				(4i) edge node[below, pos=0.35]   {}  (5i)
				(5i) edge[bend left=22] node[below]   {}  (i)

				(z) edge [loop above]  node {}(z)
				(2z) edge [bend left=20] node {} (z)
				(3z) edge [bend left=8] node {} (z)
				(5z) edge [bend left=5] node {} (z)
				(i) edge [bend right=18]node  {}(z)
				(2i) edge[bend right=6] node {} (z)
				(5i) edge [bend right=4]node {} (z) 
				(9) edge[loop right, color=white] node {$|\lambda_1|^2$} (9) 
				;
		\end{tikzpicture}}
		\vspace{-8mm}
		\caption*{\footnotesize finite-elliptic chain}

		\vspace{6mm}
		
		\captionsetup[subfigure]{labelformat=empty}	 
		\subfloat[\mbox{{\rm generic-null chain}}]
		{
			\scalebox{0.8}{ 	\begin{tikzpicture}[->,>=stealth',shorten >=1pt,auto,node distance=1.55cm,
					main node/.style={rectangle,rounded corners,draw,font=\sffamily\small,  minimum size=0.75cm}]

					\node[circle, draw] (z)[above right of=3] {$\rho_{\mathsf{z}}$} ;	
					\node[ main node] (2z)[above right of=z] {$\rho_{\mathsf{v}}$};
					\node[ main node] (3z)[right of=2z] {$\rho_{e_1}$};	
					
					\node[ main node,rounded corners] (i)[below right of=z] {$\rho_{\mathsf{m}}$} ;

					\path[every node/.style={font=\sffamily\normalsize}]
					
					(z) edge node  {} (2z)
					(2z) edge node {}  (3z)
					(3z) edge [bend left=20] node {}  (z)
					(3z) edge [loop right] node {$|\lambda_1|^2$}  (3z)
					(2z) edge [bend left=20] node {} (z)
					(i) edge[bend right=20] node  {}  (3z)
					(i) edge [bend right=16]node  {}(z)
					;
			\end{tikzpicture}}
		}
		\subfloat[\mbox{{\rm taupek-null chain}}]
		{			\scalebox{0.8}{	\begin{tikzpicture}[->,>=stealth',shorten >=1pt,auto,node distance=1.55cm,
					main node/.style={rectangle,rounded corners,draw,font=\sffamily\small,  minimum size=0.75cm}]

					\node[circle, draw] (z)[above right of=3] {$\rho_{\mathsf{z}}$} ;	
					\node[ main node] (2z)[above right of=z] {$\rho_{e_2}$};
					\node[ main node] (3z)[right of=2z] {$\rho_{e_1}$};	
					
					\node[ main node,rounded corners] (i)[below right of=z] {$\rho_{\mathsf{m}}$} ;

					\path[every node/.style={font=\sffamily\normalsize}]
					(z) edge node  {} (2z)
					(3z) edge [loop right, white] node {$xxx$}  (z)
					(3z) edge [loop right, ] node {}  (3z)
					(2z) edge [bend left=20, ] node {} (z)
					(i) edge[bend right=20] node[below]  {}  (3z)
					(i) edge [bend right=16]node  {}(z)
					;
			\end{tikzpicture}}
		} 
		\hspace{-4mm}
		%		\\ \vspace{2mm}
		\subfloat[\mbox{{\rm double-null chain}}]
		{		\scalebox{0.8}{
				\begin{tikzpicture}[->,>=stealth',shorten >=1pt,auto,node distance=1.55cm,
					main node/.style={rectangle,rounded corners,draw,font=\sffamily\small,  minimum size=0.75cm}]

					\node[circle, draw] (z)[above right of=3] {$\rho_{\mathsf{z}}$} ;	
					\node[ main node] (2z)[above right of=z] {$\rho_{\mathsf{v}}$};
					\node[ main node] (3z)[right of=2z] {$\rho_{e_1}$};	
					
					\node[ main node,rounded corners] (i)[below right of=z] {$\rho_{\mathsf{m}}$} ;

					\path[every node/.style={font=\sffamily\normalsize}]
					(3z) edge [loop right, white] node {$xxx$}  (z)
					(z) edge node  {} (2z)
					(2z) edge node {}  (3z)
					(3z) edge [bend left=20, ] node {}  (z)
					(i) edge[bend right=20] node[below]  {}  (3z)
					(i) edge [bend right=16]node  {}(z)
					;
			\end{tikzpicture}}
		}
		
		\vspace{9mm}
		
		\scalebox{0.8}{ 	\begin{tikzpicture}[->,>=stealth',shorten >=1pt,auto,node distance=1.75cm,
				main node/.style={rectangle,rounded corners,draw,font=\sffamily\small,  minimum size=0.75cm}]
				
				\node[circle, draw] (z) {$\rho_{\mathsf{z}}$} ;	
				
				\node[ main node,rounded corners] (i)[  below right of=z] {$\rho_{\mathsf{m}}$} ; 
				\path[every node/.style={font=\sffamily\normalsize}]
	
				(z) edge [loop right, ]  node[right] {}(z)
				(i) edge [loop right, ]  node[right] {}(i)		
				;
				
		\end{tikzpicture}}
		\vspace{-2mm}
		\caption*{\footnotesize unitary chain}
		\captionsetup{width=0.7\linewidth}
		\caption{Transition diagrams of the chains listed in Thm. \ref{chainsclass}}
		\label{diag2}
	\end{figure}
	
	\newpage
	
	\subsection{Numerical range}\label{secNumrange}

	Let $U$ and $\Pi$ be as in the previous subsection, i.e., $U \in \mathcal{U}(\C^3)$ and a PVM  $\Pi=\Pi(\mathsf{z})=\{\Pi_1, \Pi_2\}$ is such that $\Pi_2=\rho_{{\mathsf{z}}}$ and $\Pi_1=\mathbb{I}_3 - \rho_{{\mathsf{z}}}$  for some unit vector $\mathsf{z} \in \C^3$.  
	According to Theorem \ref{chainsclass}, the type of the  chain generated by  $\mathcal{F}_{U,\, \Pi}$ is determined  by 
	the eigenvalues of $\Pi_1U|_{\Theta}$.
	In Remark~\ref{remarknumrange} we observed that $\Pi$ enters the formula for the eigenvalues of $\Pi_1U|_{\Theta}$ only through \mbox{$\omega := \braket{\mathsf{z}|U\mathsf{z}}\in \C$}.
	In what follows, we explore how the type of the chain generated by  $\mathcal{F}_{U,\, \Pi}$ depends on~$U$ and $\omega$.

	The \emph{numerical range} of $T
	\in \mathcal{L}(\C^d)$ is defined as 
	$\mathfrak{W}(T):=\{\braket{z|Tz} \colon {z} \in \C^d,\: ||z||=1\}$. It is   well-known   that the numerical range  of a normal  operator is the convex hull of its eigenvalues,  so a polygon in $\C$.
	Therefore, the numerical range of $U\in \mathcal{U}(\C^3)$ is spanned by three points on the unit circle. Hence,  generically, $\mathfrak{W}(U)$ corresponds to a triangle inscribed in the unit circle, unless some eigenvalues of $U$ coincide, in which case $\mathfrak{W}(U)$ reduces to a chord (if exactly two eigenvalues coincide) or to a point (if all three eigenvalues coincide). Obviously, the latter case is equivalent to $U$ being (up to an overall phase) the identity on~$\C^3$. 
	
	Our first goal is to prove that every point in $\mathfrak{W}(U)$  corresponds to a family of  conjugate PIFSs, i.e., if for two unit vectors  $\mathsf{z}, \tilde{\mathsf{z}} \in \C^3$ we have $\braket{\mathsf{z}|U\mathsf{z}}=\braket{\tilde{\mathsf{z}}|U\tilde{\mathsf{z}}}$,  then \mbox{$\mathcal{F}_{U,\,\Pi(\mathsf{z})} \sim \mathcal{F}_{U,\,\Pi(\mathsf{\tilde{z}})}$}.  Throughout this (and the next) subsection  the eigenvalues of~$U$ are denoted by $\e^{\I\phi_1}, \e^{\I\phi_2}, \e^{\I\phi_3}$, where $\phi_j \in [0, 2\pi)$, $j \in \{1,2,3\}$. 
	
	\begin{lemma}	\label{komV}
		Let $\mathsf{z}, \tilde{\mathsf{z}} \in \C^3$ be unit vectors. We have 
		$\braket{\mathsf{z}|U\mathsf{z}}=\braket{\tilde{\mathsf{z}}|U\tilde{\mathsf{z}}}$ iff 
		there exists $V \in \mathcal{U}(\C^3)$ such that $UV=VU$ and $V\mathsf{z}=\tilde{\mathsf{z}}$.
	\end{lemma}
	\begin{proof}
		($\Leftarrow$) 
		We have 
		$\braket{\tilde{\mathsf{z}}|U\tilde{\mathsf{z}}}=\braket{V\mathsf{z}|UV\mathsf{z}} =\braket{V\mathsf{z}|VU\mathsf{z}}=\braket{\mathsf{z}|U\mathsf{z}}.$
		
		\vspace{1mm}
		
		\noindent ($\Rightarrow$)
		We consider the following three cases. 
		\vspace{-1mm}
		\begin{itemize}[leftmargin=4.5mm]
			\itemsep=-1mm
			\item Assume that $U$ has three different eigenvalues and fix an orthonormal eigenbasis $w_1, w_2, w_3$ of $U$ such that $Uw_j=\e^{\I\phi_j}w_j$ for $j \in \{1,2,3\}$. We have $\mathsf{z}=\sum_{j=1}^3\alpha_jw_j$ with $\alpha_j \in \C$   and $\sum_{j=1}^3|\alpha_j|^2=1$.
			Clearly, 
			$ 
			% 		\vspace{-0.75mm}
			\braket{\mathsf{z}|U\mathsf{z}}=
			|\alpha_1|^2\e^{\I\phi_1}+|\alpha_2|^2\e^{\I\phi_2}+|\alpha_3|^2\e^{\I\phi_3}
			.$
			%		$ 
			Since $	\braket{\tilde{\mathsf{z}}|U\tilde{\mathsf{z}}}=\braket{\mathsf{z}|U\mathsf{z}} $, 
			the uniqueness of (normalised) barycentric coordinates in a triangle implies that there exist $\gamma_1, \gamma_2, \gamma_3 \in \R$  such that $\tilde{\mathsf{z}}=\sum_{j=1}^3\alpha_j\e^{\I\gamma_j} w_j$. Let $V\in \mathcal{L}(\C^3)$ be such that  $Vw_j=\e^{\I\gamma_j}w_j$, $j \in \{1,2,3\}$. Clearly, $V$ is unitary and $V\mathsf{z}=\tilde{\mathsf{z}}$. Since  $U$ and $V$ have a common eigenbasis, we also have $UV=VU$.
			
			\item  Assume that $U$ has two different eigenvalues. With no loss of generality we assume that  $\e^{\I\phi_1}\neq \e^{\I\phi_2}= \e^{\I\phi_3}$. 
			There exists an orthonormal eigenbasis $\{w_1, w_2, w_3\}$ of $U$ such that $Uw_j=\e^{\I\phi_j}w_j$ ($j \in \{1,2,3\}$) and 
			$\mathsf{z}=\alpha_1w_1+\alpha_2w_2$ with $\alpha_1, \alpha_2 \in \C$ satisfying \mbox{$|\alpha_1|^2+|\alpha_2|^2=1$}.
			Since 
			$ 
			% 		\vspace{-0.75mm}
			\braket{\tilde{\mathsf{z}}|U\tilde{\mathsf{z}}}=\braket{\mathsf{z}|U\mathsf{z}}=|\alpha_1|^2\e^{\I\phi_1}+|\alpha_2|^2\e^{\I\phi_2}
			%		$
			$ 
			and the  (normalised) barycentric coordinates in a segment are unique,    there exist 
			$\gamma_1 \in \R$ and  $\beta_2, {\beta_3} \in \C$ such that $|\beta_2|^2+|\beta_3|^2=|\alpha_2|^2$ and
			$\tilde{\mathsf{z}}=\alpha_1\e^{\gamma_1\I}w_1+\beta_2 w_2+{\beta_3} w_3 $. 
			Clearly, there exists $V \in \mathcal{U}(\C^3)$ such that $Vw_1=\e^{\gamma_1\I}w_1$ and ${\alpha_2}Vw_2={\beta_2}w_2+{\beta_3}w_3$. We   see that this $V$ satisfies 
			$V\mathsf{z}=\tilde{\mathsf{z}}$. Moreover, $U$ and $V$  share a one-dimensional eigenspace, namely that spanned by  $w_1$, and every eigenvector of $V$ contained in  $\spann\{w_2, w_3\}$ is also an eigenvector of $U$,  because  $\spann\{w_2, w_3\}$ is an eigenspace of $U$. Hence, $U$ and $V$ share an eigenbasis, and so they commute.

			\item Finally, if $U=\e^{\I\phi_1}\mathbb{I}_3$, then $U$ commutes with every $V \in \mathcal{L}(\C^3)$, and so it suffices to take any $V \in \mathcal{U}(\C^3)$ satisfying $V\mathsf{z}=\tilde{\mathsf{z}}$. 
		
		\end{itemize}
	\end{proof}

	\begin{remark}
		In proving Lemma \ref{komV} we resort to the fact that 
		\mbox{$\braket{{\mathsf{z}}|U{\mathsf{z}}}$}, where   ${\mathsf{z}} \in \C^3$ and $U \in \mathcal{U}(\C^3)$, can be uniquely written as a convex combination of the eigenvalues of $U$. This is not true in 
		higher dimensions 
		and Theorem~\ref{komV} cannot be generalised to $\C^d$ with $d>3$. 
		Indeed,  fix an orthonormal basis of $\C^4$ and let $\tilde{U} \in \mathcal{U}(\C^4)$ be such that $\tilde{U}=\diag(1,-1,\I, -\I)$. 	For $z_1=\tfrac{1}{\sqrt{2}}[1, 1, 0,0]^T$ and $z_2=\tfrac{1}{\sqrt{2}}[0,0,1, 1]^T$ we have	$\braket{z_1|\tilde{U}z_1}=\tfrac 12(1-1)=0$ 
		as well as $\braket{z_2|\tilde{U}z_2}= \tfrac 12(\I-\I)= 0$.  Since the eigenvalues of $\tilde{U}$ are all distinct, only diagonal unitary matrices commute with $\tilde{U}$. However, no~diagonal matrix can map $z_1$~to~$z_2$.
	\end{remark}
	
	\begin{remark}
		From the proof of Lemma \ref{komV} 
		we can easily deduce how to construct the PVM $\Pi(\mathsf{z})$ corresponding to a given $\omega \in  \mathfrak{W}(U)$, i.e., how to find a unit vector $\mathsf{z} \in \C^3$ satisfying $\braket{\mathsf{z}|U\mathsf{z}} = \omega$. 
		First, we determine   the barycentric coordinates of $\omega$ in $\mathfrak{W}(U)$, i.e.,  the coefficients $\alpha_1, \alpha_2, \alpha_3 \in [0,1]$ such that \mbox{$\sum_{i=1}^3 \alpha_j = 1$} and 
		$\sum_{i=1}^3 \alpha_j \e^{\I\phi_j} = \omega$. Putting  $\mathsf{z} := \sum_{i=1}^3 \sqrt{\smash[b]{\alpha_j}}w_j$, where $Uw_j =\e^{\I\phi_j}w_j$ with $j \in \{1,2,3\}$, we obtain  $\braket{\mathsf{z}|U\mathsf{z}} = \omega$, as desired. 	
	\end{remark}
	
	In order to have the unit vector $\mathsf{z}\in \C^3$ determining the measurement better exposed, 
	we adjust notation for the projectors constituting  $\Pi(\mathsf{z})$ by putting
	$\Pi_2^\mathsf{z}:=\rho_{\mathsf{z}}$ and \mbox{$\Pi_1^\mathsf{z}:=\mathbb{I}_{3}-\Pi_2^\mathsf{z}$} as well as $\Theta_{\mathsf{z}}:=\spann\{\mathsf{z}\}^\bot$. Note that 
	$
	\Pi(V\mathsf{z})=V\Pi(\mathsf{z})V^*$ for $V\in \mathcal{U}(\C^3)\cup\,  \overline{\mathcal{U}}(\C^3)$, i.e.,  $
	\Pi_i^{V{\mathsf{z}}}=
	V\Pi_i^{\mathsf{z}}V^*$ for $i\in \{1,2\}$.

	\begin{theorem}\label{zzprimeconj} If  $\braket{{\mathsf{z}}|U{\mathsf{z}}}=\braket{\tilde{\mathsf{z}}|U\tilde{\mathsf{z}}}$, then $\mathcal{F}_{U,\,\Pi(\mathsf{z})} \sim \mathcal{F}_{U,\,\Pi(\mathsf{\tilde{z}})}$, where 
		${\mathsf{z}}$, $\tilde{\mathsf{z}}\in \C^3$ are unit vectors. 
	\end{theorem}
	\begin{proof}
		Let ${\mathsf{z}}, \tilde{\mathsf{z}}\in \C^3$ be unit vectors satisfying $\braket{{\mathsf{z}}|U{\mathsf{z}}}=\braket{\tilde{\mathsf{z}}|U\tilde{\mathsf{z}}}$.	It follows from Lemma~\ref{komV} that there exists $V \in \mathcal{U}(\C^3)$ such that $UV=VU$ and $V\mathsf{z}=\mathsf{\tilde{z}}$; thus, we have  
		$UV^*=V^*U$ and $\Pi(\mathsf{\tilde{z}})=V\Pi(\mathsf{z})V^*$. For $\rho \in \mathcal{S}(\C^3)$ and $i \in \{1, 2\}$  we obtain 
		%\vspace{-1mm}
		\begin{align*}
			%		\vspace{-1mm}
			{{\Lambda}_i^{U,\, \Pi(\mathsf{\tilde{z}})}}(\rho) & = 
			\Pi^\mathsf{\tilde{z}}_i U\rho\, U^* \Pi^\mathsf{\tilde{z}}_i
			\\[0.35em] 
			& =V\Pi^\mathsf{z}_iV^* U\rho\, U^* V\Pi^\mathsf{z}_iV^*
			\\[0.35em]
			& = V\Pi^\mathsf{{z}}_i U V^* \rho \, VU^*\Pi^\mathsf{{z}}_iV^*
			% 		\\[0.35em]
			% 		& 
			=	\Lambda^{V}({{\Lambda}_i^{U,\, \Pi(\mathsf{{z}})}}(\Lambda^{V^*}(\rho))).	
		\end{align*}
		Observation \ref{obsconj} assures that  $\mathcal{F}_{U,\,\Pi(\mathsf{\tilde{z}})}$
		is $V^*$-conjugate to 	$\mathcal{F}_{U,\,\Pi(\mathsf{z})}$, as desired. 
	\end{proof}

	The above corollary allows us to write  $\mathcal{F}_{U,\,\omega}$ for any $\mathcal{F}_{U,\,\Pi(\mathsf{{z}})}$ such that $\braket{{\mathsf{z}}|U{\mathsf{z}}}=\omega$.
	The main goal of this subsection is to distinguish the subsets of  $\mathfrak{W}(U)$ that correspond to particular  types of chains, i.e., to describe the partition of $\mathfrak{W}(U)$ such that $\omega_1, \omega_2 \in \mathfrak{W}(U)$ are in the same subset if and only if the chains generated by $\mathcal{F}_{U,\,\omega_1}$ and $\mathcal{F}_{U,\,\omega_2}$ fall into the same case in Theorem~\ref{chainsclass}.  
	Actually, we have already seen in Proposition \ref{unitarycond} that $\mathcal{F}_{U,\,\omega}$ generates the unitary chain iff $\omega \in \sigma(U)$, i.e., iff $\omega$ is a vertex of  $\mathfrak{W}(U)$.
	In what follows we consider one by one the  conditions on the eigenvalues of $A$ that in Theorem~\ref{chainsclass} characterise the remaining seven types of chains and express them in terms  of $\mathfrak{W}(U)$. 
	We adopt  the following notation. Let $\omega \in \mathfrak{W}(U)$. We write  $A_\omega$  for $\Pi_1^\mathsf{z}U|_{\Theta_{\mathsf{z}}}$  if $\mathsf{z} \in \C^3$ satisfies  $\braket{\mathsf{z}|U\mathsf{z}}=\omega$. By $\lambda_1(\omega)$, $\lambda_2(\omega)$ we denote  the eigenvalues of $A_\omega$, ordered as $|\lambda_1(\omega)| \geq |\lambda_2(\omega)|$, and put $\psi_\omega:=\Arg \frac{\lambda_2(\omega)}{\lambda_1(\omega)}$, provided that $0 \notin  \sigma(A_\omega)$.

	\smallskip
	
\begin{figure}[h]
		\centering 	
		\centering
		\includegraphics[scale=0.35]{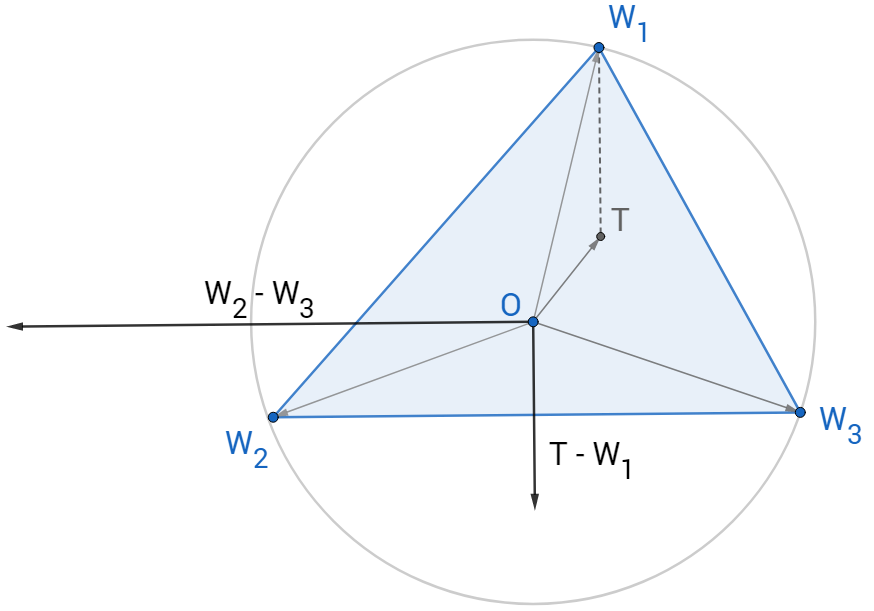}  
		\vspace{-1mm}
		\caption{$\sf\overline{W_1T}$ is contained in the altitude dropped from $\mathsf{W}_1$ to $\overline{\mathsf{W}_2\mathsf{W}_3}$.}
		\label{triangle1}
	\end{figure}

	Let us start with null chains. Recall that $\mathcal{F}_{U,\,\omega}$ generates a null chain iff $\lambda_2(\omega)=0$ and that it depends on $|\lambda_1(\omega)|$ which of the three subtypes is actually generated.  
	Also, recall that $0 \in \sigma(A_\omega)$ is equivalent to $\omega=0$, see Prop.~\ref{Ainvertibleomegazero}. Assuming $0 \in \mathfrak{W}(U)$, we deduce that $\lambda_2(0)=0$ and $\lambda_1(0)=\tr U$, see \eqref{trdet}. 
	If $\mathfrak{W}(U)$ is non-degenerate, then $0$ is  obviously its circumcentre, while $\tr U$ turns out to be its orthocentre. Indeed, putting $\mathsf{W}_j:=[\cos\phi_j, \sin\phi_j]^T \sim \e^{\I\phi_j}$, where $j\in \{1,2,3\}$, and 
	$\mathsf{T}:=\mathsf{W}_1 + \mathsf{W}_2 +\mathsf{W}_3 \sim \tr U$, we~obtain (see also Fig. \ref{triangle1})
	\vspace{-1mm}
	\begin{align*}  
		\vspace{-1mm}
		(\mathsf{T}-\mathsf{W}_1) \cdot (\mathsf{W}_2-\mathsf{W}_3)  =     
		||\mathsf{W}_2||^2- {\mathsf{W}_2 \cdot \mathsf{W}_3}+ {\mathsf{W}_3 \cdot \mathsf{W}_2}-||\mathsf{W}_3||^3
		= 
		0,
	\end{align*}
	so $\sf T$ lies at the altitude dropped from $\mathsf{W}_1$ to $\overline{\mathsf{W}_2\mathsf{W}_3}$. We easily conclude that all altitudes of $\mathfrak{W}(U)$ intersect at $\tr U$, as claimed.

	It is a basic fact of Euclidean geometry that the orthocentre and circumcentre of any triangle are linked by a~homothety with ratio $\shortminus \frac 12$ centred at this triangle's centroid, so 
	\vspace{-1mm}	
	\begin{enumerate}[(i), leftmargin=9mm]
		\itemsep=-0.5mm
		\item if the triangle is acute, then 
		both orthocentre and circumcentre lie inside the triangle;
		
		\item if the triangle is right-angled, then the orthocentre lies at the vertex of the right angle and the circumcentre lies at the centre of the hypotenuse;
		
		\item  if the triangle is obtuse, then both orthocentre and circumcentre lie outside the triangle;
	\end{enumerate}
	%\vspace{-2mm}
	Moreover, the orthocentre and circumcentre coincide iff the triangle is equilateral.
	Consequently, in terms of $\mathfrak{W}(U)$ we have
	\vspace{-1mm}	
	\begin{enumerate}[(i), leftmargin=9mm]
		\itemsep=-0.5mm
		
		\item   $\tr U \in \Int  \mathfrak{W}(U)\ \Leftrightarrow\ 0 \in \Int \mathfrak{W}(U)\ \Leftrightarrow\ \mathfrak{W}(U)$ is an acute triangle;

		$\bullet$ in particular,  $\tr U=0\ \Leftrightarrow\ \mathfrak{W}(U)$ is an equilateral triangle;
		
		\item
		$\tr U \in \sigma(U)\ \Leftrightarrow\ 0 \in \partial \mathfrak{W}(U)\ \Leftrightarrow\ \mathfrak{W}(U)$ is a right-angled triangle or  a diameter;  
		
		\item $\tr U \notin  \mathfrak{W}(U)\ \Leftrightarrow\ 0 \notin \mathfrak{W}(U)\ \Leftrightarrow\ \mathfrak{W}(U)$ is an obtuse triangle or  a single point or a chord that is not a diameter.
	\end{enumerate} 
	
	\noindent
	Therefore, for null chains we have the following 
	\begin{corollary}\label{propnull} 	$\mathcal{F}_{U,\,\omega}$ generates a null chain if and only if $\omega=0$. The subtype of the null chain generated by $\mathcal{F}_{U,\,0}$ is
		\vspace{-1mm}	
		\begin{enumerate}[{\rm (a)}, leftmargin=9mm]
			\itemsep-0.25mm
			\item  generic-null  $\Leftrightarrow$  $0<|\tr U|<1$ $\Leftrightarrow$  $\mathfrak{W}(U)$ is an acute-non-equilateral triangle;
			
			\item double-null  $\Leftrightarrow$ $\tr U=0$ $\Leftrightarrow$ $\mathfrak{W}(U)$ is the equilateral triangle;
			\item  taupek-null   $\Leftrightarrow$  $|\tr U|=1$ $\Leftrightarrow$   $\mathfrak{W}(U)$ is a right-angled triangle or a diameter.
		\end{enumerate}
	\end{corollary}
	
	\pagebreak

	Next, we investigate the types of chains that can be generated at the boundary of $\mathfrak{W}(U)$. 
	As before, by $e_j$ we denote a normalised eigenvector of $A_\omega$ associated with $\lambda_j(\omega)$, $j \in \{1,2\}$.
	
	\begin{theorem}\label{brzeg} Let $\omega \in \mathfrak{W}(U)$. We have 
		$\omega \in \partial \mathfrak{W}(U)$ if and only if   $|\lambda_1(\omega)|=1$.
	\end{theorem}

	\begin{proof}
		\noindent
		($\Leftarrow$)	Let $\omega \in \mathfrak{W}(U)$ and assume that $|\lambda_1(\omega)|=1$. 	 From Lemma \ref{lemmal1} we know that $\lambda_1(\omega)\in \sigma(U)$ and that ${{e_1}}$ is an   eigenvector of $U$ associated with $\lambda_1(\omega)$. 
		With no loss of generality we assume that  $\lambda_1(\omega)=\e^{\I\phi_1}$. From  \eqref{trdet} we obtain 
		\vspace{-1mm}
		\begin{equation}
			\label{baryccc}
			\vspace{-1mm}
			\e^{\I\phi_2}+\e^{\I\phi_3}=\lambda_2(\omega)+\omega\ \aand\  \e^{\I\phi_2+\I\phi_3}\overline{\omega}=\lambda_2(\omega).\end{equation}
		Let  $a_1, a_2, a_3\in [0,1]$  stand for the normalised barycentric coordinates of $\omega$ in $\mathfrak{W}(U)$, i.e., $\omega=\sum_{j=1}^3a_j\e^{\I \phi_j}$ and $\sum_{j=1}^3a_j=1$. From \eqref{baryccc} it follows that  
		%		\vspace{-1mm}
		$
		%		\vspace{-1mm}
		a_1(\e^{\I \phi_2}-\e^{\I \phi_1})(\e^{\I \phi_3}-\e^{\I \phi_1})=0$,  which gives  $a_1=0$
		or $\e^{\I\phi_1} \in \{\e^{\I\phi_2}, \e^{\I\phi_3}\}$, and in either of these cases we  conclude that  \mbox{$\omega \in \conv\{\e^{\I\phi_2}, \e^{\I\phi_3}\} \subset \partial \mathfrak{W}(U)$}, as desired.
		
		\smallskip
		
		\noindent	
		($\Rightarrow$)	Let   $\omega \in \partial \mathfrak{W}(U)$. With no loss of generality we assume that $ \omega=a\e^{\I\phi_1}+(1-a)\e^{\I\phi_2}$, where $a \in [0, 1]$. Observe that for  $\widetilde{\omega} :=(1-a)\e^{\I\phi_1}+a\e^{\I\phi_2}$  we have 
		$\tr U= \omega + \widetilde{\omega} +\e^{\I\phi_3} $ and $\e^{\I\phi_1+\I\phi_2}  \overline{\omega} =\widetilde{\omega}$.
		Thus, again via  \eqref{trdet}, we obtain 
		%	\vspace{-1mm}
		$$
		%	\vspace{-1mm}
		\tr A_\omega =\widetilde{\omega} + \e^{\I\phi_3}\ \aand\  \det A_\omega=\widetilde{\omega} \e^{\I\phi_3}.$$ 
		It follows easily that  $\lambda_{1}(\omega)=\e^{\I\phi_3}$ and $\lambda_{2}(\omega)=\widetilde{\omega}$; in particular, we have  $|\lambda_1(\omega)|=1$, which concludes the proof.
	\end{proof}

	Clearly, if $|\lambda_1(\omega)|=1$, then the determinant formula in \eqref{trdet} gives  $|\lambda_2(\omega)|=|\omega|$. Therefore, for the boundary of $\mathfrak{W}(U)$ we have the following classification of chain types.
	\begin{corollary} \label{obsboundary}
		Let $\omega \in \partial\mathfrak{W}(U)$. Then the Markov  chain generated by $\mathcal{F}_{U,\,\omega}$ is 
		\vspace{-1mm}	
		\begin{enumerate}[{\rm (a)}, leftmargin=10mm]
			\itemsep-0.25mm
			\item taupek-null  $\,\Leftrightarrow\,$ $\omega=0$   $\,\Leftrightarrow\,$ $\omega$ is the midpoint of the hypotenuse of  right-angled $\mathfrak{W}(U)$  or the~midpoint of $\mathfrak{W}(U)$ degenerate to a diameter  (see also Cor.  \ref{propnull}(c));
			\item   unitary   $\,\Leftrightarrow\,$ $|\omega|=1$  $\,\Leftrightarrow\,$ $\omega$ is a vertex of $\mathfrak{W}(U)$ (see also Prop. \ref{unitarycond});
			\item taupek   $\,\Leftrightarrow\,$ $0<|\omega|<1$  $\,\Leftrightarrow\,$ $\omega$ is neither  the midpoint of the hypotenuse of right-angled $\mathfrak{W}(U)$ {nor the midpoint of $\mathfrak{W}(U)$ degenerate to a diameter} nor a vertex of $\mathfrak{W}(U)$.
		\end{enumerate}
	\end{corollary}

	\begin{remark} \label{normal}
		The $(\Rightarrow)$ part of the proof of Theorem  \ref{brzeg} along with Proposition~\ref{orthbasis} imply that if $\omega=a\e^{\I\phi_1}+(1-a)\e^{\I\phi_2}$ for some  $a \in [0,1]$, then  $A_\omega$ is a normal operator with $\sigma(A_\omega)= \{\e^{\I\phi_3}, (1-a)\e^{\I\phi_1}+a\e^{\I\phi_2}\}$. Collecting the results of Proposition~\ref{orthbasis}, Lemma~\ref{lemmal1} and Theorem  \ref{brzeg}, we   deduce that the following conditions are equivalent:		
		\vspace{-1mm}	
		\begin{enumerate}[(i), leftmargin=10mm]
			\itemsep-0.25mm
			\item  $A_\omega$ is normal (unitarily diagonalisable);
			\item  an eigenvector of $U$ is  orthogonal to $\mathsf{z}$;
			\item  $\omega$ is a convex combination of at most two eigenvalues of $U$ (i.e., $\omega \in \partial \mathfrak{W}(U)$), 
			\item  $|\lambda_1(\omega)|=1$.
		\end{enumerate}
		\vspace{-1mm}
		One may want to  compare these conditions with those in  Propositions \ref{unitarycond} \& \ref{unit11}, which characterise the case of $A_\omega$ being unitary.
	\end{remark}

	\begin{remark} From \cite[Thm. 14]{KusuokaQuantum} and  Remark \ref{normal} it follows that 
		the probability measure   induced by $\mathcal{F}_{U,\,\omega}$ on the space of sequences of measurement outcomes $\{1,2\}^\N$ is non-ergodic (with respect to the shift operator) if and only if $\omega \in \partial\mathfrak{W}(U)$.
	\end{remark}

	It remains to describe the subsets of $\Int\mathfrak{W}(U)\setminus \{0\}$ corresponding to finite- and \mbox{$\infty$-elliptic} chains. 
	The first step is to characterise  $\omega \in \mathfrak{W}(U)\setminus \{0\}$ such that $A_\omega$ induces elliptic dynamics on the Bloch sphere,
	which, as we recall, is equivalent to	$|\lambda_1(\omega)|=|\lambda_2(\omega)|>0$ 
	with $\psi_\omega \neq 0$. Also, recall that elliptic dynamics is called circular if we additionally have $\psi_\omega = \pi$, which amounts to $\lambda_1(\omega)=-\lambda_2(\omega)\neq 0$.
	
	\begin{proposition}\label{ell}  If $\omega \in   \mathfrak{W}(U)\setminus \{0\}$, then 
		$A_\omega$ generates elliptic dynamics iff
		\begin{equation}
			\label{elliptcond} \frac{ (\tr U - \omega)^2}{\overline{\omega}\det U} \in [0, 4). \end{equation}
		In particular, $A_\omega$ generates circular dynamics iff $\omega=\tr U$.
	\end{proposition}
	
	\begin{proof}
		Let $\omega \in \mathfrak{W}(U)\setminus \{0\}$. 
		Recall   that $\omega \neq 0$ assures that $0 \notin \sigma(A_\omega)$ and $\det A_\omega\neq 0$.
		First, we discuss the case of $A_\omega$ generating circular dynamics. 
		Clearly,   $\lambda_{1}(\omega)=-\lambda_{2}(\omega)$ iff $\tr A_\omega=0$ iff $\omega=\tr U$, as claimed.
		Let us now characterise elliptic-non-circular dynamics, i.e., with $\psi_\omega \neq \pi$. Denote by $\Delta_\omega$
		the discriminant of the characteristic polynomial of  ${A_\omega}$, 
		i.e., $\Delta_\omega:=(\tr A_\omega)^2-4\det A_\omega$, and observe that 
		%	\vspace{-2mm}
		\begin{align}
			|\lambda_1(\omega)|=|\lambda_2&(\omega)|  \aand \psi_\omega \notin \{0, \pi\}  
			\nonumber	
			\\[0.2em] & \Longleftrightarrow  \ 
			\Delta_\omega \neq 0 \aand\Arg({(\tr A_\omega)^2}/{\Delta_\omega})=\pi
			\nonumber	
			\\[0.2em] & \Longleftrightarrow\ 
			\Arg((\tr A_\omega)^2/\det A_\omega)=0 \aand |(\tr A_\omega)^2|<4|\det A_\omega|
			\nonumber	
			\\[0.2em] & \Longleftrightarrow \ 
			{(\tr A_\omega)^2}/{\det A_\omega} \in (0, 4),\nonumber\end{align}
		and so the assertion of the proposition follows.
	\end{proof}

	\begin{remark}\label{mobtemp} 	Consider $\omega \in   \mathfrak{W}(U)\hspace{-0.35mm}\setminus\hspace{-0.35mm} \{0\}$. 
		Analogously to the proof of Proposition~\ref{ell},  one can verify that the dynamics induced on the Bloch sphere by $A_\omega$ is 	parabolic iff   $\frac{ (\tr U - \omega)^2}{\overline{\omega}\det U} =4$ and $\omega \notin \sigma(U)$, and it is loxodromic  iff $\frac{ (\tr U - \omega)^2}{\overline{\omega}\det U} \notin [0, 4]$.
		In fact, we recovered here a~well-known criterion for deciding whether a M\"obius transformation of $\hat{\C}$ is  elliptic, parabolic or loxodromic, see, e.g., \cite[Prop. 2.16]{anderson2005hyperbolic}. It is also straightforward to check that the remaining case of $\frac{ (\tr U - \omega)^2}{\overline{\omega}\det U} =4$ and $\omega \in \sigma(U)$   corresponds to $A_\omega$ being proportional to the identity map, and so to the trivial dynamics on the Bloch sphere.
	\end{remark}

	Finally, let us consider $\omega \in \Int \mathfrak{W}(U)\setminus \{0\}$ such that $A_\omega$ generates elliptic dynamics. To distinguish between  finite-elliptic and $\infty$-elliptic  chains, we need to decide whether or not $\psi_\omega$ is commensurable with $\pi$.  Putting $
	\Upsilon(\omega)	:=\min \{\psi_\omega, 2\pi-\psi_\omega\}$, by direct calculation we obtain $\cos \frac {\Upsilon(\omega)}{2}=\frac 12 {|\tr A_\omega|}/{\sqrt{|\det A_\omega|}}$, so 
	\begin{equation}\label{wzornakat}
		\Upsilon(\omega)		=2\arccos\frac{|\tr U - \omega|}{2\sqrt{{|\omega|}}}.
	\end{equation}
	Obviously,  $\psi_\omega \in \pi \mathbb{Q}$ if and only if   $\Upsilon(\omega)	 \in \pi \mathbb{Q}$.
	The next theorem summarizes the results of this subsection.
	
	\pagebreak
	
\begin{figure}[h]\centering
		\vspace{-1mm}
		\includegraphics[scale=0.16]{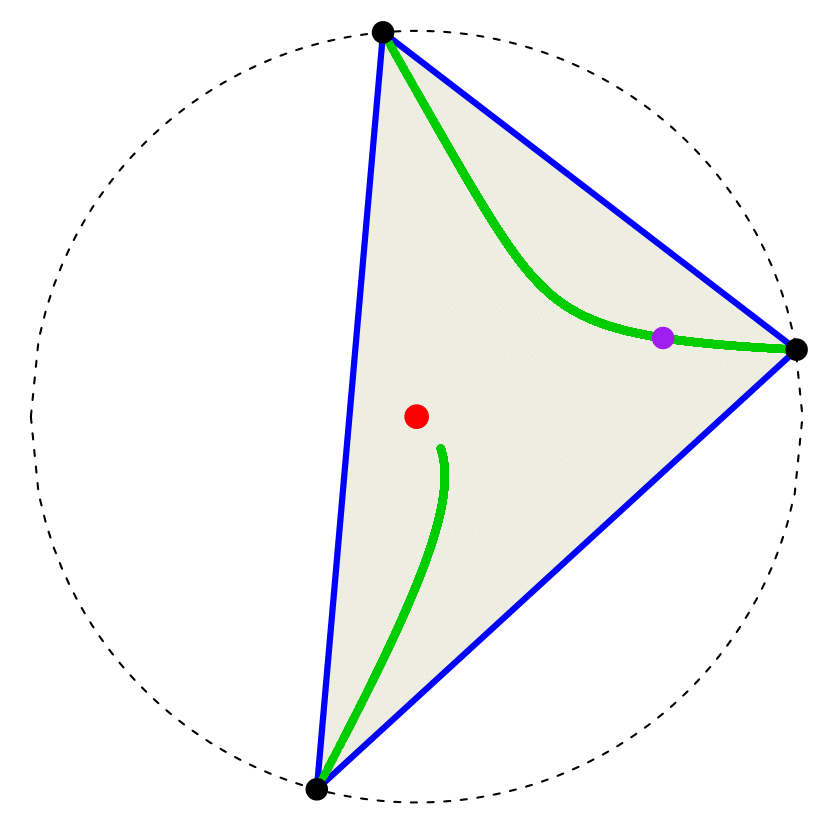}	\hspace{-2mm}
		\includegraphics[scale=0.16]{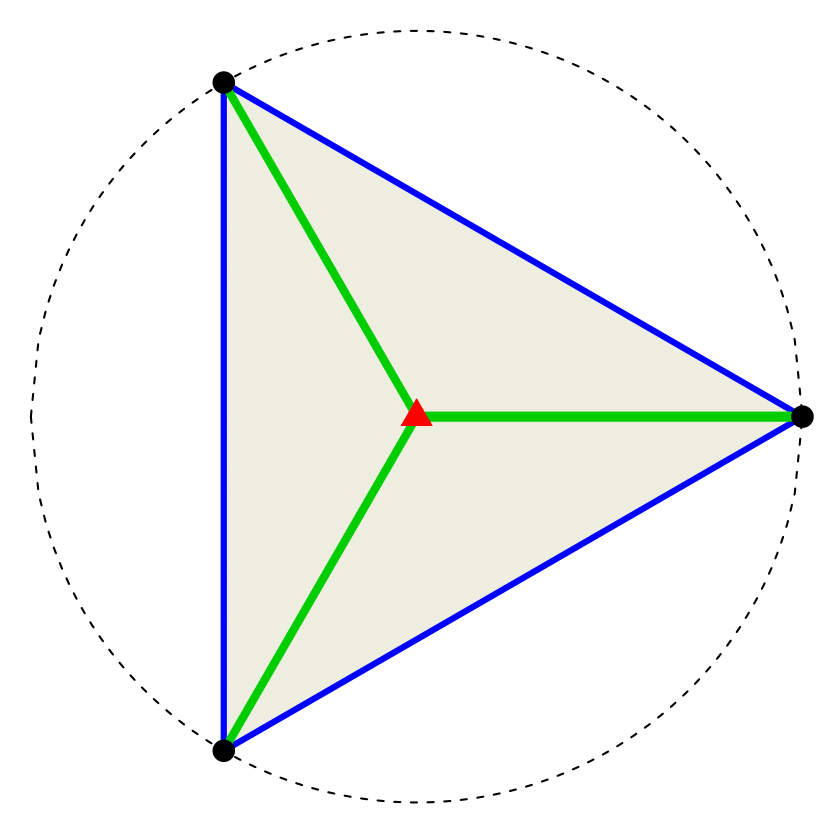}	\hspace{-2mm}
		\includegraphics[scale=0.16]{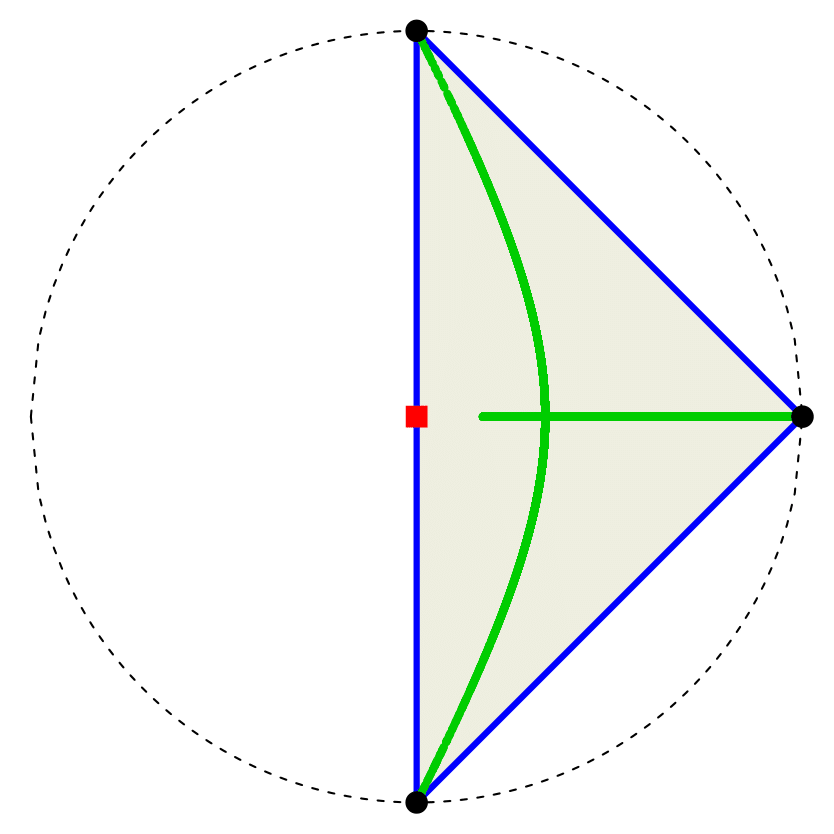}	
		
		{\color{black} \sf \vspace{-3mm} \hspace{0mm} \footnotesize (a) \small  \hspace{4.3cm}   \footnotesize (b) \small  \hspace{4.3cm}   \footnotesize (c) \small }\\ 
		\vspace{1mm}
		\includegraphics[scale=0.65]{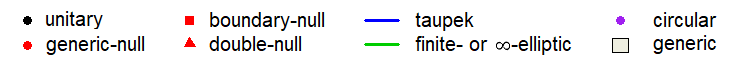}	
		\vspace{-1mm}
		\captionsetup{width=0.8\linewidth}
		\caption{Type of the Markov chain generated by $\mathcal{F}_{U,\,  \omega}$ for $\omega \in  \mathfrak{W}(U)$ in the case of  $\mathfrak{W}(U)$ being non-degenerate and (a) acute-non-equilateral;  (b)~equilateral; (c) right-angled (and isosceles). The lines corresponding to finite- or $\infty$-elliptic chains were found numerically.}\label{figtwtrojkat}
	\end{figure}

	\begin{theorem}[Thm. \ref{chainsclass} revisited,  see also  Fig. \ref{figtwtrojkat}]\label{klasyfikacjanatrojkacie}
		Let $\omega \in \mathfrak{W}(U)$. The Markov chain generated by $\mathcal{F}_{U,\,  \omega}$ is 
		\vspace{-0.5mm}
		\begin{enumerate}[leftmargin=27mm]
			\itemsep-0.1mm
			\item[\sf \underline{unitary}]     iff  $\omega\in \sigma(U)$;
			\item[\sf \underline{null\vphantom{y}} ]   iff $\omega=0$ and the subtype of this chain is
			\vspace{-1.5mm}
			\begin{enumerate}[leftmargin=2.2cm]
				\itemsep0.35mm
				\item  [\sf \underline{generic-null}]   iff $\,\mathfrak{W}(U)$ is an acute-non-equilateral triangle;
				
				\item[\sf \underline{double-null\vphantom{y}}]     iff $\,\mathfrak{W}(U)$ is an equilateral triangle;
				
				\item[\sf \underline{taupek-null}]    iff $\,\mathfrak{W}(U)$ is a right-angled triangle or a diameter;
			\end{enumerate}
			
			\vspace{-0.35mm}

			\item[\sf \underline{taupek}] iff $\omega \in \partial \mathfrak{W}(U)\setminus (\sigma(U) \cup \{0\})$;
			
			\item[\sf  \underline{finite-elliptic}]   iff $\omega\in \Int \mathfrak{W}(U)\setminus \{0\}$,  $\frac{ (\tr U - \omega)^2}{\overline{\omega}\det U} \in [0, 4)$ and  $\arccos\frac{|\tr U - \omega|}{2\sqrt{\smash[b]{|\omega|}}} \in \pi\mathbb{Q}$;
			
			$\bullet$  in particular: 
			\underline{\sf circular} iff $\omega=\tr U$;
			
			\vspace{1mm}
			
			\item[\sf  \underline{$\infty$-elliptic}]   iff $\omega\in \Int \mathfrak{W}(U)\setminus \{0\}$,  $\frac{ (\tr U - \omega)^2}{\overline{\omega}\det U} \in (0, 4)$ and  $\arccos\frac{|\tr U - \omega|}{2\sqrt{\smash[b]{|\omega|}}} \notin \pi\mathbb{Q}$;
			
			\item[\sf \underline{generic}] otherwise.
			\vspace{1mm}
			
		\end{enumerate}	
	\end{theorem}

	Theorem \ref{klasyfikacjanatrojkacie} allows us to determine  the type of the Markov chain generated by $\mathcal{F}_{U,\,\omega}$ by inspecting  $\mathfrak{W}(U)$ and $\omega$ instead of  the eigenvalues of $A_\omega$. Furthermore, taking a~look at $\mathfrak{W}(U)$, one can quickly tell which of the eight  possible chain types can be generated with help of $U$ as one varies the ray in $\C\mathbb{P}^2$ that specifies the measurement, which translates into varying $\omega$ over  $\mathfrak{W}(U)$. Clearly, the only non-trivial problem here is to decide whether elliptic chains can be generated. 
	In the next  subsection  we shall prove that the subsets of $\mathfrak{W}(U)$ where finite- and $\infty$-elliptic chains are generated are both infinite whenever $\mathfrak{W}(U)$ is non-degenerate. Also, we shall see that these two subsets are  contained in a cubic  plane curve. Therefore, if $\Int \mathfrak{W}(U) \neq \varnothing$, then $\{\omega \in \mathfrak{W}(U)\, \colon \, \mathcal{F}_{U,\,\omega}\, \textrm{ generates a generic  chain}\}$ is  indeed  generic in both  topological and measure-theoretic sense.

	\subsection{Elliptic chains}\label{secEll}
	
	Throughout this subsection we assume  that 
	$\operatorname{Int}\mathfrak{W}(U)\neq \varnothing$.
	We shall now prove that the two sets  $\{\omega \in \mathfrak{W}(U)\, \colon \, \mathcal{F}_{U,\,\omega}\, \textrm{ generates a finite-elliptic  chain}\}$ and  \mbox{$\{\omega \in \mathfrak{W}(U) \,\colon\, \mathcal{F}_{U,\,\omega} \,\textrm{ generates an $\infty$-elliptic chain}\}$} are both infinite. 
	Let us adopt the following notation. Put   $\tilde{f}(z):=\tfrac{(\tr U - z)^2}{\overline{z}\det U}$, $z \in \C\setminus \{0\}$, and 
	\begin{align*}
		\mathcal{E}:=&\,\big\lbrace \omega \in \operatorname{Int}\mathfrak{W}(U) \colon A_\omega \textrm{ generates elliptic dynamics}\big\rbrace\\[0.1em] 
		= & \, 
		\big\lbrace
		\omega \in \operatorname{Int} \mathfrak{W}(U)  \colon  \tilde{f}(\omega) \in [0,4) 
		\big\rbrace.
	\end{align*}
	Additionally, we consider $f(z):=\frac{ (\tr U - z)^2z}{\det U}$, $z \in \C$, and 
	$$
	{\mathcal{C}}:=\big\lbrace z \in \C \colon 
	\Im f(z) =0  \big\rbrace.$$ 
	Clearly,  $\mathcal{C}$ is a  cubic plane curve. Observe  that 
	$\tilde{f}(z)=f(z)/|z|^2$, and so $\Im f(z)=0$ iff $\Im \tilde{f}(z)=0$ or $z =0$. Hence, in particular,   $\mathcal{E} \subset   {\mathcal{C}}$.
	Further,    note that \mbox{$\{0, \tr U\} \subset \mathcal{C}$}, i.e.,   $\mathcal{C}$ contains both the circumcentre and the orthocentre of $\mathfrak{W}(U)$. It follows easily that the eigenvalues of $U$, i.e., 
	the vertices of $\mathfrak{W}(U)$, also  belong to
	$\mathcal{C}$. Indeed, it suffices to note that 
	\begin{align}
		f(\e^{\I\phi_1}) 
		=(\e^{\I\phi_2}+\e^{\I\phi_3})^2\e^{-\I\phi_2 -\I\phi_3}
		=2+2\cos(\phi_2-\phi_3) \in [0,4] \subset \R, \label{wierzchrzecz}
	\end{align}
	where, as before, {$\e^{\I\phi_1}, \e^{\I\phi_2}, \e^{\I\phi_3}$ with $\phi_1, \phi_2, \phi_3 \in [0, 2\pi)$ stand for the eigenvalues of $U$.}
	%\smallskip
	
	We now show that a vertex $\mu $ of $\mathfrak{W}(U)$ is an accumulation point of $\mathcal{C} \cap  \operatorname{Int} \mathfrak{W}(U)$ if the (interior) vertex angle of $\mathfrak{W}(U)$ at  $\mu$  is not right.	Clearly, considering  the numerical range of $\diag(\e^{\I\phi_1}, \e^{\I\phi_2}, \e^{\I\phi_3})$, we see that the vertex angle at  $\e^{\I\phi_1}$ is right iff $\phi_2 =\phi_3 \pm \pi$. By $B(z,r)$  we denote  the (Euclidean) open ball in $\C$ of radius $r>0$ centred at $z\in \C$.

	\begin{proposition}\label{oneellipticlem}
		Let $\mu \in \sigma(U)$. If the interior vertex angle of $\mathfrak{W}(U)$  at $\mu$ is not right, then
		$\mathcal{C} \cap \operatorname{Int}\mathfrak{W}(U)\cap  B(\mu,r)  \neq \varnothing$ for every ${r} >0$.
	\end{proposition}
	\begin{proof}
		{By Prop.~\ref{exconjpifs}(i)  we may, with no loss of generality, so adjust the overall phase  that $U \sim \diag(1, \e^{\I\varphi_1}, \e^{\I\varphi_2})$, $\varphi_1, \varphi_2 \in [0, 2\pi)$, and the  vertex angle of $\mathfrak{W}(U)$ at $1$ is not right.}  Fix $r>0$. 
		We   show that  $\mathcal{C} \cap \operatorname{Int}\mathfrak{W}(U)\cap  B(1,r)\neq \varnothing$.
		For $\varepsilon \in (0,1)$ we  consider  
		$$\gamma_\varepsilon \colon [0,\varepsilon] \ni t \mapsto (1-\varepsilon)+(\varepsilon-t)\e^{\I\varphi_1}+t\e^{\I\varphi_2} \in \mathfrak{W}(U).$$ 
		It can easily be seen that  $\gamma_{{\varepsilon}}([0,{\varepsilon}])$  corresponds to 	a segment in $\mathfrak{W}(U)$ which  
		connects the side between   $1$ and $\e^{\I\varphi_1}$ with that  between   $1$ and $\e^{\I\varphi_2}$, and which is parallel to the side   between $\e^{\I\varphi_1}$ and $\e^{\I\varphi_2}$.
		
		Clearly, there exists
		${\delta} \in (0, 1)$ such that  $\gamma_{{\varepsilon}}([0,\varepsilon]) \subset B(1,r)$ if $\varepsilon \in (0, {\delta})$.
		Indeed, let $\varrho \in (0, r]$ satisfy $\varrho < \min\{|1-\e^{\I\varphi_1}|, |1-\e^{\I\varphi_2}|\}$, i.e., $\partial B(1,\varrho)$ intersects both the side of $\mathfrak{W}(U)$ connecting $1$ and $\e^{\I\varphi_1}$ as well as that connecting $1$ and $\e^{\I\varphi_2}$. For $j \in \{1,2\}$  denote
		$\delta_j:={\varrho}/{|1-\e^{\I\varphi_j}|}$. 
		Putting ${\delta}:=\min\{\delta_1, \delta_2\}$ and letting $\varepsilon \in (0, {\delta})$,
		we obtain 
		\begin{align*}
			|1-\gamma_{{\varepsilon}}(0)|&=|1-(1-\varepsilon)-\varepsilon\e^{\I\varphi_1}|
			%			\\ & 
			=\varepsilon|1-\e^{\I\varphi_1}| 
			%\\ &
			={\varepsilon}\varrho/{\delta_1} 
			%\\ & 
			<\varrho 
		\end{align*}
		so 	$\gamma_{{\varepsilon}}(0) \in B(1,\varrho)$. Likewise,  $\gamma_{{\varepsilon}}(\varepsilon) \in B(1,\varrho)$. Hence,  $\gamma_{{\varepsilon}}([0, \varepsilon]) \subset B(1,\varrho) \subset B(1,r)$,  as claimed.

		Moreover, for every $\varepsilon \in (0, 1)$ we have  $\gamma_\varepsilon((0,\varepsilon)) \subset \operatorname{Int}\mathfrak{W}(U)$. Therefore, to conclude that 
		$  \mathcal{C} \cap \operatorname{Int}\mathfrak{W}(U) \cap B(1, r) \neq \varnothing$, it suffices to show that 
		$\mathcal{C} \cap \gamma_{{\varepsilon}}((0,{\varepsilon}))    \neq \varnothing$ for some 
		${\varepsilon} \in (0, \delta)$, where  $\delta$ is as above.	
		To this end, let us   consider $(0,1) \ni \varepsilon \mapsto  \gamma_\varepsilon(0) =  (1-\varepsilon)+\varepsilon\e^{\I\varphi_1},$ which, obviously, is a parametrisation of the side of $\mathfrak{W}(U)$ between $1$ and $\e^{\I\varphi_1}$. By direct computation we obtain 
		\[
		\Im f(\gamma_\varepsilon(0))=2\varepsilon(1-\varepsilon)(1-\cos \varphi_1) \left[ \sin(\varphi_2-\varphi_1)+    \varepsilon
		( \sin(\varphi_1-\varphi_2)+\sin \varphi_2) \right].
		\]
		Note that $\sin(\varphi_2-\varphi_1) \neq 0$, because  $\varphi_1  \neq  \varphi_2$ since    $\operatorname{Int}\mathfrak{W}(U)\neq \varnothing$, and   $\varphi_1 \neq \varphi_2 \pm \pi$ since the vertex angle at $1$ is not right. 
		Consequently, there exists $\varepsilon_1\in (0,\delta)$ such that  for $\varepsilon \in (0, \varepsilon_1)$ we have  $\sgn\Im f(\gamma_\varepsilon(0))=\sgn \sin(\varphi_2-\varphi_1)$.
		Analogously, for 
		$(0,1) \ni \varepsilon \mapsto  \gamma_\varepsilon(\varepsilon)= (1-\varepsilon)+\varepsilon\e^{\I\varphi_2}$ 
		there exists $\varepsilon_2\in (0,\delta)$ such that $\sgn\Im f(\gamma_\varepsilon(\varepsilon))=\sgn \sin(\varphi_1-\varphi_2)$
		for $\varepsilon \in (0, \varepsilon_2)$. 
		
		Finally, letting ${\varepsilon}  \in (0, \min\{\varepsilon_1, \varepsilon_2\})$, we have  
		$
		\sgn\Im f(\gamma_\varepsilon(0))=-\sgn\Im f(\gamma_\varepsilon(\varepsilon))$.
		Since 	$\Im \circ f \circ \gamma_{{\varepsilon}} \colon [0,{\varepsilon}] \to \R$ is a continuous function which  takes on values of different signs at the endpoints of the interval constituting its domain,  
		Bolzano's (intermediate value) theorem  assures the existence of  $t \in (0, \varepsilon)$ such that $\Im f(\gamma_\varepsilon(t))=0$.
		Therefore, $ \mathcal{C} \cap \gamma_\varepsilon((0,\varepsilon))  \neq \varnothing$,  which concludes the proof. 
	\end{proof}

	\begin{corollary}\label{branch}
		Since $\mathcal{C} \cap \operatorname{Int}\mathfrak{W}(U)$ is a semi-algebraic set, Theorem  \ref{oneellipticlem} and the Curve Selection Lemma (see, e.g., \cite[\S 3]{milnor1968singular} or \cite[Thm. 2.5.5]{bochnak2013real})   imply that 
		for each $\mu \in \sigma(U)$ there exists  a~smooth arc $g \colon  [0,1]    \to   \mathcal{C}$ such that  $g(0)=\mu$ and   $g(t) \in  \operatorname{Int}\mathfrak{W}(U)$ for  $t \in (0,1]$.
	\end{corollary}

	Next, we show that  $\mathcal{E}$ coincides with $\mathcal{C}$ in the vicinity of $\mu \in \sigma(U)$, provided that  the (interior) vertex angle of $\mathfrak{W}(U)$ at  $\mu $  is not right.
	
	\begin{proposition}\label{oneelliptic2}
		Let $\mu \in \sigma(U)$. If the interior vertex angle of $\mathfrak{W}(U)$  at $\mu$ is not right, then there exists $r>0$ such that   $\mathcal{C}\cap \operatorname{Int}\mathfrak{W}(U)\cap B(\mu, r) \subset  \mathcal{E}$.
	\end{proposition}
	\begin{proof}
		Let the overall phase be so adjusted that $U \sim \diag(1, \e^{\I\varphi_1}, \e^{\I\varphi_2})$, where $\varphi_1, \varphi_2 \in [0,2\pi)$, and the  vertex angle of $\mathfrak{W}(U)$ at $1$ is not right.
		First, we show that
		$\tilde{f} (1) \in (0,4)$. Indeed, 
		from \eqref{wierzchrzecz} we obtain
		$	
		\tilde{f}(1) = f(1) 
		= 2+2\cos(\varphi_1-\varphi_2).
		$
		%\end{align*}
		Again, since  $\operatorname{Int}\mathfrak{W}(U)\neq \varnothing$, we have $\varphi_1   \neq \varphi_2 $, and  from the assumption that  the  vertex angle of $\mathfrak{W}(U)$ at  $1$ is not the right angle we have 
		$\varphi_1 \neq \varphi_2 \pm \pi$.
		Therefore,  $\cos(\varphi_1-\varphi_2) \in (-1,1)$, and so $\tilde{f}(1) \in (0,4)$, as claimed.
		%		\item 
		Clearly, $\tilde{f}$ is continuous at~$1$, and thus so is $\Re \circ \tilde{f}$; hence,  there exists $r \in (0,1)$ such that $\Re \tilde{f}(B(1,{r})) \subset (0,4)$.
		
		Let $\omega \in \mathcal{C} \cap \operatorname{Int}\mathfrak{W}(U) \cap B(1, {r})$. It follows that   $\omega \neq 0$  and $\Re \tilde{f}(\omega) \in (0,4)$  since  $\omega \in B(1, {r})$ with $r<1$, while  $\omega  \in \mathcal{C}$ guarantees that $\Im \tilde{f}(\omega)=\Im f(\omega) = 0$.
		Therefore, $\omega \in  \operatorname{Int}\mathfrak{W}(U)\setminus\{0\}$ and  $\tilde{f}(\omega) \in (0,4)$, which implies that 
		$\omega \in  \mathcal{E}$, as desired. 
	\end{proof}
	
	\smallskip

	Finally,   recall from \eqref{wzornakat} that 	if $\omega \in \mathcal{E}$, then  the (smaller) angle between the eigenvalues of~$A_\omega$ is given by 
	$$\Upsilon(\omega) =  2 \arccos\frac{|\tr U - \omega |}{2\sqrt{|	 \omega	|}}.$$
	\begin{proposition}\label{psieinterval}
		There is a non-trivial interval   contained in $\Upsilon(\mathcal{E})$.
	\end{proposition}	
	\begin{proof}
		Here we adjust the overall phase so that $\det U=1$. Recall that $\tilde{f}(\mathcal{E}) \subset [0,4)$. Let  $\omega \in \mathcal{E}$ and observe that 
		$$		\Upsilon(\omega)=2 \arccos \left( \tfrac 12  
		\sqrt{{\tilde{f}(\omega)} } \right).$$
		Clearly, it suffices to show that  a~non-trivial interval is contained in $\tilde{f}(\mathcal{E})$.
		Recall that 
		Cor.~\ref{branch} \& Prop.~\ref{oneelliptic2} assure that
		there exists $\mu \in \sigma(U)$ and a~smooth arc  $g \colon  [0,1]  \to   \mathfrak{W}(U)$ such that  $g(0)=\mu$ and $g(t) \in \mathcal{E}$ for $t \in (0,1]$. 
		Since  $\tilde{f}$ is continuous, it actually suffices to show   that $ \tilde{f} \circ g\colon (0,1] \to [0,4) \subset \R$ is not constant.  
		
		Fix $a \in \R$. 
		We will show that there are finitely many solutions to $\tilde{f}(z)=a$ in $\C$.  
		Clearly, for $z  \in \C \setminus\{0\}$ this equation can be equivalently written as $(\tr U   - z)^2=a\overline{z}$. 
		Letting $x, y \in \R$ be such that $z = x+\I y$,
		we obtain 			\begin{align*}
			\left\lbrace\begin{array}{ll}
				(\Re \tr U - x)^2 - (\Im \tr U - y)^2 = ax
				\\[0.2em]
				2(\Re \tr U - x)(\Im \tr U - y) = -ay
			\end{array}\right.	
		\end{align*}  		
		Each of these equations describes a hyperbola (possibly degenerated to two intersecting lines). 
		The hyperbola given by the upper equation has asymptotes with slopes equal to $\pm 1$, while the other hyperbola has asymptotes parallel to the axes. Hence, these hyperbolas have  at most four points in common, and so the 
		equation $\tilde{f}(z)=a$  holds at finitely many  points, as claimed.  
		In consequence, $\tilde{f} \circ g$ cannot be constant on $(0,1]$. It follows that  \mbox{$\{\tilde{f}(g(t))\colon t \in (0,1] \}$} contains a non-trivial interval, and thus so does $\Upsilon(\mathcal{E})$, as desired.
	\end{proof}

	Let $\omega \in \mathcal{E}$ and recall that the chain generated by $\mathcal{F}_{U,\,\omega}$ is finite-elliptic  iff $\Upsilon(\omega) \in \pi\mathbb{Q}$; otherwise, this  chain is $\infty$-elliptic, see Thm. \ref{klasyfikacjanatrojkacie}.
	It follows from Prop. \ref{psieinterval} that  
	$\Upsilon(\mathcal{E})\cap (\pi\mathbb{Q})$
	and 
	$\Upsilon(\mathcal{E})\setminus (\pi\mathbb{Q})$
	are both infinite, so we have the following
	
	\begin{theorem}
		If	$\,\operatorname{Int}\mathfrak{W}(U)\neq \varnothing$, then 
		$\{\omega \in \mathfrak{W}(U) \colon  \mathcal{F}_{U,\,\omega} \textrm{ generates a finite-elliptic  chain}\}$
		and  
		$\{\omega \in \mathfrak{W}(U) \colon \mathcal{F}_{U,\,\omega} \textrm{ generates an $\infty$-elliptic chain}\}$ 
		are both infinite.
	\end{theorem}

	\begin{remark}\label{secMuss}

		Let us now take a closer look at the cubic curve $\mathcal{C}$ in  which $\mathcal{E}$ is contained.  
		It turns out that if the polynomial defining $\mathcal{C}$ is irreducible, then $\mathcal{C}$ is the  \emph{Musselman (third) cubic}, which was introduced in \cite{Musselman}
		and catalogued in \cite{MusselmanCatalogue} as K028.
		Indeed, let the overall phase be  so adjusted   that $\det U=1$. Then $\mathcal{C}$ is the zero-set of the following polynomial
		\begin{equation}\label{musselmaneq}
			\mathcal{M}(x,y)=	y^3-3x^2y+2\beta x^2+4\alpha xy-2\beta y^2-2\alpha\beta x+\beta^2y-\alpha^2y,
		\end{equation} 
		where $\alpha:=\Re \tr U$, $\beta:=\Im \tr U$.
		Substituting in \eqref{musselmaneq} the Cartesian coordinates $(x,y)$ with the barycentric coordinates $(a, b, c)$, $a+b+c=1$,  
		via the conversion formula 
		\[
		\begin{array}	{ll}
			x=a\cos\phi_1 + b\cos\phi_2 +c\cos\phi_3, \\
			y=a\sin\phi_1\, + b\sin\phi_2\, + c\sin\phi_3,
		\end{array} 	
		\]
		where $\e^{\I\phi_1}, \e^{\I\phi_2}, \e^{\I\phi_3}$ denote  the eigenvalues of $U$, we arrive at the barycentric equation of the Musselman cubic as given in \cite{MusselmanCatalogue}.

		The Musselman cubic is  an \emph{equilateral cubic}, i.e., 
		it has three   asymptotes that concur and the (smaller) angle of intersection of every two of these asymptotes is equal to ${\pi}/{3}$, see also Fig. \ref{musselman}(a)--(c).  This can be easily deduced by examining the cubic terms in \eqref{musselmaneq}. Namely,    the slopes of the  asymptotes   satisfy the equation $s^3=3s$ ($s \in \R$), which implies that these slopes read $s_1=0=\tan 0$, $s_2=\sqrt{3}=\tan \frac \pi 3$, $s_3=-\sqrt{3}=\tan  \frac {2\pi} 3$.
		The point of the concurrence of  the asymptotes turns out to be  the midpoint between the centroid and orthocentre of $\mathfrak{W}(U)$ \cite{MusselmanCatalogue}.  
		The Musselman cubic has a \emph{node} (an \emph{ordinary double point}, i.e., a point where exactly two branches intersect and they have distinct tangent lines) at the orthocentre of $\mathfrak{W}(U)$, i.e., at $\tr U$, where two of its branches intersect orthogonally  \mbox{\cite[p. 96]{MusselmanGibert}}.

		\begin{figure}[b]
			\vspace{-1mm}
			\includegraphics[scale=0.25]{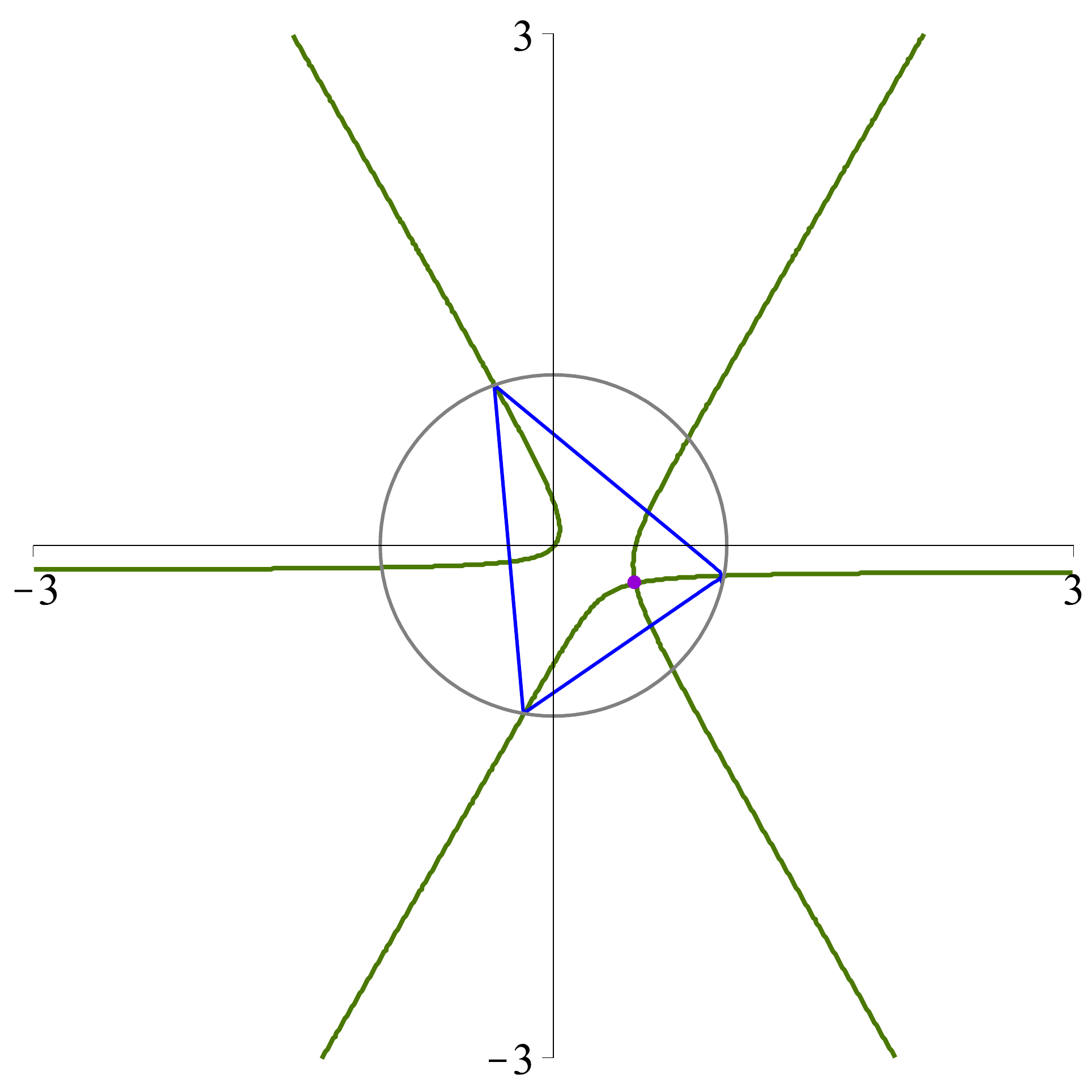}
			\includegraphics[scale=0.25]{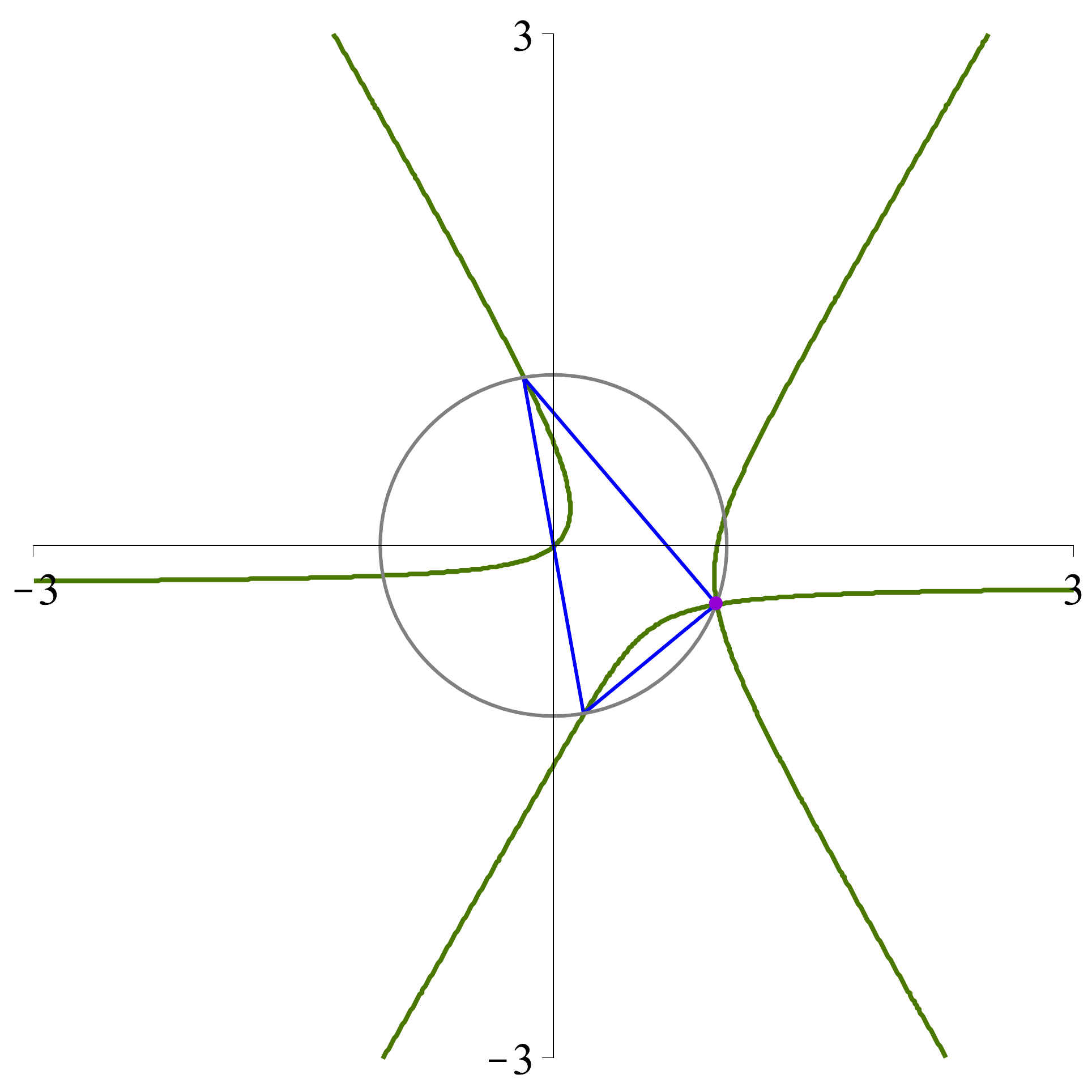}
			\includegraphics[scale=0.25]{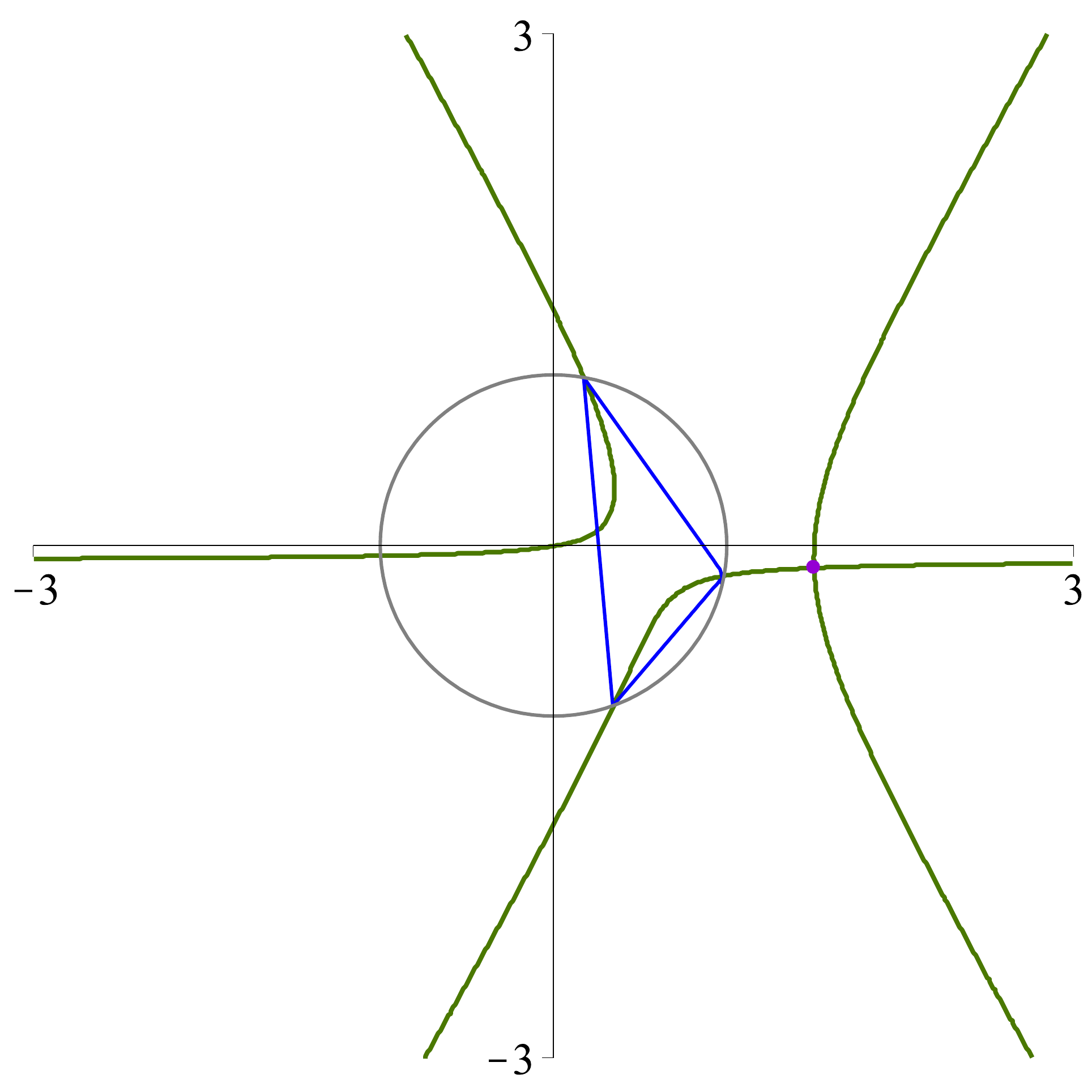}\\
			{	\vspace{-3mm} \color{black}  
				\flushleft	\footnotesize (a) \small $\tr U \in \operatorname{Int}\mathfrak{W}(U)$ 
				\hspace{17mm}   
				\footnotesize (b) \small $\tr U \in \partial\mathfrak{W}(U)$ 	
				\hspace{17mm}   
				\footnotesize (c) \small $\tr U \notin \mathfrak{W}(U)$ 
				\centering
			}\\ 
			\vspace{4mm}
			\includegraphics[scale=  0.25]{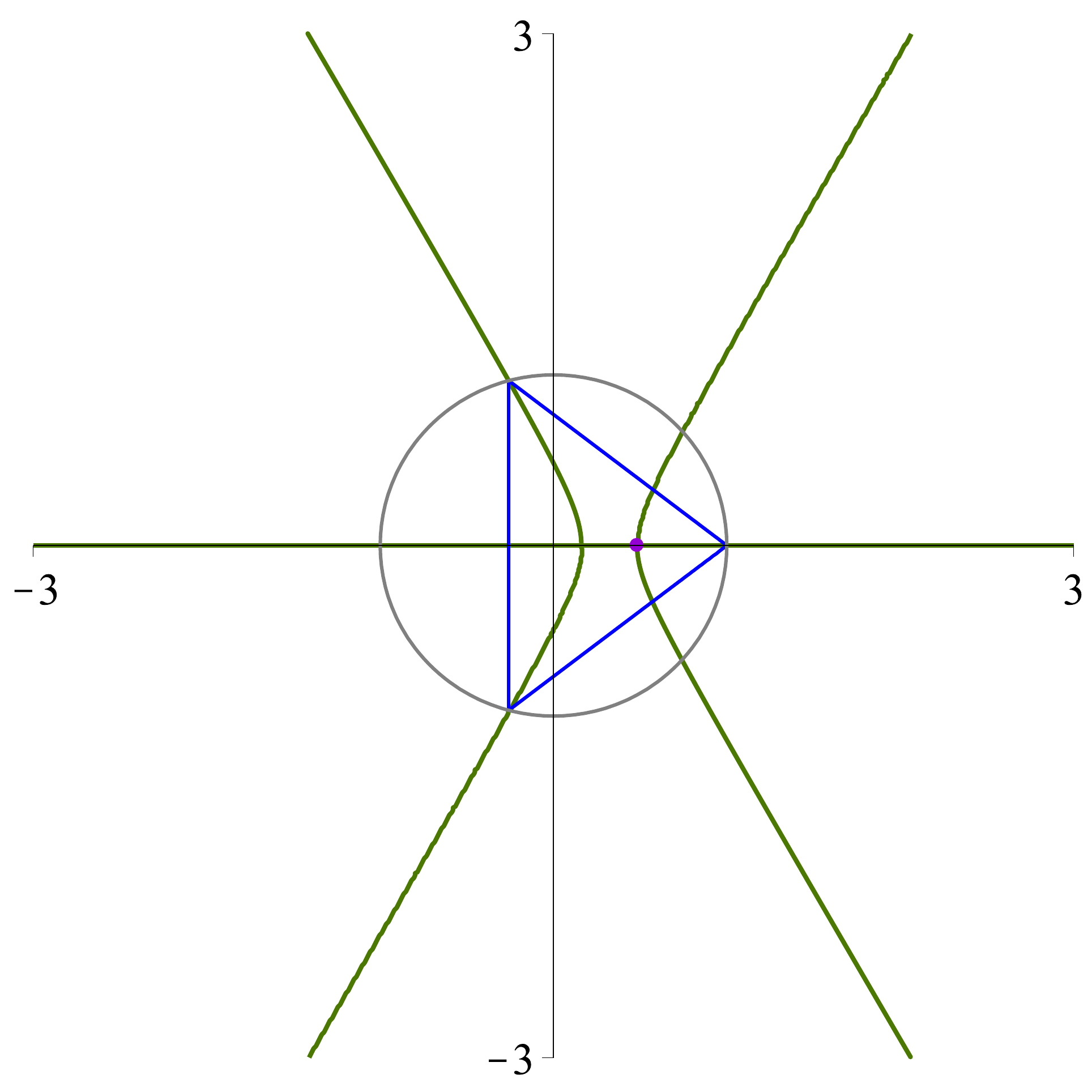}
			\includegraphics[scale=  0.25]{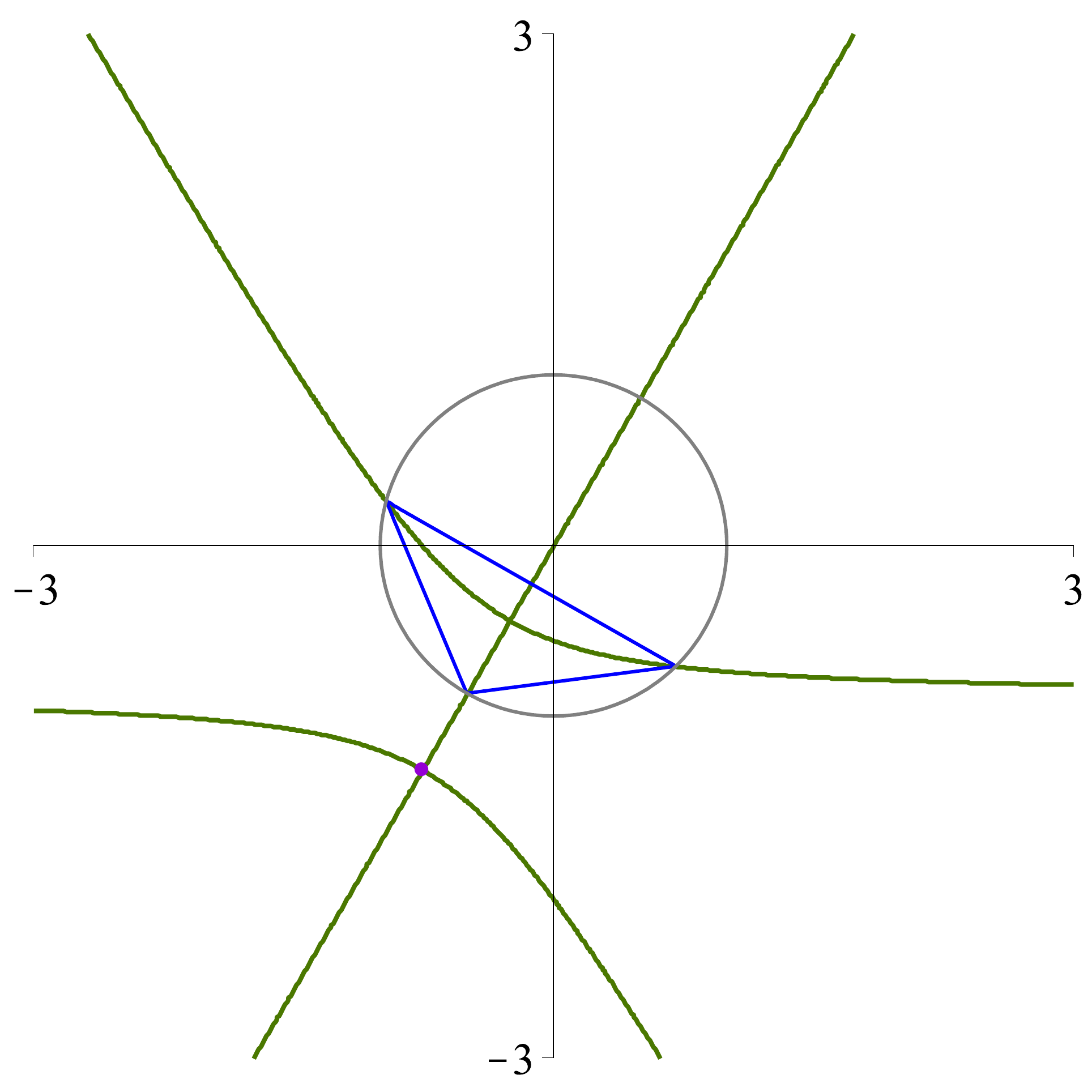}
			\includegraphics[scale=  0.25]{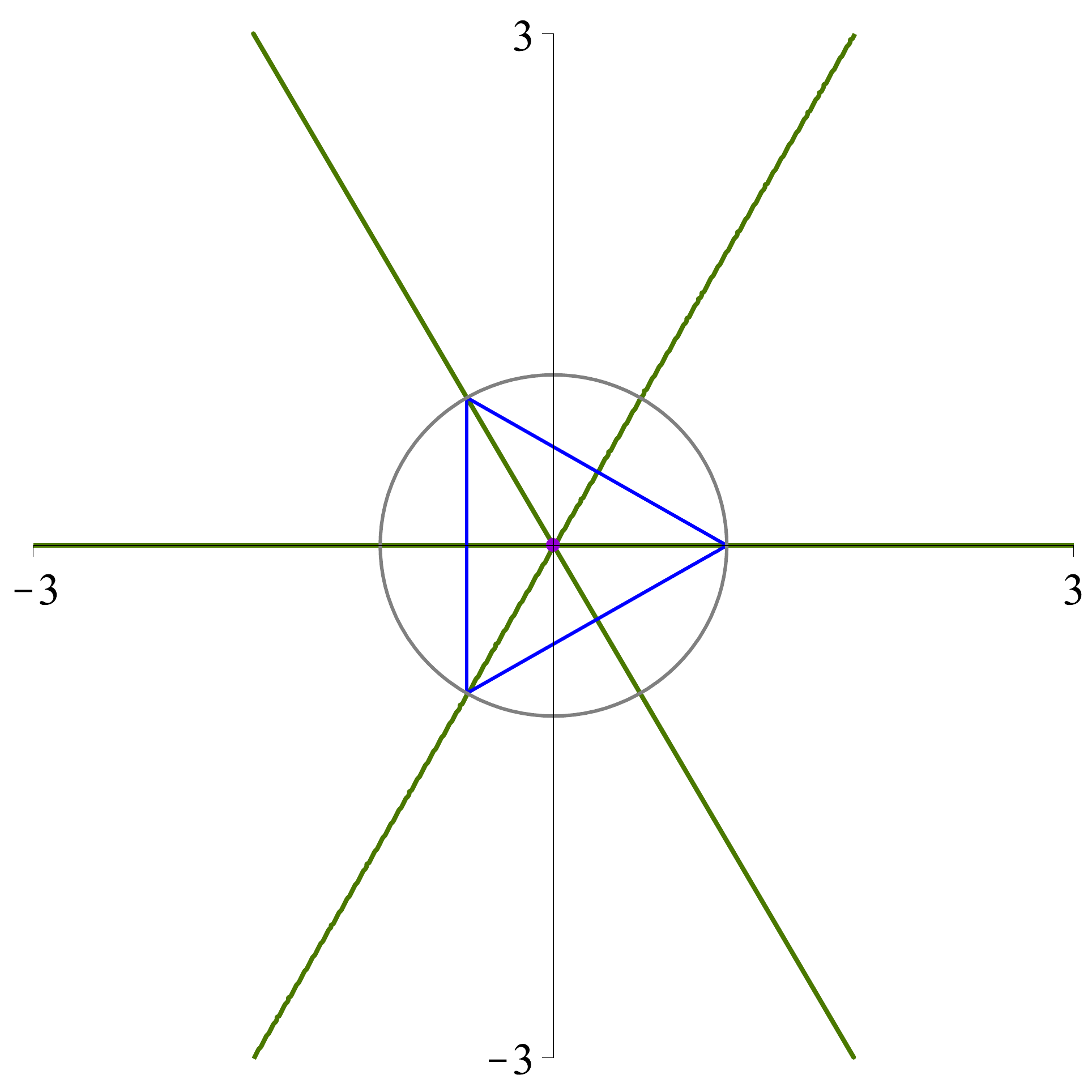}\\
			{\flushleft	\vspace{-3mm} \color{black}  
				\footnotesize (d) \small$\alpha \neq 0, \beta=0$ 
				%		\hspace{1.25cm}  
				\hspace{22mm} 
				\footnotesize (e) \small $\alpha \neq 0, \beta=\alpha\sqrt{3}$ 
				\hspace{18mm} 
				\footnotesize (f) \small $\alpha=\beta=0$
				\hspace{0.2cm} 
				\centering}
			\captionsetup{width=0.85\linewidth}
			\caption{	$\mathcal{C}$ (in green) along with $\partial\mathfrak{W}(U)$ (in blue) and $\tr U$ (the purple dot): \\ (a)--(c)  Musselman (third) cubics in the generic case  of non-isosceles $\mathfrak{W}(U)$ \\ (d)--(f)  degenerate cubics in the case  of isosceles $\mathfrak{W}(U)$
			}\label{musselman}	
		\end{figure}
		
		Let us further inspect $\mathcal{M}(x,y)$ and discuss the potential degenerations of $\mathcal{C}$. 	
		Namely, 
		$$
		\tfrac{\partial \mathcal{M}}{\partial x}(x,y)=\tfrac{\partial \mathcal{M}}{\partial y}(x,y)=0\  \Longleftrightarrow\  (x,y)\in \big\lbrace(\alpha, \beta), (\tfrac \alpha 3,\tfrac \beta 3)\big\rbrace \sim \big\lbrace \!\tr U, \tfrac13\!\tr U\big\rbrace.$$
		We already know that $\tr U \in \mathcal{C}$, i.e., $\mathcal{M}(\alpha,\beta)=0$. As for the other point, we get \mbox{$\mathcal{M}(\frac \alpha 3,\frac \beta 3)=0$} iff  $\beta\in \{0, \alpha\sqrt{3}, -\alpha\sqrt{3}\}$, which turns out to be  equivalent to the fact that   $\mathfrak{W}(U)$ is isosceles. 
		Obviously, $(\alpha, \beta)$ and $(\frac \alpha 3,\frac \beta 3)$ coincide iff $\alpha=\beta=0$ iff $\mathfrak{W}(U)$ is equilateral.

		Thus, if $\mathfrak{W}(U)$ is  isosceles  but not equilateral, then $\mathcal{C}$ has two  singular points. It follows that $\mathcal{C}$ degenerates into the union of a line and a hyperbola, see Fig.~\ref{musselman}(d)-(e);
		indeed, it can be easily verified that if $\alpha \neq 0$ and $\beta=0$, then 
		\begin{equation}\label{hiperb1}
			\mathcal{M}(x,y)=y(3{x}^{2}-{y}^{2}-4\alpha x+ {\alpha}^{2}),
		\end{equation}
		and so $\mathcal{C}$ consists of the $x$-axis and of the hyperbola given by the equation  
		\begin{equation}\label{hiperb2}
			\frac{(x-2\alpha/3 )^2}{(\alpha/3 )^2} - \frac{y^2}{ ( {\alpha}/{\sqrt{3}} )^2}=1.
		\end{equation}
		Observe that the transverse axis of this hyperbola coincides with the $x$-axis and its vertices are located at $\alpha = \tr U$ and $\tfrac \alpha3  = \frac 13\!\tr U$. 
		Similarly, one can verify that if $\alpha \neq 0$ and 
		$\beta= \pm \alpha\sqrt{3}$, then  $\mathcal{C}$ consists of the median of base of the triangle corresponding to  $\mathfrak{W}(U)$ and of a hyperbola  whose transverse axis coincides with this median and which has vertices at $\tr U$ and $\frac 13\!\tr U$.
		Lastly, if the triangle is equilateral ($\alpha = \beta=0$), then
		$$
		\mathcal{M}(x,y)=y(\sqrt{3}x-y)(\sqrt{3}x+y),$$ and so 
		$\mathcal{C}$ degenerates into the union of three triangle medians,
		see Fig. \ref{musselman}(f).

		In consequence, $\mathcal{M}(x,y)$ is reducible whenever   
		the triangle corresponding to $\mathfrak{W}(U)$ is isosceles. We now argue that the converse  also holds. If $\mathfrak{W}(U)$ is not isosceles, then $(\alpha, \beta)$ is the only singular point of $\mathcal{C}$ and this point is a node since $\det \mathcal{H}_\mathcal{M}(\alpha, \beta) 
		=-4(\alpha^2+\beta^2)<0$, where 
		$\mathcal{H}_\mathcal{M}(\alpha, \beta)$ stands for the Hessian matrix of $\mathcal{M}$ at~$(\alpha,\beta)$.
		Clearly, a reducible cubic curve is the union of an irreducible conic and a~line or of three lines, see also  \cite[p. 677]{weinberg1988} or \cite[p. 70]{gibson1979singular}.  
		It follows that a reducible cubic has exactly one singular point which is a node iff this cubic degenerates into a~parabola and a line parallel to the symmetry axis of this parabola, or into a hyperbola and a line parallel to an asymptote of this hyperbola. 
		In what follows we exclude both these possibilities, which allows us to conclude that  $\mathcal{M}(x,y)$ is irreducible if $\mathfrak{W}(U)$ is not isosceles, as claimed. 
		
		{Namely, assume that the Musselman cubic   degenerates into a conic and a line, i.e., that  
			$$\mathcal{M}(x,y) = (Ax^2+Bxy+Cy^2+Dx+Ey+F)(Gx+Hy+J),$$
			where $A,B,C,D,E,F,G,H,J \in \R$.
			Comparing the right-hand side coefficients with those of $\mathcal{M}$, see \eqref{musselmaneq}, we deduce that   
			\begin{align*}
				\left\lbrace\begin{array}{ll}
					AG = 0 \\
					CH = 1 \\
					CG + BH = 0 \\
					BG + AH = -3
				\end{array}\right.	
			\end{align*} 
			In what follows we will repeatedly use the obvious fact that $A = 0$ or $G = 0$, as well as  $C \neq 0$ and $H \neq 0$. 
			\begin{enumerate}
				\item Assume that  the conic is a parabola, i.e.,  
				$B^2 = 4AC$. 
				\begin{itemize}\itemsep = 1mm
					\item  If $A = 0$, then also $B = 0$, which contradicts $BG + AH = -3$.
					
					\item  If $G = 0$, then $BH = 0$. It follows that $B = 0$, thus also $AC = 0$, and so $A= 0$. This again contradicts $BG + AH = -3$. 
				\end{itemize}
				
				\item Assume that the conic is a hyperbola, i.e., 
				$B^2  > 4AC$. Also, assume that  the line $Gx+Hy+J=0$ is parallel to an asymptote of this hyperbola, i.e., that $-G/H$ is equal to the slope of an asymptote, which reads $\tfrac1{2}(-B \pm \sqrt{B^2-4AC})/C$. 
				\begin{itemize}\itemsep = 1mm
					\item 	If $A = 0$, then the slopes of the asymptotes are equal to $0$ and $-B/C$. If $G/H = 0$, then $G = 0$, which contradicts $BG + AH = -3$. 
					If  $G/H = B/C$, then $G = B  =0$ since we also have $CG + BH = 0$. This contradicts $B^2 > 4AC$.
					
					\item 	If $G = 0$, then also $B = 0$ since $CG + BH = 0$. The asymptotes have slopes $\pm \sqrt{ - A/C}$. If $0=G/H = \pm \sqrt{-A/C}$, then $A = 0$, which   contradicts  $B^2 > 4AC$.  
				\end{itemize}
		\end{enumerate}}
		
	\end{remark}	
	
	\smallskip	
	
	\subsection{{The Crutchfield \& Wiesner example}} 
	\label{secCW}
	In this final subsection we examine the ball \& point   system whose evolution is governed by $U \in \mathcal{U}(\C^3)$   represented  in the standard  basis $e_1, e_2, e_3$ of $\C^3$ as
	\vspace{1mm}
	$$
	\vspace{1mm}
	U \sim 
	\resizebox{2.75cm}{!}{$ 
		\begin{bmatrix}
			\tfrac 1{\sqrt{2}} &\tfrac 1{\sqrt{2}}&0
			\\ \noalign{\medskip}0&0&-1\\ \noalign{\medskip}-\tfrac 1{\sqrt{2}}&\tfrac 1{\sqrt{2}}&0
		\end{bmatrix}
		$}.
	$$
	This  unitary matrix   has been discussed extensively by  Crutchfield \& Wiesner in a~series of papers \cite{CruWie08,Monetal11,Wie10,CruWie08a,CruWie10}. The   measurements they have considered are the PVMs of the form  $\Pi(\mathsf{z})=\{\rho_{\mathsf{z}},\, \mathbb{I}_3-\rho_{\mathsf{z}}\}$ with $\mathsf{z}$ taken from the standard basis  of $\C^3$. This has led to two different chains, namely those corresponding to $\omega  = \braket{\mathsf{z}|U\mathsf{z}} \in \{0,\frac{1}{\sqrt{2}}\}$, because   $\braket{e_1|Ue_1}=\frac{1}{\sqrt{2}}$ and $\braket{e_2|Ue_2}=\braket{e_3|Ue_3}=0$.
	
	\smallskip
	
\begin{figure}[b]
		\captionsetup{width=0.7\linewidth}	
		\vspace{-10mm}	
		\centering
		\includegraphics[scale=0.7125]{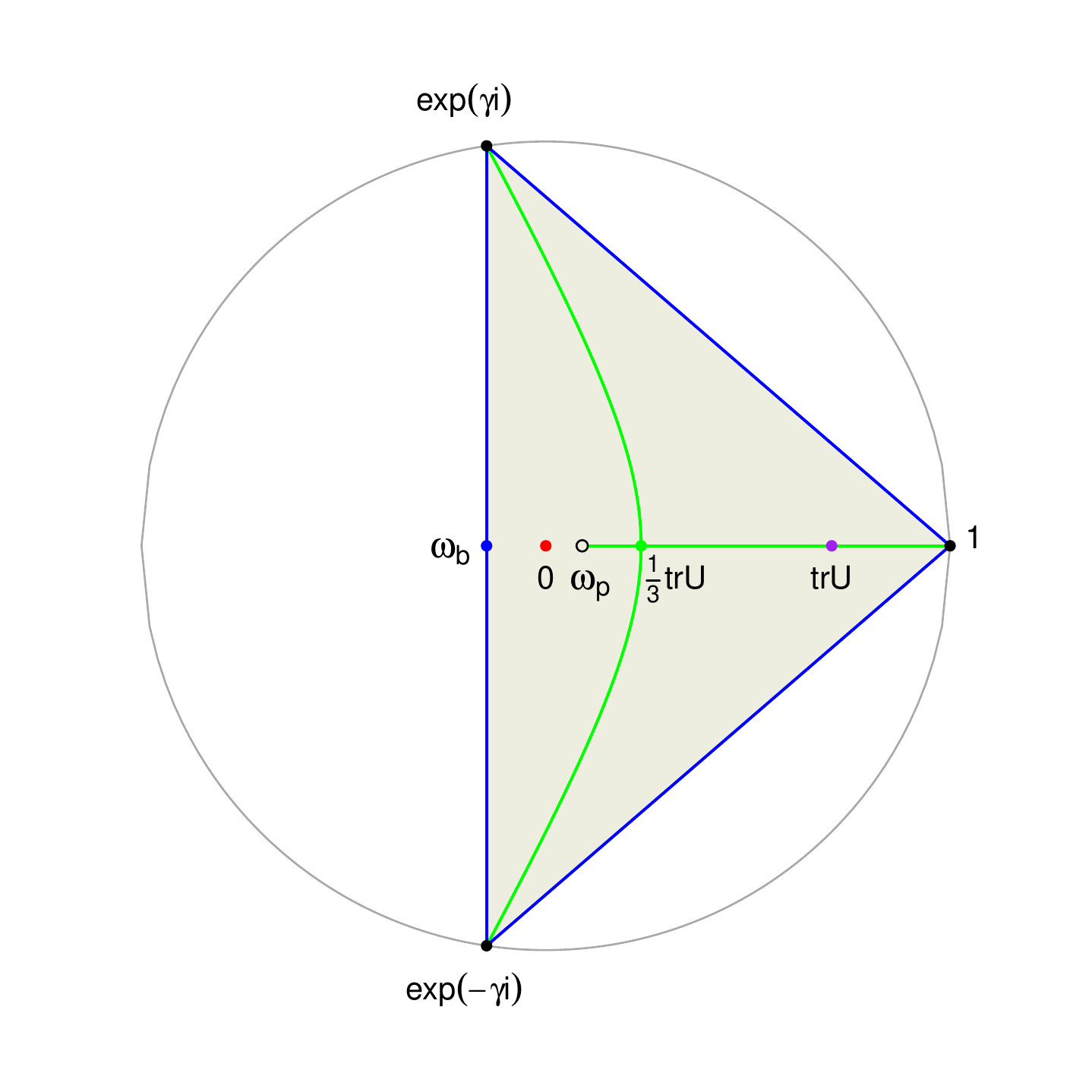}
		
		\vspace{-5mm}
		
		\includegraphics[scale=0.6]{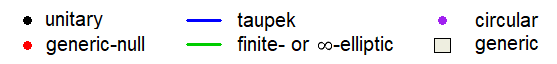}

		\caption{Numerical range of the unitary matrix introduced by Crutchfield \& Wiesner along with the types of Markov chains that can be generated by the ball \& point system whose evolution is governed by this unitary.}
		\label{numrangeUmuss1}
	\end{figure}

	Since  $\sigma(U) =\{1, \exp({\I\gamma}),\exp({-\I\gamma})\}$ with $\gamma$ such that  $\cos\gamma=\tfrac 14(\sqrt{2}-2)$, which gives $\gamma\approx98.42^\circ$, it follows that $\mathfrak{W}(U)$ is an acute-non-equilateral isosceles triangle. 
	Note that $\det U=1$ and $\tr U=\frac{1}{\sqrt{2}}$. Hence,   $\mathcal{F}_{U,\,0}$ generates a generic-null chain and   $\mathcal{F}_{U,\,\frac{1}{\sqrt{2}}}$ generates a circular chain. 
	Let us determine the subset $\mathcal{E}$ of $\mathfrak{W}(U)$, where elliptic chains are generated. 
	Let $\omega = x + \I y \in \operatorname{Int}\mathfrak{W}(U)$ with $x, y\in \R$. From \eqref{hiperb1} and  \eqref{hiperb2} we know that  ${(\tr U-\omega)^2}/{\overline{\omega}}\in \R$ holds on the \mbox{$x$-axis} (apart from the origin) and  on the hyperbola given by 
	\begin{equation}\label{hyperbole}
		18(x-\tfrac{\sqrt{2}}{3})^2-6y^2=1,
	\end{equation}
	which  intersects the $x$-axis at  $( \tfrac{\sqrt{2}}2,0) \sim \tr U$ and  $(\tfrac{\sqrt{2}}6,0) \sim \frac 13 \tr U$. 
	First, we examine the \mbox{$x$-axis}. Let $\omega \in \mathfrak{W}(U)\cap \R$ and put 
	$\Delta_\omega:= {{ \omega}}^{2} - (4+\sqrt {2}){\omega}+\tfrac 12$. It follows easily  that
	${(\tr U-\omega)^2}/{\overline{\omega}}\in [0,4)$ iff 
	$\Delta_\omega <0$ iff $\omega\in (\omega_p,1]$, where 
	$\omega_p:=\tfrac{1}{2}\sqrt{2}+2-\sqrt{\smash[b]{4+2\sqrt{2}}}$. We conclude that $\mathcal{F}_{U,\,\omega}$ generates an elliptic chain iff  $\omega \in (\omega_p,1)$.  
	Next, we examine the hyperbola. Let $\omega = x + \I y$ with $x, y\in \R$. Then ${(\tr U-\omega)^2}/{\overline{\omega}}\in [0, 4)$ 
	is equivalent~to
	\begin{equation*}
		0 \leq x^3-\sqrt{2}x^2+\tfrac 12 x+y^2(\sqrt{2}-3x) < 4 (x^2+y^2)
	\end{equation*}
	which holds on the hyperbola   \eqref{hyperbole} iff 
	$x \in (\tfrac{\sqrt{2}}2-2,\, \tfrac 13 \tr U \,]\cup \{\tr U\}$. Restricting to 
	$\mathfrak{W}(U)$, we get $x \in (\omega_b,\, \tfrac 13 \tr U\,]\cup \{\tr U\}$, where $\omega_b:=\cos \gamma \approx - 0.146$. 
	Hence, $\mathcal{E}$ consists of the  segment $(\omega_p, 1)$  in the \mbox{$x$-axis} and of the part of a~hyperbola branch contained between the vertices $\exp(\pm \I \gamma)$. 
	In Fig.~\ref{numrangeUmuss1} we show all chain types that can be generated by $\mathcal{F}_{U,\,\omega}$ for $\omega \in \mathfrak{W}(U)$.

	Let us compute the eigenvalues $\lambda_1$, $\lambda_2$ of $A_\omega$ along the $x$-axis.
	Let \mbox{$\omega \in \mathfrak{W}(U)\cap \R = [\omega_b, 1]$}. The  discriminant of the 
	characteristic polynomial of $A_\omega$, which reads
	\begin{align*}
		p_A(\lambda)
		&  
		=\lambda^2-(\tr U - \omega) \lambda + \det U \overline{\omega} 
		\\[0.25em] 
		&
		=\lambda^2-\big(\tfrac{1}{\sqrt{2}}-\omega\big) \lambda + \omega
		,\end{align*}
	is given by $\Delta_\omega$. 
	It follows that
	\begin{enumerate}[(i),leftmargin=12.5mm]
		\itemsep= -0.5mm
		
		\item  if $\omega\in [\omega_b,\omega_p)$, then $\Delta_\omega>0$ and 
		$\lambda_{1,2}=\tfrac{1}{2\sqrt{2}}-\tfrac 12 \omega \pm \sqrt{\Delta_\omega}$; in particular, $|\lambda_{1}|\neq |\lambda_{2}|$;

		\item  if $\omega=\omega_p$, then $\Delta_\omega=0$ and  $\lambda_1 =	\lambda_2= \frac {1}{\sqrt{2}}\sqrt{\smash[b]{2+\sqrt{2}}}-1$;

		\item  if $\omega\in (\omega_p,1]$,
		then $\Delta_\omega<0$ and  
		$\lambda_{1,2}=\tfrac{1}{2\sqrt{2}}-\tfrac 12 \omega \pm \I
		\sqrt{|\Delta_\omega|}$;
		in particular, $\lambda_{1}=\overline{\lambda_{2}}$, and so $|\lambda_{1}| = |\lambda_{2}|$;
	\end{enumerate}

	The  following table summarises  how
	the properties of the system, in particular the type of dynamics   induced on the Bloch sphere, depend on $\omega$ varying between $\omega_b$ and $1$.
	
	\vspace{-1mm}	
	
	\begin{table}[H]
		\renewcommand{\arraystretch}{1.33}
		\scalebox{1.0}{
			\begin{tabular}{  c | c |c|c }
				&  & dynamics   &  type of chain    \\[-1.5mm]
				value of	$\omega$	& eigenvalues of $A_\omega$ &  $\hphantom{i}$induced by $A_\omega\hphantom{i}$  & generated by $\mathcal{F}_{U,\,\omega}$      \\
				\hline  
				\hline
				$\hphantom{i}\omega_b\approx -0.146\hphantom{i}$  & $\lambda_1=1$, $\lambda_2=\omega_b$ & loxodromic   &  taupek  
				\\ \hline
				$(\omega_b, 0)$   & $\lambda_2<0<\lambda_1$ & loxodromic   &  generic 
				\\[-1mm] 
				& $|\lambda_1|\neq|\lambda_2|$ &  &     
				\\
				\hline
				$0$ & $\lambda_1=\tr U$, $\lambda_2=0$ & singular   & generic-null 
				\\\hline
				$(0,\omega_p)$  & $0<\lambda_2<\lambda_1$  & loxodromic  &  generic
				\\\hline
				$\omega_p \approx 0.094$& $\lambda_1=\lambda_2$ & parabolic   &  generic 
				\\\hline
				$(\omega_p,\tr U)$  & $\lambda_1=\overline{\lambda_2}$ & elliptic   &   finite- or $\infty$-elliptic 
				\\\hline
				$\tr U =\tfrac 12 {\sqrt{2}}$  & $\hphantom{i}\lambda_1=-\lambda_2=\I2^{-1/4}\hphantom{i}$ & circular    & circular  
				\\\hline	
				$(\tr U,1)$  & $\lambda_1=\overline{\lambda_2}$ & elliptic    &   $\hphantom{i}$finite- or $\infty$-elliptic$\hphantom{i}$
				\\\hline
				$1$  & $\lambda_1=\overline{\lambda_2}=\e^{\I\gamma}$ & elliptic  & unitary 
		\end{tabular} }
	\end{table}

	\section{Acknowledgments}
	The author is grateful to Wojciech Słomczyński for numerous  comments that substantially improved this manuscript. 
	
	{  
		\renewcommand*{\bibfont}{\small }
		\setlength{\bibitemsep}{0.25\baselineskip}         
		\printbibliography
	}

\end{document}